%% file: elsarticle-template-num.tex
\documentclass[preprint,12pt,nopreprintline]{elsarticle}

\usepackage{amssymb}
\usepackage{amsmath}
\usepackage{bm}
\usepackage{cleveref}
\usepackage{braket}
\usepackage{quantikz}
\usepackage{tabularx}
\usepackage{booktabs}
\usepackage{pgfplots}
\usepackage{xcolor}
\usepackage{gensymb}
\usepackage{soul}

\definecolor{colour1}{HTML}{045275}
\definecolor{colour2}{HTML}{089099}
\definecolor{colour3}{HTML}{7CCBA2}
\definecolor{colour4}{HTML}{FCDE9C}
\definecolor{colour5}{HTML}{F0746E}
\definecolor{colour6}{HTML}{DC3977}
\definecolor{colour7}{HTML}{7C1D6F}

\definecolor{TUMOrange}{RGB}{227,114,34}
\definecolor{TUMGreen}{RGB}{162,173,0}
\definecolor{RevPurple}{RGB}{112,48,160}
\definecolor{TUMBlack}{RGB}{0, 0, 0}
\newif\ifshowrevisions
\showrevisionsfalse
\ifshowrevisions
  \newcommand{\revone}[1]{{\color{TUMOrange}#1}}
  \newcommand{\revtwo}[1]{{\color{TUMGreen}#1}}
  \newcommand{\revthree}[1]{{\color{TUMBlack}#1}}
\else
  \newcommand{\revone}[1]{#1}
  \newcommand{\revtwo}[1]{#1}
  \newcommand{\revthree}[1]{#1}
\fi

\usetikzlibrary{patterns,calc,plotmarks}
\usepgfplotslibrary{groupplots,fillbetween,statistics}

\begin{document}
\setstcolor{TUMOrange}
\begin{frontmatter}



\title{A Quantum-Classical Surrogate Model for the Collision Operator of the Lattice Boltzmann Method}


\author[label1]{Lukas C. Birk}
\author[label2]{David M. Wawrzyniak}
\author[label1]{Josef M. Winter}
\author[label1]{Steffen J. Schmidt}
\author[label1]{Thomas Indinger}
\author[label3]{Christian F. Janßen}
\author[label1,label2]{Nikolaus A. Adams}

\affiliation[label1]{organization={Technical University of Munich, School of Engineering and Design, Department of Engineering Physics and Computation, Chair of Aerodynamics},
            addressline={Boltzmannstr. 15}, 
            city={Garching},
            postcode={85748}, 
            state={Bavaria},
            country={Germany}}

\affiliation[label2]{organization={Technical University of Munich, Munich Institute of Integrated Materials, Energy and Process Engineering},
            addressline={Lichtenbergstr. 4a}, 
            city={Garching},
            postcode={85748}, 
            state={Bavaria},
            country={Germany}}

\affiliation[label3]{organization={Siemens Digital Industries Software},
            addressline={5800 Granite Parkway}, 
            city={Plano},
            postcode={75024}, 
            state={TX},
            country={United States}}

\begin{abstract}
We introduce a hybrid approach utilising a quantum machine learning surrogate model to approximate the non-linear collision dynamics of the LBM. 
It offloads the non-unitary operations that challenge pure quantum solvers.
The surrogate's expressivity relies on variational quantum circuits (VQCs) that implement partial Fourier series, while data re-uploading extends the representable frequency spectrum.
\revtwo{
Unlike previous approaches that use a fixed relaxation parameter, the relaxation step is embedded analytically in the circuit, allowing the learned equilibrium to be used for any relaxation parameter without retraining.
}\par
We reassess standard VQC metrics, including expressibility, entanglement, and effective dimension, by relating them directly to task-specific surrogate performance and identifying the architectural parameters that determine approximation accuracy.
The proposed surrogate is validated against the classical BGK collision operator using benchmark problems, including the Taylor--Green vortex for energy dissipation and the double shear layer for shear-driven instabilities and nonlinear flow evolution.
\revtwo{Our results show that the surrogate closely reproduces the BGK reference on both benchmarks, using a single set of trained circuits across all relaxation parameters.} 
These findings suggest that hybrid quantum-classical strategies offer a practical path toward realising the potential of quantum computing in fluid engineering.
\end{abstract}



\begin{keyword}
Quantum Computing \sep Quantum Computational Fluid Dynamics \sep Lattice Boltzmann Method \sep Collision Operator


\end{keyword}

\end{frontmatter}


\section{Introduction}
\label{chapter:Introduction}
Significant progress has been achieved in \revthree{developing} Noisy Intermediate-Scale Quantum (NISQ) devices.
NISQ devices typically consist of approximately 10 to 100 qubits, with interactions restricted by the underlying hardware topology rather than being fully connected.
Furthermore, these devices have limited error correction, resulting in significant computational errors.
Despite these limitations, NISQ devices still have the potential to demonstrate the advantages of quantum computing.
Well-established quantum algorithms (QAs) that achieve a proven advantage over classical counterparts include Shor’s factorisation algorithm \cite{Shor.1997}, Grover’s search algorithm \cite{Grover.1996}, and the Harrow–Hassidim–Lloyd (HHL) algorithm \cite{Harrow.2009} for solving linear systems of equations.
These successes motivate extending QA techniques to more complex computational problems, including computational fluid dynamics (CFD).\par
Previous approaches to solving the \revthree{Navier--Stokes} equation on a quantum architecture include the work of Ray et al. \cite{Ray.2019}, who reformulated the problem as a binary optimisation task on an adiabatic annealing-based quantum computer.
Steijl et al. \cite{Steijl.2020, Steijl.2018} explored a hybrid quantum-classical computer approach using a particle-mesh vortex-in-cell method, where vorticity transport is coupled with Poisson equations for velocity solved via the quantum Fourier transform.
Gaitan \cite{Gaitan.2020} introduced another approach that formulates a steady one-dimensional inviscid nozzle flow as an ordinary differential equation (ODE) via spatial discretisation and solves it with a quantum ODE algorithm.
Later, Oz et al. \cite{Oz.2022} adopted Gaitan's algorithm to address the 1D Burgers equation.\par
An alternative to macroscopic continuum CFD is the Lattice Boltzmann Method (LBM).
It discretises space, time, and velocity on a lattice, where particle distributions propagate with discrete lattice velocities and interact.
The potential use of quantum lattice-gas models first emerged in the 1990s, with pioneering research by Yepez \cite{Yepez.2001} and Berman et al. \cite{Berman.2002}. 
The variables at each lattice node are encoded on dedicated qubits, which interact within a large parallel array of quantum processors, exchanging information with other lattice nodes via classical communication channels.
A more recent approach by Todorova et al. \cite{Todorova.2020} applies a reservoir‑based time‑integration method \cite{Alouges.2008} to the collisionless Boltzmann equation, thereby reducing memory and measurement demands while still relying on multiple realisations to attain the desired level of detail.
Schalkers et al. \cite{Schalkers.2024b} advanced the field further with a QA for the collisionless Boltzmann equation, where inspiration from the quantum Draper Adder \cite{Draper.2000} enables a significant reduction in the number of two-qubit gates required for the streaming step.
Moreover, they introduced a novel velocity-vector encoding and a quantum specular reflection step at boundaries.\par
Budinski \cite{Budinski.2021} proposed the first quantum LBM algorithm for solving the advection-diffusion equation on one- and two-dimensional lattices.
In a subsequent paper, Budinski \cite{Budinski.2022} extended the approach to a vorticity-streamfunction formulation for the linear advection-diffusion equation.
Wawrzyniak et al. \cite{Wawrzyniak.2025} introduced a QA for the advection-diffusion equation LBM, enabling the computation of multiple time steps by using dynamic circuits.
This approach combines unitary evolution with mid-circuit measurements to adaptively update the quantum state without state reinitialisation.\par
Quantum computers are naturally adept at executing linear operations, owing to the fundamentally unitary structure of quantum mechanics.
While streaming between lattice nodes in the LBM is linear, the local collision term introduces non-linear dependencies. 
Because gate-based \revthree{quantum computers employ} unitary operators, a growing body of research is directed towards incorporating non-linearities into quantum algorithms.
Sanavio et al. \cite{Sanavio.2024} used Carleman linearisation \cite{Carleman.1932} to embed the non-linear lattice Boltzmann dynamics into a higher-dimensional linear system.
For Kolmogorov-like flows at Reynolds numbers up to 100, they found that Carleman linearisation yields acceptable accuracy.
Building on this, they designed an explicit quantum circuit for a \revthree{single time step} collision operator with fixed depth independent of the number of lattice sites, thereby mapping the non-linear collision into a structured sequence of gates. 
However, this depth still amounts to on the order of ten thousand quantum gates per \revthree{time step,} so the scheme remains out of reach for present-day hardware despite providing a fully linear framework for non-linear fluid dynamics.\par
\revone{
Jennings et al. \cite{Jennings.2026} critically examined such Carleman-based schemes. They derived fault-tolerant resource estimates for an algorithm that embeds all time steps of the incompressible LBE into a single linear system.
Their analysis shows that the Carleman expansion diverges beyond a threshold Reynolds number and that any quantum advantage is at most polynomial and limited to scalar observables, such as the drag force, whereas it is lost for full flow-field readout.
Wang et al. \cite{Wang.2025} circumvented the Carleman embedding by replacing the non-linear BGK collision with a node-level lattice-gas ensemble, whose collision acts linearly but whose state space grows exponentially with the number of discrete velocities.
Although transport and collision proceed coherently over many time steps, an additional H-step is required to keep the ensemble near equilibrium.\par
}
Another emerging concept, \revthree{variational quantum computing,} has recently gained significant attention for its potential to efficiently tackle non-linear problems.
Variational Quantum Algorithms are hybrid approaches in which a loss function ${L(\vec{\theta})}$ is evaluated on a quantum computer. 
The optimisation of variational parameters $\vec{\theta}$ is executed on a classical computer.
Noise is mitigated by maintaining shallow quantum circuit depths, offering a more immediate path to achieving quantum advantage.\par
Itani \cite{Itani.2025} trained a \revthree{variational quantum circuit (VQC)} to map from pre- to post-collision populations.
The distribution functions are encoded either as their fixed-point binary representation or as a trigonometric function in the amplitude $\ket{f_i}$, together with a velocity-direction register, $\ket{\bm{e}_i}$, and a grid-position register, $\ket{\bm{x}}$.
The collision unitary is composed of trained rotation gates and CNOT gates. 
Symmetry is enforced by applying rotation- and reflection-transformed versions of the circuit in superposition using ancillary qubits, and subsequently tracing out the ancilla register to obtain the averaged transformation.
Conservation of mass and momentum is imposed through the training loss function.
After each \revthree{time step,} the local distribution functions are measured and used to reprepare the quantum state for the subsequent \revthree{time step.} 
However, the learned collision operator remains accurate only above a certain velocity threshold. 
Below this threshold, the relative prediction error exceeds the magnitude of the collision update $f_i^{*} - f_i$.\par
Lacatus et al. \cite{Lacatus.2025} likewise realised the BGK collision operator with a trained VQC, using a rooted-density encoding where amplitudes encode $\sqrt{f_i/\rho}$ and measurements of the velocity register recover $f_i$ as probabilities.
Constructed from globally applied single-qubit rotations with trainable angles and Ising-type entangling layers, the collision operator is exactly \revthree{$D_4$-equivariant} at the layer level, with mass conservation inherently preserved by the unitary operations. 
By contrast, momentum conservation is enforced via penalty terms in the loss function.
For subsequent time steps, the post-collision distribution functions are extracted through computational basis measurements.
Both approaches demonstrate that learned unitary collision circuits can reproduce the non-linear BGK relaxation with controllable accuracy in the low-Reynolds-number regime; however, this capability is shown only for a fixed relaxation time ($\tau=1$), and extension to other relaxation parameters requires retraining.\par
In the following, we introduce a novel hybrid quantum machine learning surrogate model tailored for NISQ hardware to approximate the complete post-collision distribution of the lattice Boltzmann method.
The circuit adheres to established NISQ-friendly design principles \cite{Preskill.2018,Cerezo.2021}.
It employs a variational ansatz consisting of angle-encoding layers and parameterised entangling layers built exclusively from single-qubit rotation and CNOT gates, thereby keeping circuit depth shallow and decoherence minimal.
At inference time, only the expectation of a single-qubit observable is required, avoiding the exponential cost of full state-vector readout and making the approach tractable on current hardware.
Central to our framework is the angle-encoding scheme, whose Fourier-analytic structure provides a systematic and physically motivated basis for representing the distribution functions.
\revtwo{
The surrogate is not intended to accelerate the BGK collision, whose closed form is cheaper than any surrogate, classical or quantum.
Instead, it addresses the missing component of an end-to-end quantum LBM, showing that the non-linear equilibrium with an explicit relaxation can be represented by a shallow circuit.
}\par
This work makes the following contributions.
\revtwo{
To our knowledge, we propose the first hybrid quantum-classical surrogate that reconstructs the complete BGK collision by learning the equilibrium and imposing the relaxation exactly for arbitrary relaxation parameters.
}
Established VQC analysis \revthree{tools, namely expressibility, entanglement and effective dimension,} are applied to the circuit design process, and their predictive validity is critically assessed for fluid-dynamics applications.
A problem-specific performance analysis further investigates the surrogate's interpolation and extrapolation \revthree{behaviour and its} sensitivity to circuit depth and width, identifying the key architectural parameters that govern accuracy.
To the best of our knowledge, this work also presents the first three-dimensional high-fidelity quantum LBM surrogate simulation, validated on the Taylor--Green vortex benchmark for two relaxation parameters, and demonstrates the applicability of the framework to moderate Reynolds number flows through simulations of the \revthree{double shear layer.\par}
The paper is organised as follows: \cref{chapter:Methodology} recalls the LBM and quantum circuit learning foundations; \cref{chapter:VQC} develops the variational circuit design and its Fourier-series interpretation; the circuit analysis and model extensions are detailed in \cref{chapter:Analysis,chapter:Extensions}; numerical validation is presented in \cref{chapter:Results}; and conclusions together with an outlook are given in \cref{chapter:Conclusion}. 
\section{Methodology}
\label{chapter:Methodology}
\subsection{The Lattice Boltzmann Method}
The LBM originates from kinetic theory, where the fundamental quantity is the particle distribution function $f(\bm{x}, \bm{\xi}, t)$, representing the density of particles with velocity $\bm{\xi}$ at position $\bm{x}$ and time $t$.
The Boltzmann equation (BE) \cite{Boltzmann.1872,Chapman.1990} governs its evolution as
\begin{equation} \label{eq:2.1}
    \frac{\partial f}{\partial t} + \xi_\beta \frac{\partial f}{\partial x_\beta} + \frac{F_\beta}{\rho} \frac{\partial f}{\partial \xi_\beta} = \Omega(f) \text{,}
\end{equation}
where $\Omega(f)$ is the collision operator accounting for the redistribution of $f$ due to particle interactions.
Discretising the phase space of the BE \cite{Abe.1997,He.1997} yields the lattice Boltzmann equation (LBE) \cite{McNamara.1988,Succi.1991,Mohamad.2011,Kruger.2017}
\begin{equation} \label{eq:2.2}
    f_i (\bm{x} + \bm{c}_i \Delta t, t + \Delta t) - f_i (\bm{x}, t) = \Delta t \, \Omega_i (\bm{x}, t) \text{,}
\end{equation}
where $f_i(\bm{x}, t)$ is the population associated with discrete velocity $\bm{c}_i$.
The specific choice of velocity set (e.g., D2Q9 or D3Q19) determines the lattice topology; the exact values of $\bm{c}_i$ and weights $w_i$ can be found in Succi \citep{Succi.2001}.
The macroscopic moments, density $\rho$ and momentum $\rho \bm{u}$, are recovered as
\begin{equation} \label{eq:2.3}
    \rho = \sum_i f_i, \,\,\,\,\,\,\,\,\,\,\,\,\,\,\,\,\,\,\,\, \rho\bm{u} = \sum_i f_i \, \bm{c}_i \text{.}
\end{equation}
The collision operator $\Omega_i$ in its exact form involves a high-dimensional integral over all possible pre- and post-collision velocities, with non-linear dependencies on the particle distribution functions and the collision cross-section.
To obtain a tractable numerical formulation while preserving the essential relaxation behaviour, simplified models are introduced.
The Bhatnagar–Gross–Krook (BGK) approximation \cite{Bhatnagar.1954} formulates $\Omega_i$ as
\begin{equation} \label{eq:2.4}
    \Omega_i = -\frac{1}{\tau} \,\, (f_i - f^{eq}_i) \text{,}
\end{equation}
representing the relaxation of $f_i$ towards its equilibrium state $f_i^{eq}$ with relaxation time $\tau$.
The equilibrium distribution function with macroscopic velocity field $\bm{u}$, weights $w_i$, and speed of sound $c_s$ is given by
\begin{equation} \label{eq:2.5}
    f^{eq}_i (\bm{x}, t) = w_i \rho \left[ 1 + \frac{\bm{u} \cdot \bm{c}_{i}}{c_s^2} + \frac{(\bm{u} \cdot \bm{c}_{i})^2}{2 c_s^4} - \frac{\bm{u} \cdot \bm{u}}{2 c_s^2} \right].
\end{equation}
The LBE is solved in two steps. In the collision step, each population undergoes a local update
\begin{equation} \label{eq:2.6}
    f_{i}^{*}(\bm{x},t) = f_i (\bm{x}, t) - \frac{\Delta t}{\tau} (f_i(\bm{x}, t) - f_i^{eq}(\bm{x}, t)),
\end{equation}
where $f_{i}^{*}(\bm{x},t)$ denotes the post-collision distribution.
\revthree{The relaxation parameter $\omega = \Delta t/\tau$ determines the kinematic viscosity $\nu = c_s^2 \left(1/\omega - 1/2\right) \Delta t$.}
In the streaming step, post-collision populations are propagated along $\bm{c}_i$
\begin{equation} \label{eq:2.7}
    f_{i}(\bm{x} + \bm{c}_{i}\Delta t, t + \Delta t) = f_{i}^{*}(\bm{x},t).
\end{equation}
\subsection{Quantum Circuit Learning}
Quantum Circuit Learning (QCL), first proposed by Mitarai et al. \cite{Mitarai.2018}, leverages parameterised quantum circuits to learn functions from data through a hybrid quantum–classical training loop.
The algorithm then learns predictions $y(\bm{x}_i, \bm{\theta})$ to approximate the target value $f(\bm{x}_i)$ by tuning $\bm{\theta} = (\theta_1, \ldots, \theta_M)$.\par
A typical \revthree{VQC} consists of three components: an encoding unitary $U(\bm{x})$, a trainable unitary $U(\bm{\theta})$, and a measurement of some observable $\hat{O}$.\par
The encoding unitary $U(\bm{x})$ embeds classical information into the Hilbert space. 
Angle Encoding (AE) \cite{Nielsen.2010} maps classical features to qubit rotation angles, assigning up to three independent parameters per qubit.
An arbitrary single-qubit unitary can be decomposed as \cite{Nielsen.2010}
\begin{equation} \label{eq:2.8} \renewcommand{\arraystretch}{1.5}
    U (\alpha, \beta, \gamma, \delta) = e^{i \delta}\begin{bmatrix} e^{-i \frac{\alpha}{2}} & 0 \\ 0 & e^{i \frac{\alpha}{2}}  \end{bmatrix} \begin{bmatrix} \cos \frac{\beta}{2} & -\sin \frac{\beta}{2} \\ \sin \frac{\beta}{2} & \cos \frac{\beta}{2} \end{bmatrix} \begin{bmatrix} e^{-i \frac{\gamma}{2}} & 0 \\ 0 & e^{i \frac{\gamma}{2}}  \end{bmatrix},
\end{equation}
where $\alpha, \beta, \gamma, \delta \in \mathbb{R}$.
The global phase $e^{i \delta}$ is omitted, given its invariance under quantum measurement. 
The remaining matrices represent rotations on the Bloch sphere.
To capture higher-order correlations between input features, entangling layers parameterised by tunable angles $\bm{\theta}$ are incorporated. Their connectivity affects the circuit's expressivity, depth, and trainability.\par
The circuit output is obtained as the expectation of a Hermitian observable $\hat{O}$, which can act on one or multiple qubits
\begin{equation} \label{eq:2.9}
    y(\bm{x}_i, \bm{\theta}) = \bra{\psi_0} U(\bm{x})^{\dagger} U(\bm{\theta})^{\dagger} \hat{O} U(\bm{\theta}) U(\bm{x}) \ket{\psi_0}, 
\end{equation}
where $\ket{\psi_0} = \ket{0}^{\otimes n}$ is the $n$-qubit ground state, with all qubits initialised in the computational basis state $\ket{0}$.
The expectation of the observable thus yields a real quantity with direct physical interpretation.
\section{Variational Quantum Circuit Design}
\label{chapter:VQC}
The proposed quantum circuit is grounded in a Fourier-series interpretation, providing a principled view of quantum circuit learning as a truncated spectral approximation. 
To ensure near-term feasibility, the design relies on established hardware-efficient ansätze tailored to NISQ devices, employing shallow circuits, a small number of qubits, angle-based encoding, and hardware-native quantum operations consisting of single-qubit rotations and two-qubit CNOT gates.
A single-qubit observable yields continuous outputs suitable for hybrid quantum-classical optimisation while remaining compatible with current hardware constraints.
The individual components of the circuit and their theoretical and practical implications are discussed in detail below.
\subsection{A Fourier Series Perspective on Quantum Circuit Learning}
\label{section:Fourier}
Schuld et al. \cite{Schuld.2021} demonstrated that when data is encoded through gates of the form $\mathcal{G} = e^{-ixH}$, where $H$ is some Hamiltonian, the encoded function naturally takes the form of a Fourier series
\begin{equation} \label{eq:3.1}
    f_{\bm{\theta}}(\bm{x}) = \sum_{\bm{\omega} \in \Omega} c_{\bm{\omega} }(\bm{\theta}) e^{i \bm{\omega} \bm{x}} .
\end{equation}
Here, the frequency spectrum $\Omega \subset \mathbb{R}^N$ is determined by the eigenvalues of the data-encoding Hamiltonians, while the circuit determines the Fourier coefficients $c_{\bm{\omega}}(\bm{\theta})$.\par
For a variational quantum circuit, defined as
\begin{equation} \label{eq:3.2}
    \revthree{U(\bm{x}, \bm{\theta}) =  U^{(2)}(\bm{\theta}^{(2)}) \mathcal{G}(\bm{x}) U^{(1)}(\bm{\theta}^{(1)}),}
\end{equation}
encoding a data feature once via a single-qubit \revthree{rotation} restricts the model to representing a Fourier series with a single frequency \cite{Ostaszewski.2021}.
\begin{figure}[!tb]%
    \centering%
    \resizebox{8cm}{!}{%
    \begin{tabular}{ c }
        $(a)$ Line Ansatz (LA) \\
        \\
        \begin{quantikz} 
        \lstick{$\ket{0}$} & \gate{L^{(0)}_{W}} & \gate{L^{(1)}_{SW}} & \gate{L^{(2)}_{SW}} & \ \ldots \ & \gate{L^{(L)}_{SW}}
        \end{quantikz}
        \\
        \\
        \resizebox{10cm}{!}{%
        \begin{quantikz} 
        \ \ldots \ & \gate[]{S(x_1)} \gategroup[steps=5,style={dashed, inner xsep=2pt},background,label style={anchor=mid,yshift=0.1cm}]{{$L^{(1)}_{SW}$}} & \gate[]{W(\bm{\theta}^{(1,1)})} & \ \ldots \ & \gate[]{S(x_N)} & \gate[]{W(\bm{\theta}^{(1,N)})} & \ \ldots \
        \end{quantikz}
        }
        \\
        \\
        \\
        $(b)$ Parallel Ansatz (PA) \\
        \\
        \begin{quantikz} 
        \lstick{$\ket{0}$} & \gate[wires=3]{W(\bm{\theta}^{(0)})} & \gate[]{S(x_0)} \gategroup[3,steps=2,style={dashed, inner xsep=2pt},background,label style={anchor=mid,yshift=0.1cm}]{{$L^{(1)}_{SW}$}} & \gate[wires=3]{W(\bm{\theta}^{(1)})} & & \ \ldots\ & & \gate[]{S(x_0)} \gategroup[3,steps=2,style={dashed, inner xsep=2pt},background,label style={anchor=mid,yshift=0.1cm}]{{$L^L_{SW}$}} & \gate[wires=3]{W(\bm{\theta}^{(L)})} \\
        \wireoverride{n} & \wireoverride{n} & \wireoverride{n} \vdots & \wireoverride{n} & \wireoverride{n} & \wireoverride{n} & \wireoverride{n} & \wireoverride{n} \vdots & \wireoverride{n} \\
        \lstick{$\ket{0}$} & & \gate[]{S(x_N)} & & & \ \ldots\ & & \gate[]{S(x_N)} & 
        \end{quantikz}
        \\
        \\
        \\
        $(c)$ Super-Parallel Ansatz (SPA)\\
        \\
        \begin{quantikz} 
        \lstick{$\ket{0}$} & \gate[wires=7]{W(\bm{\theta}^{(0)})} & \gate[]{S(x_0)} \gategroup[7,steps=2,style={dashed, inner xsep=2pt},background,label style={anchor=mid,yshift=0.1cm}]{{$L^1_{SW}$}} & \gate[wires=7]{W(\bm{\theta}^{(1)})} & & \ \ldots\ & & \gate[]{S(x_0)} \gategroup[7,steps=2,style={dashed, inner xsep=2pt},background,label style={anchor=mid,yshift=0.1cm}]{{$L^L_{SW}$}} & \gate[wires=7]{W(\bm{\theta}^{(L)})} \\
        \wireoverride{n} & \wireoverride{n} & \wireoverride{n} \vdots & \wireoverride{n} & \wireoverride{n} & \wireoverride{n} & \wireoverride{n} & \wireoverride{n} \vdots & \wireoverride{n} \\
        \lstick{$\ket{0}$} & & \gate[]{S(x_N)} & & & \ \ldots\ & & \gate[]{S(x_N)} & \\
        \wireoverride{n} & \wireoverride{n} & \wireoverride{n} \vdots & \wireoverride{n} & \wireoverride{n} & \wireoverride{n} & \wireoverride{n} & \wireoverride{n} \vdots & \wireoverride{n} \\
        \lstick{$\ket{0}$} & & \gate[]{S(x_0)} & & & \ \ldots\ & & \gate[]{S(x_0)} &  \\
        \wireoverride{n} & \wireoverride{n} & \wireoverride{n} \vdots & \wireoverride{n} & \wireoverride{n} & \wireoverride{n} & \wireoverride{n} & \wireoverride{n} \vdots & \wireoverride{n} \\
        \lstick{$\ket{0}$} & & \gate[]{S(x_N)} & & & \ \ldots\ & & \gate[]{S(x_N)} &
        \end{quantikz}
    \end{tabular}
    }
    \caption{Quantum circuit ansätze for an $N$-qubit circuit: $(a)$ Line,  $(b)$ Parallel, $(c)$ Super-Parallel. The initial entangling layer $W(\bm{\theta}^{(0)})$ is followed by $L$ layers, each comprising encoding gates $S(x_n)$ and entangling layers $W(\bm{\theta}^{(l)})$.}%
    \label{fig:re-uploading_ansatz}%
\end{figure}%
To enrich the model's representational capacity, we can extend the accessible frequency spectrum in two ways. 
In a single-layer setting $L=1$, the encoding gate can be repeated $K$ times in parallel on separate qubits.
Alternatively, a multi-layer architecture \revthree{$L > 1$} effectively repeats the encoding operation $K=L$ times in series.
In both cases, the degree of the truncated Fourier expansion scales with the number of data re-encodings $K$ \cite{Schuld.2021, GilVidal.2020}.\par
This data re-uploading strategy was first proposed by Pérez-Salinas et al. \cite{PerezSalinas.2020} to construct a universal quantum classifier.
Building on this idea, Casas et al. \cite{Casas.2023} introduced three distinct circuit architectures.
The Line Ansatz (LA) encodes all data dimensions into a single qubit (\cref{fig:re-uploading_ansatz} $(a)$).
Each layer $L_{SW}^{(l)}$ encodes $N$ data features as per
\begin{equation} \label{eq:3.3}
    L^{(l)}_{SW}(\bm{x}, \bm{\theta})_{LA} = \prod_{n=1}^{N} S(x_n) \,\, W (\bm{\theta}^{(l,n)}).
\end{equation}
While the data features $S(\bm{x})$ encoded in $L^l_{SW}$ do not change, each entangling layer $W(\bm{\theta}^{(l,n)})$ is parameterised with its own parameter set.
\revtwo{Multi-dimensionality arises from the non-commuting sequence of encoding and trainable rotations on a single qubit.
The accessible frequency spectrum grows with the number of encoding repetitions, while the trainable rotations control the Fourier coefficients.}\par
Similarly, the Parallel Ansatz (PA) applies a distinct encoding operator $S(x_n)$ for each data feature (\cref{fig:re-uploading_ansatz} $(b)$).
However, in the PA, these encoding operations are applied simultaneously across multiple qubits
\begin{equation} \label{eq:3.4}
    L^{(l)}_{SW}(\bm{x}, \bm{\theta})_{PA} = \left( \bigotimes_{n=1}^{N} S(x_n) \right) \,\, W (\bm{\theta}^{(l)}).
\end{equation}
Thus, encoding $N$ data features requires $N$ \revthree{qubits, but all features are encoded within a single layer.}
\revtwo{For identical encoding generators and the same number of encodings per feature, the LA and PA access the same frequency spectrum and differ mainly in their available Fourier coefficients \cite{Schuld.2021, Casas.2023}.}\par
The Super-Parallel Ansatz \revthree{(SPA)} extends the PA circuit by incorporating multiple encoding blocks (\cref{fig:re-uploading_ansatz} $(c)$).
Instead of encoding each data feature once per layer, the SPA consists of $P$ encoding blocks.
Entangling operators $W(\bm{\theta}^{(l)})$ are then applied collectively to all qubits in the quantum register, irrespective of the encoding block structure.
Each layer is thus defined as
\begin{equation} \label{eq:3.5}
    L^{(l)}_{SW}(\bm{x}, \bm{\theta})_{SPA} = \left( \bigotimes_{p=1}^{P} \left( \bigotimes_{n=1}^{N} S(x_n) \right) \right) \,\,W (\bm{\theta}^{(l)}).
\end{equation}
\revtwo{The larger Hilbert space gives more freedom to tune the Fourier coefficients independently, which increases expressivity.
The cost is greater circuit depth, more trainable parameters and higher susceptibility to barren plateaus.}
\subsection{Encoding Layer}
A variational quantum circuit with $L$ encoding gates maps a scalar input $x \in \mathbb{R}$ to a quantum state whose output is a finite Fourier series, characterised by a frequency spectrum $\Omega$.
These frequencies are determined by $s_l$ distinct eigenvalues of the $l$-th encoding Hamiltonian $H_l$.
Inserting resolutions of the identity into each encoding layer's eigenbasis, the observable oscillates at frequencies of the form $\Lambda_i - \Lambda_j$. 
Here, $\Lambda_i = \lambda^{i_1}_{1} + \cdots + \lambda^{i_L}_{L}$ accumulates one eigenvalue per layer along independently chosen ket and bra paths, giving
\begin{equation} \label{eq:3.7}
    \Omega = \left\{\, \Lambda_i - \Lambda_j \;\middle|\; 
    i,\, j \in \bigtimes_{l=1}^{L} \, \revthree{\{1, \dots, s_l\}} \,\right\},
\end{equation} 
where $\bigtimes_{l=1}^{L} \{1, \dots, s_l\}$ is the Cartesian product of the per-layer eigenvalue index sets, so that each multi-index $i = (i_1, \dots, i_L)$ independently selects one of the $s_l$ eigenvalues in each layer $l$. 
For single-qubit Pauli rotation gates, which have eigenvalues $\lambda = \pm 1/2$, the frequencies \revthree{$\Lambda_i - \Lambda_j \in \{-L, \dots, L\}$,} so the number of distinct frequencies is $2L+1$ regardless of the number of index pairs.
This shows that naively repeating the same encoding gate does not enrich the frequency spectrum but may instead introduce redundancies \cite{Landman.2022}.\par
Equipping each layer with a trainable scaling parameter $\kappa_l \in \mathbb{R}$ breaks this degeneracy \cite{PerezSalinas.2020, Casas.2023, Shin.2023}, so that the rotation angle of the $l$-th encoding gate becomes $\theta = \kappa_l\, x_n$ and the accessible spectrum of a single input component $x_n$ generalises to
\begin{equation} \label{eq:3.8}
    \Omega_n(\bm{\kappa}) = \left\{\, \sum_{l=1}^{L} \kappa_l
    \left(\lambda_l^{i_l} - \lambda_l^{j_l}\right)
    \;\middle|\; i,\, j \in \bigtimes_{l=1}^{L} \{1, \dots, s_l\} \,\right\} \subset \mathbb{R}.
\end{equation}
Since the per-layer differences $d_l \equiv \lambda_l^{i_l} - \lambda_l^{j_l}$ take values in $\{-1, 0, +1\}$ for Pauli rotations, choosing the components of $\bm{\kappa}$ to be linearly independent over $\mathbb{Q}$, every $d_l$-pattern could lead to a distinct frequency $\omega_n$, expanding each component spectrum up to as many as $3^L$ frequencies.\par
If we encode an $N$-component feature $\bm{x} = (x_0, \dots, x_{N-1})$, the joint frequency spectrum is then the $N$-fold Cartesian product
\begin{equation} \label{eq:3.9}
    \Omega(\bm{\kappa}) = \Omega_0(\bm{\kappa}) \times \cdots \times \Omega_{N-1}(\bm{\kappa}) \subset \mathbb{R}^{N},
\end{equation}
whose elements are frequency vectors $\bm{\omega} = (\omega_0, \dots, \omega_{N-1}) \in \mathbb{R}^N$, of which there are now up to $3^{LN}$.
However, a larger frequency support does not by itself imply greater expressivity, as the trainable block controls far fewer coefficients than the support contains, leaving frequencies not independently tunable \cite{Schuld.2021, Mhiri.2025}.
\subsection{Entangling Layer}
Entangling layers control how quantum basis functions are combined and determine the accessible Fourier coefficients $c_{\omega}$ \cite{Wiedmann.2024}.
Without sufficient entanglement, the circuit's output is constrained to products of marginal, low-order terms.
Excessive expressivity can cause the loss landscapes to become flat, a well-known phenomenon termed \revthree{`barren} plateaus', rendering the model untrainable with classical optimisers \cite{McClean.2018, OrtizMarrero.2021,Holmes.2021,Holmes.2022}.\par
The two entangling strategies used are the Basic Entangling Layer (BEL) and the Strongly Entangling Layer (SEL), as implemented in PennyLane \cite{pennylane}.
\begin{figure}[htb]%
    \centering%
    \resizebox{8cm}{!}{%
    \begin{tabular}{ c }
        $(a)$ BEL \\
        \begin{quantikz}
            & \gate{R(\alpha^{(1)}_1)} \gategroup[4,steps=5,style={dashed, inner xsep=2pt},background,label style={anchor=mid,yshift=0.1cm}]{{$L^{(1)}_{\text{BEL}}$}} & \ctrl{1} & & & \targ{} & \ \ldots \ & \gate{R(\alpha^{(L)}_1)} \gategroup[4,steps=5,style={dashed, inner xsep=2pt},background,label style={anchor=mid,yshift=0.1cm}]{{$L^{(L)}_{\text{BEL}}$}} & \ctrl{1} & & & \targ{} & \\
            & \gate{R(\alpha^{(1)}_2)}  & \targ{} & \ctrl{1} & & & \ \ldots \ & \gate{R(\alpha^{(L)}_2)}  & \targ{} & \ctrl{1} & & & \\
            & \gate{R(\alpha^{(1)}_3)}  & & \targ{} & \ctrl{1} & & \ \ldots \ & \gate{R(\alpha^{(L)}_3)}  & & \targ{} & \ctrl{1} & &\\
            & \gate{R(\alpha^{(1)}_4)}  & & & \targ{} & \ctrl{-3} & \ \ldots \ & \gate{R(\alpha^{(L)}_4)} & & & \targ{} & \ctrl{-3} &
        \end{quantikz}
        \\
        \\
        \\
        $(b)$ SEL \\
        \begin{quantikz}
            & \gate{R(\alpha^{(1)}_1, \beta^{(1)}_1, \gamma^{(1)}_1)} \gategroup[4,steps=5,style={dashed, inner xsep=2pt},background,label style={anchor=mid,yshift=0.1cm}]{{$L^{(1)}_{\text{SEL}}$}} & \ctrl{1} & & & \targ{} & \ \ldots \ & \gate{R(\alpha^{(L)}_1, \beta^{(L)}_1, \gamma^{(L)}_1)} \gategroup[4,steps=5,style={dashed, inner xsep=2pt},background,label style={anchor=mid,yshift=0.1cm}]{{$L^{(L)}_{\text{SEL}}$}} & \ctrl{2} & & \targ{} & & \\
            & \gate{R(\alpha^{(1)}_2, \beta^{(1)}_2, \gamma^{(1)}_2)}  & \targ{} & \ctrl{1} & & & \ \ldots \ & \gate{R(\alpha^{(L)}_2, \beta^{(L)}_2, \gamma^{(L)}_2)}  & & \ctrl{2} & & \targ{} & \\
            & \gate{R(\alpha^{(1)}_3, \beta^{(1)}_3, \gamma^{(1)}_3)}  & & \targ{} & \ctrl{1} & & \ \ldots \ & \gate{R(\alpha^{(L)}_3, \beta^{(L)}_3, \gamma^{(L)}_3)}  & \targ{} & & \ctrl{-2} & & \\
            & \gate{R(\alpha^{(1)}_4, \beta^{(1)}_4, \gamma^{(1)}_4)}  & & & \targ{} & \ctrl{-3} & \ \ldots \ & \gate{R(\alpha^{(L)}_4, \beta^{(L)}_4, \gamma^{(L)}_4)} & & \targ{} & & \ctrl{-2} &
        \end{quantikz}
    \end{tabular}
    }
    \caption{Quantum circuits illustrating two entangling-layer architectures: (a) the Basic Entangling Layer (BEL) and (b) the Strongly Entangling Layer (SEL). Each entangling layer may consist of multiple repetitions of the corresponding BEL or SEL structure.}%
    \label{fig:basic_entangler_layer}%
\end{figure}%
A BEL (\cref{fig:basic_entangler_layer} $(a)$) applies one single-qubit rotation per qubit, followed by a ring of CNOT gates connecting each qubit to its nearest neighbour cyclically. 
The total number of trainable angles is $NL$, where $L$ is the number of layers and $N$ the number of qubits. 
Non-diagonal rotations, such as $R_X$ and $R_Y$, are required to generate entanglement, whereas $R_Z$ introduces only relative phases.
A SEL (\cref{fig:basic_entangler_layer} $(b)$) follows the same structure but applies three rotations around orthogonal axes per qubit, yielding $3NL$ trainable angles. 
\revthree{It additionally includes a range parameter $r$, the distance between the control and target qubits of the CNOT gates, which varies from layer to layer (PennyLane default).}
\subsection{Observable}
Instead of accessing the full probability distribution of the quantum state, the model outputs the expectation of a single-qubit observable.
Estimating a single expectation requires only repeated measurements in the computational basis, whereas reconstructing the complete state or probability distribution scales exponentially with the number of qubits and requires additional measurement bases and post-processing.
The expectation of the Pauli-$Z$ acting on the least significant qubit defines a real-valued, bounded observable that maps the quantum state produced by the circuit to a scalar output in the interval $[-1, 1]$.
This expectation corresponds to $\braket{Z_{i}} = P(0) - P(1)$, where $P(0)$ and $P(1)$ are the probabilities of the qubit being measured in $\ket{0}$ or $\ket{1}$.
\subsection{Quantum Surrogate Model for the Collision Step}
The proposed quantum circuit framework allows us to define a quantum surrogate model for the BGK collision \revthree{step (\cref{eq:2.6}).}
The rotation angles are initially parameterised as uniformly random values $\theta_{a}^{l} \in [-\pi, \pi]$.
Since a single-qubit observable is used, each distribution function $f_i$ requires its own trained parameters $\bm{\theta}_i$, evaluated in a separate circuit while sharing the input parameters $\bm{u}$ across all $f_i$ at a single grid point.
As discussed later in \cref{chapter:OrbitCompression}, this per-direction overhead can be substantially reduced from one per distribution function to one per orbit.
The trained model is then subsequently applied to all grid points.\par
Training data follow the structured synthetic methodology of Corbetta et al. \cite{Corbetta.2023}.
The quantities \revtwo{$\rho$ and $u_0$} are sampled from respective uniform distributions. 
The velocity $\bm{u}$ is determined by the azimuth \revthree{$\varphi \sim U(0, 2 \pi)$} and polar angle \revthree{$\vartheta \sim U(0, \pi)$}, such that
\begin{equation} \label{eq:3.12}
    \revthree{\bm{u} = u_0 (\sin\vartheta \cos\varphi, \sin\vartheta \sin\varphi, \cos\vartheta) \text{.}}
\end{equation}
The equilibrium distribution function $f_i^{eq}(\rho, \bm{u})$ is then evaluated using the sampled macroscopic density and velocity.
To conserve mass and momentum, the non-equilibrium perturbation $f_i^{\prime,\text{neq}} \sim \mathcal{N}(0, \sigma^2)$ with mean $0$ and variance $\sigma^2$ is transformed as
\begin{equation} \label{eq:3.14}
    f_i^{\text{neq}} = f_i^{\prime, \text{neq}} - \revtwo{\frac{1}{Q}} \rho^{\prime} - \revtwo{\frac{1}{K}} \bm{c}_i \cdot (\rho^{\prime} \bm{u}^{\prime}) \text{,}
\end{equation}
where the auxiliary quantities are computed as
\begin{equation} \label{eq:3.15}
    \revthree{\rho^{\prime} = \sum_{i}^{Q} f_i^{\prime, \text{neq}} \text{,} \qquad}
    \revthree{\rho^{\prime} \bm{u}^{\prime} = \sum_{i}^{Q} f_i^{\prime, \text{neq}} \bm{c}_i \text{.}}
\end{equation}
\revtwo{Here, $Q$ denotes the number of discrete velocities and $K$ describes the number of velocities with a nonzero component along a given axis ($K = 6$ for D2Q9, $K = 10$ for D3Q19).}
Finally, the post-collision distribution \revthree{$f_i^{*}$ is computed from $f_i = f_i^{eq} + f_i^{\text{neq}}$ with the BGK update (\cref{eq:2.6})} for \revtwo{$\omega = 1$}\revthree{, for which $f_i^{*} = f_i^{eq}$}.
\revtwo{Since $f_i^{eq}$ is linear in $\rho$, the circuit learns the normalised equilibrium $f_i^{eq}/\rho$ from the velocity $\bm{u}$ alone (\cref{tab:hyperparameters}).
During the simulation, $\rho$ and $\bm{u}$ are computed classically from the pre-collision populations (\cref{eq:2.3}), and the circuit output is rescaled by the local density.}
\section{Circuit Design Analysis} 
\label{chapter:Analysis}
The introduced variational quantum circuit architecture establishes a flexible framework for approximating nonlinear equations within the LBM.
To establish informed design principles for solving non-linear relations, we analyse the circuit ansatz variants in terms of costs, expressibility, and entanglement structure.
We further investigate the data re-uploading scheme by focusing on its ability to learn the three-dimensional equilibrium distribution function as a representative non-linear target.
\subsection{Circuit Costs}
\label{section:Costs}
\begin{figure}[htb]%
    \centering%
    \resizebox{\textwidth}{!}{%
    \begin{tabular}{ c c c }
        \includegraphics{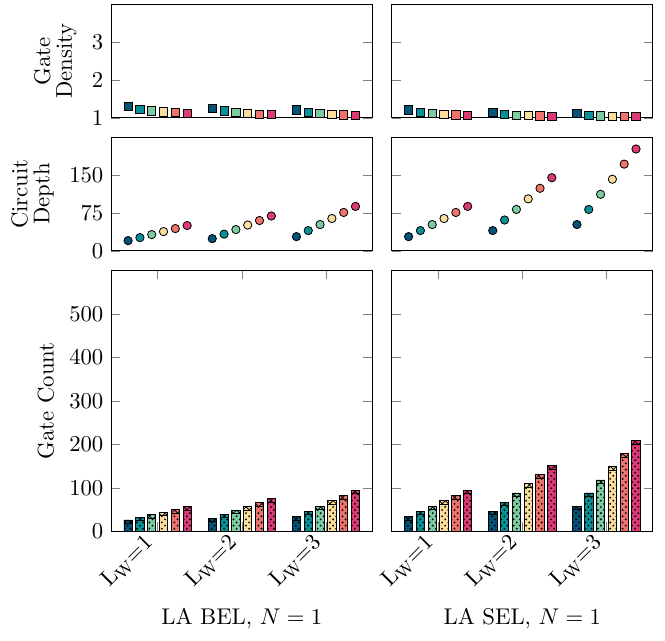}%
        &
        \includegraphics{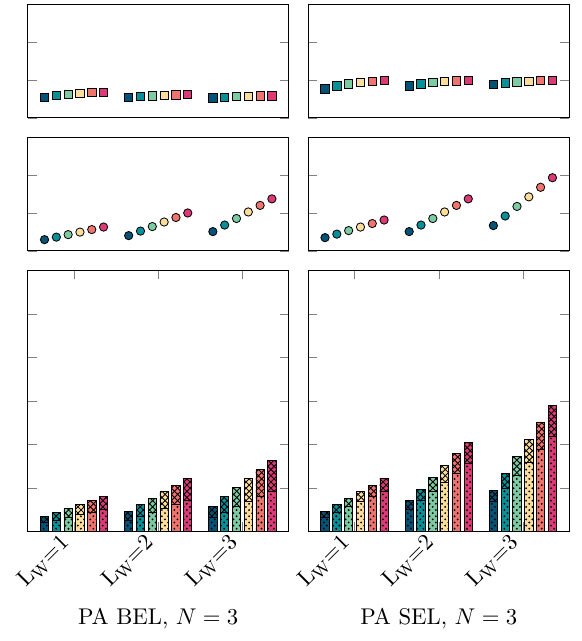}%
        &
        \includegraphics{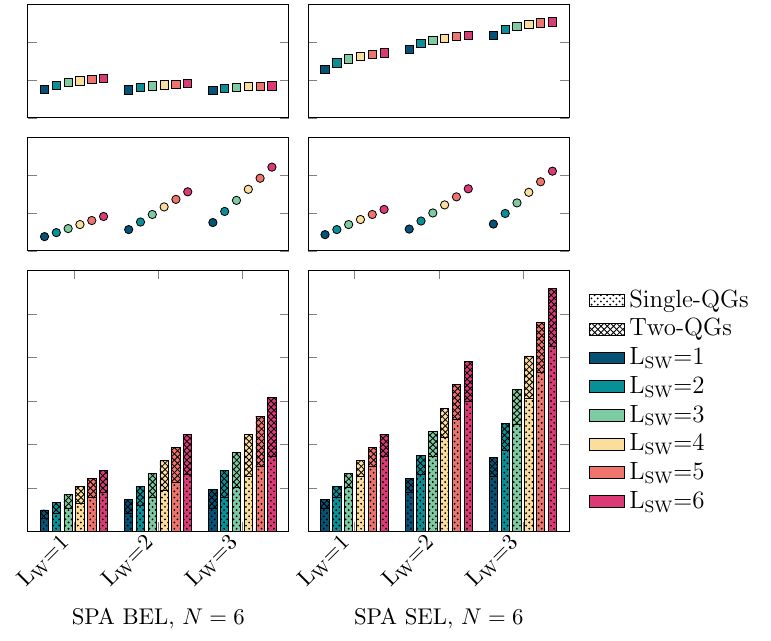}%
    \end{tabular}
    }%
    \caption{
        Gate counts (single-qubit, two-qubit, total), circuit depths, and gate density of VQC variants \revthree{across} varying $L_{\text{W}}$ and $L_{\text{SW}}$.
        The x-axis is categorised by the circuit ansatz (\revtwo{LA,} PA or SPA), the entangling layer type (BEL or SEL), where the trailing integers ($1,2,3$) denote the number of internal entangling layers $L_{\text{W}}$ applied.
        The colour coding indicates the scaling factor for the number of encoding gates ($L_{\text{SW}}$), ranging from $1$ to $6$.
    }%
    \label{fig:costs}%
\end{figure}

\revthree{In addition to the ansätze presented, we} vary the VQC structure \revthree{by} the number of internal entangling layers $L_{\text{W}} \in \{1, 2, 3\}$ and the number of \revthree{encoding-entangling} repetitions \revthree{$L_{\text{SW}} \in \{1, 2, \ldots, 6\}$, as shown in \cref{fig:costs}.}
These parameters directly control the number of trainable parameters, circuit depth, and overall gate count.
$L_{\text{W}}$ governs the depth and parameterisation within each entangling block, whereas $L_{\text{SW}}$ determines the number of alternating encoding and entangling layers and thus the overall circuit depth.\par
\revtwo{Across all variants, gate count and circuit depth grow linearly with $L_{\text{W}}$ and $L_{\text{SW}}$, with SEL requiring more single-qubit gates than BEL owing to its three rotations per qubit (\cref{fig:costs}).
The gate density (ratio of gate count to circuit depth) quantifies gate-level parallelism and is highest for the SPA with SEL. 
This reflects the layer-dependent CNOT range, which allows CNOT gates to be executed in parallel and thus keeps the depth below that of a sequential CNOT chain for $L_{\text{W}} \geq 2$.
The LA itself contains no two-qubit gates. 
The eight CNOT gates shown for the LA stem from the BGK relaxation extension (\cref{fig:relax_circuit}), which is included in all variants and discussed in \cref{chapter:Extensions}.}\par
\revtwo{All circuits are trained on a classical state-vector simulator, whose cost is negligible at these qubit counts.
On quantum hardware, the parameter-shift rule \cite{Mitarai.2018, Schuld.2019b} requires two executions per trainable gate, so each training sample requires $N_{\text{grad}} + 1$ circuit executions (\cref{fig:costs_train_meas} (a)), with $N_{\text{grad}} = 2\,(N_{\text{rot}} + N_{\text{enc}})$ for $N_{\text{rot}}$ trainable rotations and $N_{\text{enc}}$ encoding gates and one additional forward pass.
\begin{figure}[htb]%
    \centering%
    \resizebox{13.5cm}{!}{%
        \includegraphics{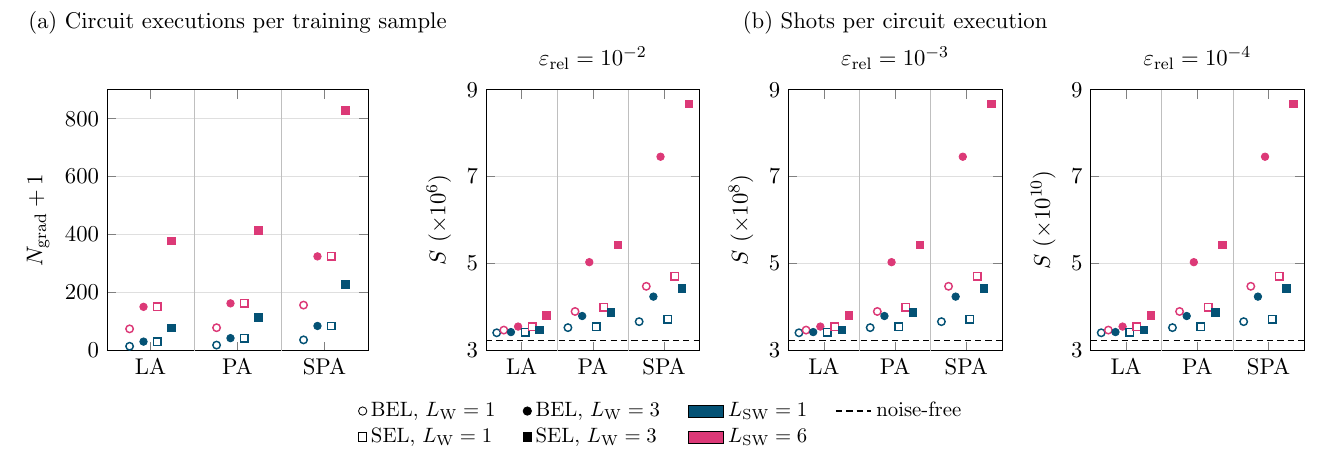}%
    }%
    \caption{\revthree{(a) Circuit executions per training sample, $N_{\text{grad}} + 1$, under the parameter-shift rule and (b) shots per circuit execution $S$ (\cref{eq:cost_shots}) on noisy hardware for target relative accuracies.
    Marker shape denotes the entangler, fill $L_{\text{W}}$, colour $L_{\text{SW}}$ (1 or 6) and dashed lines the noise-free shot count.}}%
    \label{fig:costs_train_meas}%
    \label{fig:TMCosts}%
\end{figure}%
\Cref{fig:costs_train_meas} (a) shows that $N_{\text{grad}} + 1$ grows with $L_{\text{W}}$ and $L_{\text{SW}}$.
LA and PA require nearly the same number of executions, whereas the SPA doubles it by duplicating the encoding and processing gates.\par
The shots per execution $S$ are determined by the target accuracy and the gate errors.
We model the latter by attenuating the expectation value to $f_i^{\text{noisy}} = F f_i$ with the circuit fidelity
\begin{equation} \label{eq:fidelity}
    F = (1 - p_{1\text{q}})^{n_{1\text{q}}} (1 - p_{2\text{q}})^{n_{2\text{q}}}
\end{equation}
for $n_{1\text{q}}$ single- and $n_{2\text{q}}$ two-qubit gates (\cref{fig:costs}) with error rates $p_{1\text{q}}$ and $p_{2\text{q}}$.
As each shot yields $\pm 1$, the sample mean has variance $(1 - (f_i^{\text{noisy}})^2)/S$, so a relative precision $\varepsilon_{\text{rel}}$ requires
\begin{equation} \label{eq:cost_shots}
    S \gtrsim \frac{1 - F^2 f_i^2}{\varepsilon_{\text{rel}}^2 \, F^2 f_i^2}
\end{equation}
shots, where $F = 1$ gives the noise-free limit.
Neglecting coherent and readout errors, this estimates the noise-induced shot overhead rather than device-specific performance.\par
\Cref{fig:costs_train_meas} (b) evaluates \cref{eq:cost_shots} for the face-adjacent D3Q19 population $f_1 \approx w_1 = 1/18$ and median IBM Heron error rates $p_{1\text{q}} = 3.0 \times 10^{-4}$ and $p_{2\text{q}} = 2.7 \times 10^{-3}$ \cite{Yoshioka.2025}.
Since $S \propto \varepsilon_{\text{rel}}^{-2}$, each additional order of magnitude in accuracy increases the number of shots by a factor of $100$, whereas the relative overhead due to gate noise is independent of the target accuracy.
The total shot count for training one discrete velocity is the product of the executions per sample (\cref{fig:costs_train_meas} (a)), the shots per execution (\cref{fig:costs_train_meas} (b)), the number of training samples and the number of epochs.}
\subsection{Expressibility}
The expressibility of a VQC refers to its ability to generate quantum states that approximate the uniform Haar distribution over the state space. 
An approach to quantify expressibility, proposed by Sim et al. \cite{Sim.2019}, is based on comparing the distribution of pairwise fidelities between states produced by the VQC with that of Haar-random states.
A divergence measure, here the Kullback--Leibler (KL) divergence \cite{Kullback.1951}, is used to assess the discrepancy between these two probability distributions.
For two independently drawn Haar-random states $\ket{\psi_i}$ and $\ket{\psi_j}$ in a Hilbert space of dimension $d = 2^N$, the probability density function of fidelity $F_{ij} = \left|\braket{\psi_i \vert \psi_j}\right|^2 \in [0,1]$ is known to be \cite{Zyczkowski.2005}
\begin{equation} \label{eq:4.3}
    p_{\text{Haar}}(F) = (d-1)(1-F)^{d-2}, 
\end{equation} 
which is integrated over each bin $[b_k, b_{k+1}]$ to obtain the expected Haar probabilities.
\revthree{Fidelities are evaluated only for pairs of distinct samples, $\ket{\psi_a}$ and $\ket{\psi_b}$ with $a \neq b$.}
Hence, the expressibility is quantified by the KL divergence
\begin{equation} \label{eq:4.4}
    \revthree{D_{\text{KL}}\!\left(\hat{P}_{\text{VQC}} \,\Vert\, P_{\text{Haar}}\right) = \sum_{k} \hat{P}_{\text{VQC},k} \,\ln \frac{\hat{P}_{\text{VQC},k}}{P_{\text{Haar},k}},}
\end{equation}
where smaller values of $D_{\text{KL}}$ indicate a higher expressibility.\par
\begin{figure}[htbp]%
    \centering%
    \resizebox{\textwidth}{!}{%
    \begin{tabular}{ c c c }
        \includegraphics{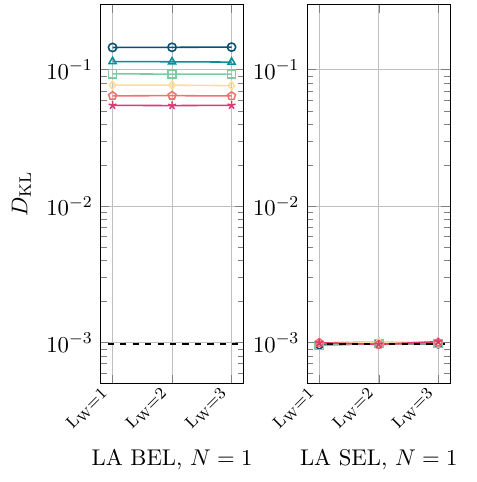}%
        &
        \includegraphics{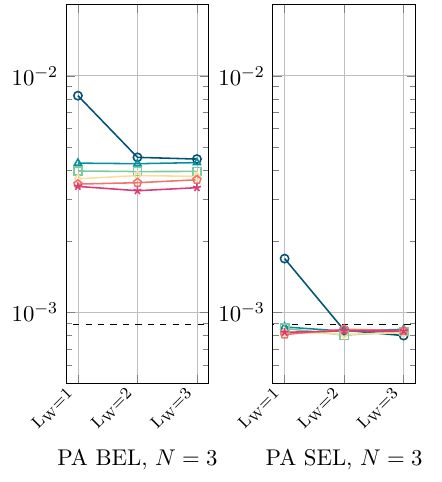}%
        &
        \includegraphics{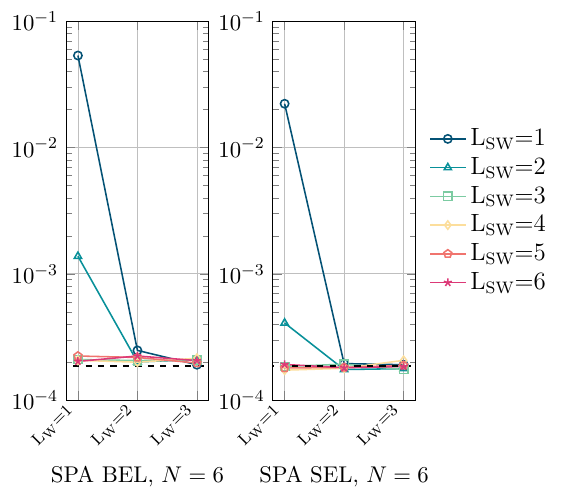}%
    \end{tabular}
    }%
    \caption{
        The KL divergence $D_{\mathrm{KL}}$ is computed for various VQC variants (see \cref{fig:costs} for notation). 
        Since the Haar reference fidelity distribution varies with $N$, \revthree{compare $D_{\mathrm{KL}}$ results only within the same} $N$.
        \revthree{The black dashed line marks the finite-sample Haar floor, obtained by applying the same sampling and binning procedure to Haar-random states.}
    }%
    \label{fig:expressibility}%
\end{figure}

The expressibility in \cref{fig:expressibility} is determined using a sample size of $10^4$ and $100$ bins.
Circuit parameters $\bm{\theta}$ are uniformly sampled from \revthree{$[-\pi, \pi]$, the scaling parameters $\bm{\kappa}$ from $[-1, 1]$,} and the encoding inputs $\bm{x}$ from \revthree{$[-0.3, 0.3]$.}
$D_{\text{KL}}$ computed for circuits with different qubit counts are not directly comparable, because they are defined on probability distributions over different outcome spaces with different cardinality ($2^{N}$).\par
\revtwo{Across all architectures, $L_{SW}$ is the primary driver of expressibility, while additional $L_{W}$ layers contribute only marginally beyond the first.
For the LA and PA, the SEL reaches the Haar floor at shallow depth, whereas the BEL remains well above it.
The SPA variants saturate at lower layer counts and reach the Haar floor for $L_{SW} \geq 3$.
Here, the SEL shows no expressibility advantage over the BEL, suggesting that the additional parameters are redundant when the Hilbert space is already well covered by the larger qubit count.}
\subsection{Entanglement}
\begin{figure}[htbp]%
    \centering%
    \resizebox{0.70\textwidth}{!}{
    \begin{tabular}{ c c }
        \includegraphics{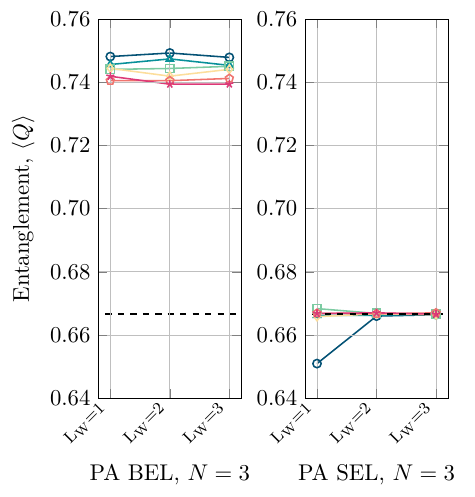}%
        &
        \includegraphics{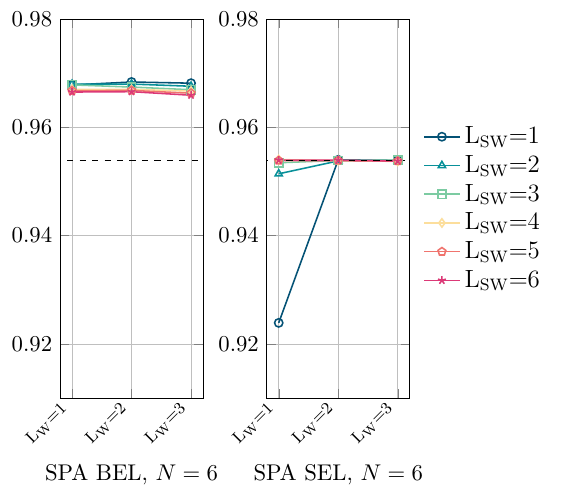}%
    \end{tabular}
    }%
    \caption{\revthree{Entanglement $\langle Q \rangle$ of the PA and SPA variants for varying $L_{W}$ and $L_{SW}$} (see \cref{fig:costs} for notation). The black dashed line shows the mean $\langle Q \rangle_{\text{Haar}}$ for random pure states.}%
    \label{fig:entanglement}%
\end{figure}

The \revthree{Meyer--Wallach} (MW) measure \cite{Meyer.2002}, adopted by Sim et al. \cite{Sim.2019} to define the entangling capability of a VQC, quantifies global entanglement by averaging the mixedness of each qubit's reduced state $\rho_n$ to
\begin{equation} \label{eq:4.5}
    Q(\ket{\psi}) = 2 \left( 1 - \frac{1}{N} \sum_{n=1}^{N} \text{tr} \rho_n^2 \right),
\end{equation}
where $\text{tr} \, \rho_n^2$ is the purity of qubit $n$.
The entanglement of a VQC is then defined as the average of $Q$ over an ensemble $S$ of randomly sampled parameter vectors \cite{Sim.2019}:
\begin{equation} \label{eq:4.6}
    \langle Q \rangle = \frac{1}{|S|} \sum_{\bm{\theta}_i \in S} Q \big( \ket{\psi_{\bm{\theta}_i}} \big),
\end{equation}
and the Haar reference value for $N$-qubit pure random states is \cite{Scott.2003}
\begin{equation} \label{eq:4.7}
    \langle Q \rangle_{\text{Haar}} = \frac{2^N - 2}{2^N + 1}.
\end{equation}
Having a higher or lower entanglement than the Haar value tells us whether the circuit tends to generate more or less multipartite entanglement than a fully pure random state.
In \cref{fig:entanglement}, we show the entanglement $\langle Q \rangle$ with a chosen sample size of $S = 10^4$ \revthree{with $\bm{\theta}$, $\bm{\kappa}$ and $\bm{x}$ sampled as for the expressibility.}
\revtwo{As a single-qubit circuit, the LA generates no entanglement and is therefore omitted here.}\par
\revthree{For PA-BEL, the entanglement is nearly independent of $L_{W}$ and $L_{SW}$ and saturates above $\langle Q \rangle_{\text{Haar}}$.}
This saturation arises from the BEL's single-axis restriction. 
Once the CNOT chain has generated the maximum entanglement compatible with a lower-dimensional submanifold of \revthree{the $N$-qubit state space,} further layers cannot increase the average MW measure.
$R_Y$ rotations generate only real-valued amplitudes on $\ket{0}$, confining the circuit to a submanifold lacking full phase freedom and causing the entanglement to undershoot the Haar value, whereas $R_X$ rotations introduce complex phases that, combined with CNOT gates, push it above.
In contrast, the SEL implements a generic single-qubit unitary per qubit via three independent rotation axes, and progressively approximates a unitary 2-design as depth increases, explaining the observed convergence toward $\langle Q \rangle_{\text{Haar}}$.
\revthree{For the PA-SEL, only $L_{SW} = L_{W} = 1$ remains below the Haar value, which is reached for all deeper circuits.}
The SPA variants show qualitatively similar behaviour, with SEL converging monotonically to the Haar benchmark as early as $L_{W} = 1$.
\subsection{Learnable \revthree{Capacity}}
\begin{figure}[htbp]%
    \centering%
    \resizebox{\textwidth}{!}{%
    \begin{tabular}{ c c c }
        \includegraphics{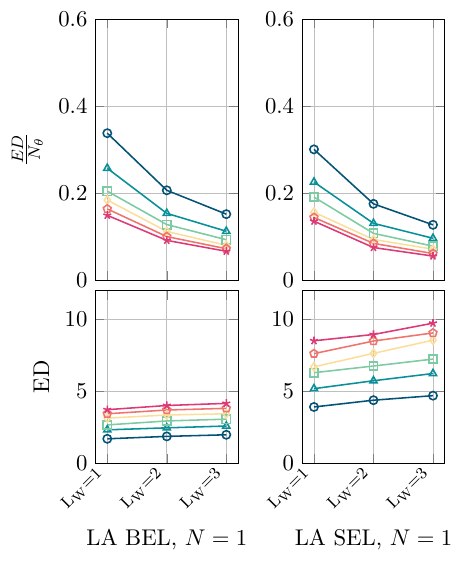}%
        &
        \includegraphics{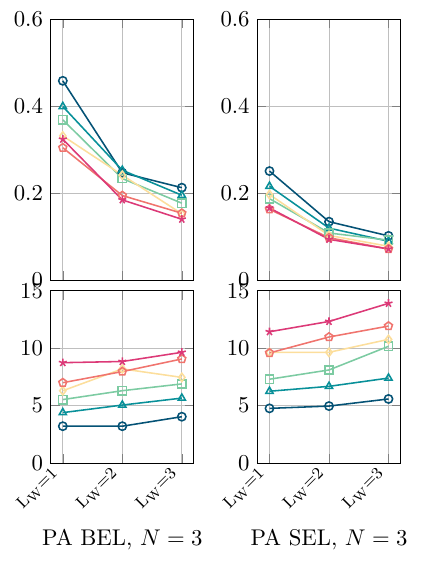}%
        &
        \includegraphics{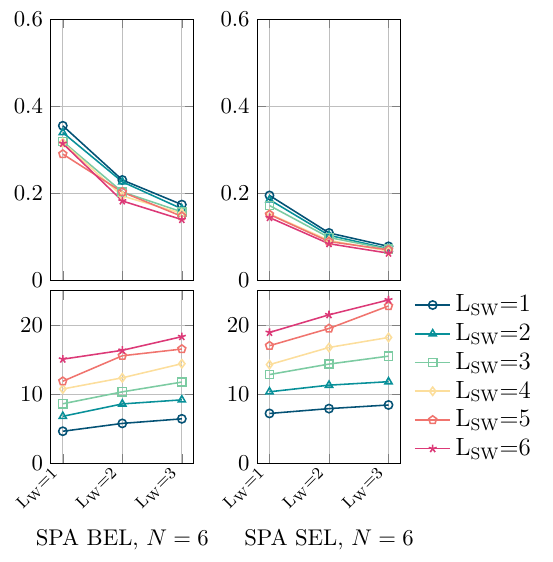}%
    \end{tabular}   
    }%
    \caption{ED and parameter-normalised ED for the \revthree{LA,} PA and SPA architectures with \revthree{$N=1$,} $N=3$ and $N=6$ qubits.}%
    \label{fig:effective_dimension}%
\end{figure}

Abbas et al. \cite{Abbas.2020, Abbas.2021} introduced the Effective Dimension (ED) $d_{\text{eff}}$ as a measure of how many parameters of a model can be independently trained given a finite number of data samples.
\revthree{A $d_{\text{eff}}$ close to the number of parameters indicates independent directions, whereas a} small $d_{\text{eff}}$ signals that many parameters are redundant and the Fisher Information (FI) spectrum is concentrated near zero, a condition explicitly linked to barren plateaus \cite{Abbas.2020}.
The ED of a statistical model $M_{\Theta}$ at sample size $n_s$ is defined as \cite{Abbas.2020,Abbas.2021,Berezniuk.2020} 
\begin{equation} \label{eq:4.8}
    d_{\text{eff}, n_s} (M) = \frac{2 \log \bigg( \frac{1}{V_{\Theta}} \int_{\Theta} \sqrt{\det (I_d + \frac{n_s}{2 \pi \log n_s} \hat{F}(\bm{\theta}))} d\bm{\theta} \bigg)}{\log \big( \frac{n_s}{2 \pi \log n_s} \big)},
\end{equation}
where $V_{\Theta}$ is the volume of the $d$-dimensional parameter space $\Theta \subset \mathbb{R}^d$, $I_d$ is the $d \times d$ identity matrix, and $\hat{F}(\bm{\theta})$ is the normalised FI matrix
\begin{equation} \label{eq:4.9}
    \hat{F}_{ij} = d \frac{V_{\Theta}}{\int_{\Theta} \text{tr}\,F(\bm{\theta})\, d \bm{\theta}} F_{ij} (\bm{\theta}),
\end{equation}
which rescales $F$ so that its average trace equals the number of parameters.
The FI matrix is estimated from $n_k$ i.i.d.\ samples $(\bm{x}_k, \bm{y}_k)$ drawn from the model’s conditional distribution $p(\bm{x}, \bm{y}; \bm{\theta})$ as
\begin{equation} \label{eq:4.10}
    \tilde{F} (\bm{\theta}) = \frac{1}{n_k} \sum_{k=1}^{n_k} \frac{\partial}{\partial \bm{\theta}} \log p(\bm{x}_k, \bm{y}_k; \bm{\theta}) \left(\frac{\partial}{\partial \bm{\theta}} \log p(\bm{x}_k, \bm{y}_k; \bm{\theta})\right)^{T} \in \mathbb{R}^{d \times d}.
\end{equation}
\revtwo{In \cref{fig:effective_dimension}, we show the ED for various architectures, evaluated using $500$ randomly sampled parameter vectors ($\bm{\theta} \in [-\pi, \pi]$, $\bm{\kappa} \in [-1, 1]$), $1000$ input samples $\bm{x} \in [-0.3, 0.3]$, and a theoretical dataset size of $n_s = 10^5$.} 
\revtwo{
Consistent with the expressibility and entanglement analysis, the raw ED grows nearly monotonically with $L_{SW}$ and $L_{W}$, and the multi-axis SEL, which approaches the Haar benchmarks, attains a higher absolute ED than the single-axis BEL across all architectures.
Normalised by the parameter count, however, the ranking reverses and $ED/N_{\bm{\theta}}$ decreases with depth, as the threefold parameter count of the SEL's and each additional layer contributes progressively fewer independent directions in function space.
A decreasing $ED/N_{\bm{\theta}}$ thus reflects a growing share of Fisher eigenvalues close to zero, which Abbas et al. \cite{Abbas.2020} associate with flatter loss landscapes and an increased susceptibility to barren plateaus.
Additional layers and SEL rotations, therefore, enlarge the learnable capacity at the expense of parameter efficiency.
}
\subsection{Problem-specific Performance}
\begin{figure}[htbp]%
    \centering%
    \begin{tabular}{c}
        {\scriptsize \revthree{(a) Line ($N=1$)}} \\
        \resizebox{!}{0.28\textheight}{\includegraphics{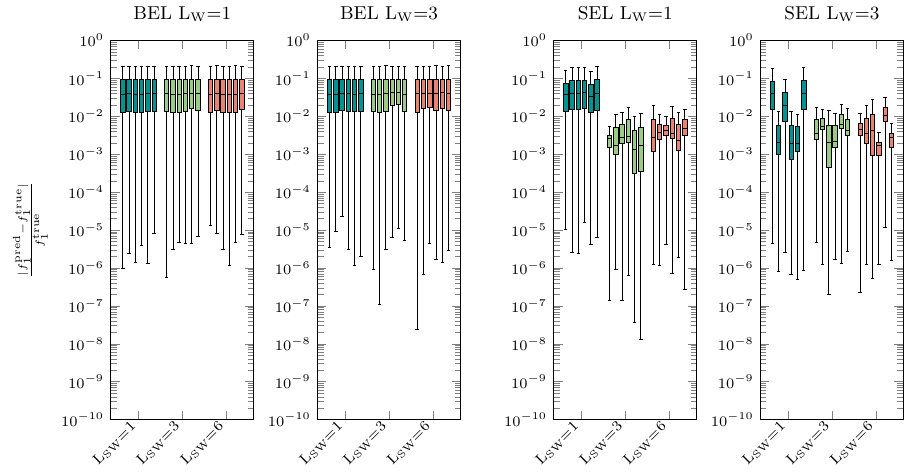}} \\[2pt]
        {\scriptsize \revthree{(b) Parallel ($N=3$)}} \\
        \resizebox{!}{0.28\textheight}{\includegraphics{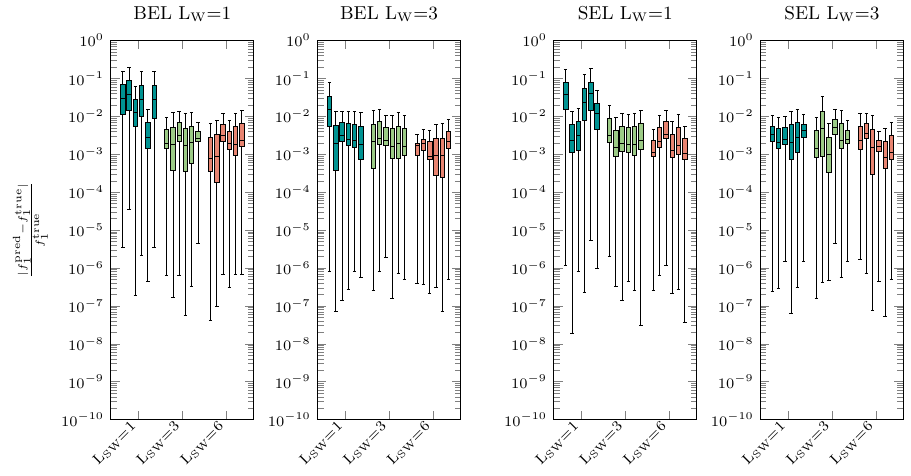}} \\[2pt]
        {\scriptsize \revthree{(c) Super-Parallel ($N=6$)}} \\
        \resizebox{!}{0.28\textheight}{\includegraphics{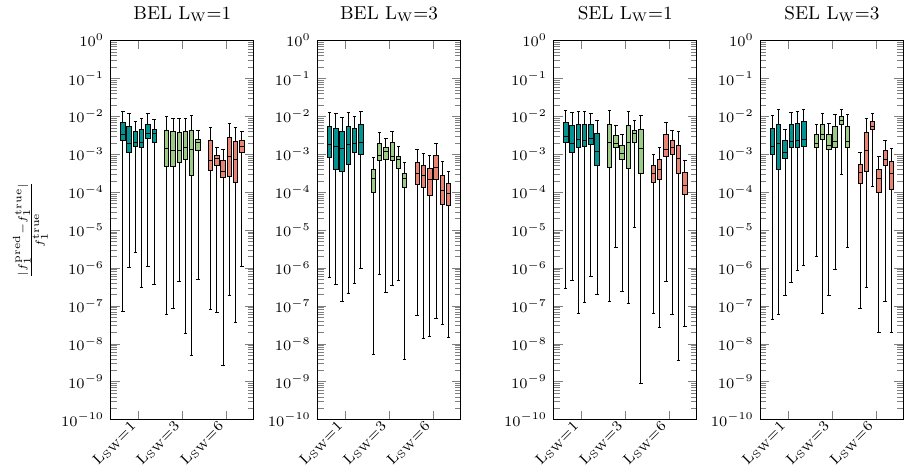}}
    \end{tabular}
    \caption{
    \revthree{Box-and-whisker plots illustrating the relative error distributions for six independent training runs of each circuit configuration.} 
    The central line denotes the median, the box spans the interquartile range from the 25th to the 75th percentiles, and the whiskers indicate the full range of the data.
    }%
    \label{fig:re_variants}%
\end{figure}

\revtwo{
The task-agnostic properties analysed above characterise what a circuit can represent in principle, but not whether this capacity is accessible for a given target.
We therefore assess how well they predict the practical performance of the LA, PA and SPA variants by training the equilibrium distribution function $f^{eq}_{1}$ of the D3Q19 lattice along $\bm{c}_1 = (1,0,0)^T$.
Each variant is trained in six independent runs with different random initialisations of $\bm{\theta}$, using the setup summarised in \cref{tab:hyperparameters}.
The resulting relative error distributions on the held-out test set $\mathcal{D}_\text{test}$ are shown in \cref{fig:re_variants}.
}
\begin{table}[htb]
    \centering
    \resizebox{13.5cm}{!}{%
        \begin{tabular}{ll|ll|ll}
            \toprule
            \textbf{\revthree{Data set}} & Value & \textbf{Hyperparameters} & Value & \textbf{\revthree{Weights}} & \revthree{Value} \\
            \midrule
            $\rho_{\min}$ & $0.98$ & Optimiser & \revthree{Adam ($\beta_1{=}0.9$, $\beta_2{=}0.999$)} & \revthree{$\bm{\theta}$} & \revthree{$\in [-\pi, \pi]$} \\
            $\rho_{\max}$ & $1.02$ & $f_{\text{loss}}$ & L2 & \revthree{$\bm{\kappa}$} & \revthree{$\in [-1, 1]$} \\
            \revthree{$|\bm{u}|_{\max}$} & $0.1$ & $\mathcal{D}_\text{size}$ & $10^{5}$ & & \\
            \revthree{$\sigma_{\text{neq}}$} & $10^{-3}$ & $\mathcal{D} \rightarrow \mathcal{D}_\text{train}, \mathcal{D}_\text{val}, \mathcal{D}_\text{test}$ & $(0.8, 0.1, 0.1)$ & & \\
            $c_s$ & $\sqrt{1/3}$ & $B_{\text{size}}$ & $32$ & & \\
            \revthree{Input} & \revthree{$\bm{u}$} & $\eta_{\text{init}}$ & $10^{-3}$ & & \\
            \revthree{Target} & \revthree{$f_1^{\text{eq}}/\rho$, $\bm{c}_1 = (1,0,0)^T$} & \revthree{$\eta_{\text{scheduler}}$} & \revthree{$\times 0.1$ every 3 epochs} & & \\
            & & \revthree{Epochs (max.)} & \revthree{$9$} & &\\
            & & \revthree{Precision} & \revthree{float64} & &\\
        \end{tabular} 
    }
    \caption{\revthree{Data set, hyperparameters and weights} used for training and evaluating the VQC variants.}
    \label{tab:hyperparameters}
\end{table}
\revtwo{
Accuracy improves from LA to PA and further to SPA, i.e. with circuit width rather than any single task-agnostic property.
Expressibility proves necessary but not sufficient for accurate learning.
The single-axis LA-BEL, whose expressibility saturates above the Haar floor (\cref{fig:expressibility}), fails to learn the target, whereas the LA-SEL reaches the Haar floor and succeeds.
The PA-SEL reaches the Haar floor at the same time as the LA-SEL, yet is more accurate.
As both access the same frequency spectrum (\cref{section:Fourier}), the difference lies in independently tunable Fourier coefficients that control the measured observable.
The amount of entanglement does not determine accuracy.
The LA-SEL learns the target without generating any entanglement, and within the PA, BEL and SEL reach comparable accuracy, although only SEL converges to $\langle Q \rangle_{\text{Haar}}$ (\cref{fig:entanglement}).
The multi-qubit PA and SPA, which can generate entanglement, are nevertheless more accurate than the LA. Since circuit width also increases the number of independently tunable Fourier coefficients, this gain cannot be attributed to entanglement alone.\par
Among the analysed properties, the absolute ED is most informative, as it reproduces the architecture ranking (\cref{fig:effective_dimension}).
It fails, however, to predict the comparison between entangling layers. The higher ED of SEL does not translate into higher accuracy, since the ED is averaged over uniformly sampled parameters, whereas accuracy depends on the region of parameter space reached by the optimiser.
This region is further shaped by trainability, which none of the task-agnostic properties captures.
Shallow LA circuits partly converge to a common error plateau, consistent with spurious local minima in underparametrised models \cite{Larocca.2023}, while deep and highly expressive circuits such as the SPA-SEL with $L_W = 3$ show the largest run-to-run variation.
\revtwo{We hypothesise that this reflects the link between expressibility and barren plateaus \cite{McClean.2018, Holmes.2021, Holmes.2022, OrtizMarrero.2021, Ragone.2024}.}
The task-agnostic properties thus serve as necessary conditions for pre-selecting candidate architectures, while the final choice requires training on the target problem.\par
For the three-dimensional D3Q19 model, we therefore adopt the SPA-BEL3, which achieves accuracy comparable to the SPA-SEL3 with a third as many rotation parameters and a lower risk of untrainable runs.
For the two-dimensional D2Q9 model, the target depends only on two velocity components, and the PA-SEL1 offers sufficient accuracy at a considerably lower circuit cost (\cref{fig:costs}).
The LA remains an option for hardware with severely restricted qubit connectivity, as it requires no entangling gates within the encoding circuit.\par
All variants show error distributions spanning several orders of magnitude, so that the accuracy depends strongly on the input.
To identify this dependence and assess behaviour beyond the training data, \cref{fig:I-E-polation} evaluates the selected D3Q19 surrogate as a function of velocity magnitude, within and beyond the training range.
}
\begin{figure}[htb]%
    \centering%
    \resizebox{13cm}{!}{%
        \def\combinedpath{graphics/InterExtraPolation/data/}%
        \input{graphics/InterExtraPolation/data/inter_extra_polation.tikz}%
    }
    \caption{
        \revthree{Interpolation and extrapolation of the learned D3Q19 population $f_1$ (SPA-BEL3-L6), trained on $|\bm{u}| \le 0.1$ (grey) and evaluated up to $|\bm{u}| = 0.2$.}
        \revthree{(a) Relative error over $|\bm{u}|$ for random flow directions.}
        \revthree{Non-linear part of $f_1$, learned (solid) vs. exact (dashed), for velocities parallel (b), perpendicular (c) and diagonal (d) to $\bm{c}_1 = (1,0,0)$.}
        \revthree{The learned curves contain all orders of $\bm{u}$ produced by the circuit beyond the exact constant and linear terms.}
        \revthree{In (d) the exact non-linear part vanishes due to the cancellation of the two quadratic terms.}
        }%
    \label{fig:I-E-polation}%
\end{figure}%
\revtwo{Within the training range, the relative error grows steadily with $|\bm{u}|$.
\Cref{fig:I-E-polation}(b)-(d) isolate the non-linear part of the learned equilibrium.
The circuit captures the positive curvature parallel to $\bm{c}_1$, the negative curvature perpendicular to it, and the cancellation along the space diagonal.
Towards the edge of the training range and beyond it, the cancellation is preserved, but (b) and (c) begin to deviate asymmetrically from the exact solution, which shows the presence of odd powers of $\bm{u}$ beyond the linear term.
Their even part, i.e. the curvature, remains within about $3\,\%$ of the exact value up to $|\bm{u}| = 0.2$.
The odd powers originate from the Fourier structure of the circuit (\cref{section:Fourier}).
The terms $\sin(\kappa u_n)$ and $\cos(\kappa u_n)$ contain powers of $u_n$ of all orders, which training balances only inside the training range.
The training range must therefore enclose the velocities occurring in the simulation.
This holds for the Taylor--Green vortex ($|\bm{u}| \le 0.1$), while in the double shear layer the vortex roll-up locally accelerates the flow to around the upper bound of the training range (\cref{fig:dsl_re}), where the error in \cref{fig:I-E-polation}(a) increases only moderately.
}
\section{Model Extensions}
\label{chapter:Extensions}
Approximating each component of the distribution function independently neglects the coupled structure of the lattice Boltzmann collision operator, since global conservation and symmetry constraints involve all discrete populations simultaneously.
This motivates the explicit enforcement of conservation and symmetry constraints in learned collision operators. 
After each iteration, we implement these hard constraints as post-processing applied to the measured outputs. 
These constraints are only enforced during case simulations and do not need to be incorporated into the training process.\par
The BGK relaxation operator, expressed as a weighted sum governed by the relaxation parameter $\omega$, enables an explicit treatment directly within the quantum circuit. 
Its linear structure eliminates the need for variational learning, reducing algorithmic complexity, and can be interpreted as a direct extension of the data re-uploading scheme.
\subsection{Conservation of Mass and Momentum}  
\label{section:Conservation}
For LBM, mass and momentum conservation are collision invariants.
However, approximation errors introduced by the \revthree{VQCs} may lead to violations of mass and momentum conservation, requiring explicit corrections.
Itani \cite{Itani.2025} \revthree{incorporates} mass and momentum conservation as explicit penalty terms in the training loss, while Lacatus et al. \cite{Lacatus.2025} rely on the unitary structure of the circuit to preserve mass conservation by construction and enforce momentum via a penalty term.
Here, we adopt and extend the algebraic post-processing correction of Corbetta et al. \cite{Corbetta.2023} to the three-dimensional D3Q19 lattice, originally developed for classical neural network surrogates.
This approach is used to correct the lattice distributions generated by the \revthree{VQC.\par}
The algebraic correction $\Omega_c$ is then
\begin{equation} \label{eq:5.6}
     \bm{f}^* = \Omega_{c} (\bm{f}) = U \bm{f} + V \Omega (\bm{f}) \text{,}
\end{equation}
with
\begin{equation} \label{eq:5.7}
    \revthree{U = \mathbf{T}^{-1} I_1 \mathbf{T} \,\,\,\,\,\,\,\,\,\, \text{and} \,\,\,\,\,\,\,\,\,\, V = \mathbf{T}^{-1} I_2 \mathbf{T} \text{,}}
\end{equation}
where \revthree{$\Omega(\bm{f})$ denotes the post-collision populations predicted by the surrogate, $\mathbf{T}$} is an invertible transformation matrix, and $I_1$ and $I_2$ are complementary diagonal projectors.
The specific choice of $U$ and $V$ is not unique and can be adapted depending on which distribution indices are targeted for correction.
\revtwo{
Here, we choose $U$ and $V$ such that the correction becomes an orthogonal projection.
This is achieved by complementing the conserved-moment rows of $\mathbf{T}$ with a basis of the population vectors that carry neither mass nor momentum.
The effect of mass and momentum correction, denoted CONS, in flow simulations is shown in \cref{app:surrogate_diagnostics}.
}
\subsection{Rotation and Reflection Equivariance}
A key requirement for a collision operator is rotational isotropy, which ensures that macroscopic fluid dynamics remain independent of the lattice orientation.
While Lacatus et al. \cite{Lacatus.2025} achieve equivariance by construction at the circuit layer level, the isotropy condition can be enforced by a group averaging method, based on the concept of Corbetta et al. \cite{Corbetta.2023}.
Itani \cite{Itani.2025} \revthree{extends} this idea to a quantum circuit implementation: ancilla qubits are initialised in superposition, so that all symmetry-transformed forward passes are evaluated simultaneously. 
The subsequent averaging is then performed implicitly through a partial trace over the ancilla register.\par
Here, we instead enforce equivariance through classical group averaging as a post-processing step; the formal condition is stated as follows.
In case of a three-dimensional velocity set, a collision operator $\Omega(f)$ is equivariant if it commutes with the symmetry operations $\sigma$ of the lattice
\begin{equation} \label{eq:5.8} 
    \revthree{\Omega( \sigma \bm{f}) = \sigma\, \Omega (\bm{f}) \text{,} \,\,\,\,\,\,\,\,\,\, \forall \sigma \in O_h \text{.}}
\end{equation}
The relevant symmetry group for a three-dimensional cubic lattice collision is the \revthree{full octahedral group $O_h$,} which comprises all 48 rotation and reflection symmetries of the cube.
Symmetry equivariance can be imposed as a hard constraint using
\begin{equation} \label{eq:5.9}
    \revthree{\bm{f}^* = \Omega_{\text{SYM}} (\bm{f}) = \frac{1}{48} \sum_{\sigma \in O_h} \sigma^{-1} \,\, \Omega (\sigma \bm{f}) .}
\end{equation}
The collision operator is evaluated for all 48 symmetry-transformed inputs, with inverse transformations applied and averaged to yield the final output.
It is important to note that all distribution functions at a given lattice node must be approximated before averaging.
In two-dimensional settings, the method applies unchanged with the summation restricted to the dihedral group \revthree{$D_4$ of the square (8 elements)} and a normalisation factor of $\tfrac{1}{8}$.
\revtwo{The effect of symmetry averaging, denoted SYM, in flow simulations is shown in \cref{app:surrogate_diagnostics}.}
\subsection{Directional Orbit Compression}
\label{chapter:OrbitCompression}
A further consequence of the lattice symmetry structure is that discrete velocity directions can be grouped into orbits under the action of the symmetry group.
For both the D2Q9 and D3Q19 velocity sets, this partitioning yields three distinct orbits: the rest population, the axis-aligned (face) directions, and the diagonal (edge) directions.
Since distributions within the same orbit are physically equivalent up to a rotation or reflection, one can enforce a shared collision model across all members of an orbit.
In practice, this is achieved by rotating the macroscopic input (e.g. the velocity vector $\mathbf{u}$) into the canonical frame of the orbit representative before evaluating the circuit.
This orbit compression reduces the number of independently trained collision circuits from one per velocity direction to the number of orbits.\par
However, directional orbit compression (DOC) alone does not guarantee \revthree{rotation and reflection} symmetry.
\revthree{Sharing parameters across} symmetry-related directions ensures that the same function is applied \revthree{to each, but} it does not constrain \revthree{the function's dependence on the} inputs.
Orbit compression must therefore be combined with the symmetry averaging of \cref{eq:5.9} to enforce \revthree{the full lattice symmetry exactly.}
\revtwo{A quantitative comparison of the symmetry error with and without DOC is given in \cref{app:doc_symmetry}.}
\subsection{BGK Relaxation}
The BGK relaxation step (see \cref{eq:2.4}) can be embedded directly into the quantum circuit by encoding both the pre-collision distribution $f_i$ and the relaxation parameter $\omega$ as $R_y$ rotation angles, and applying a controlled-SWAP (Fredkin) gate to form the weighted post-collision mixture (see \cref{fig:relax_circuit}).
\begin{figure}[htbp]%
    \centering%
    \resizebox{5cm}{!}{%
        \begin{quantikz}
            &\qwbundle{N-1} & & \gate[wires=2]{VQC} &            & \\
            &               & &                     & \swap{2}   & \\
            &               & & \gate{R_y(\theta_{f_{i}})}    & \targX{}   & \meter{}\\
            &               & & \gate{R_y(\theta_{\tilde{\omega}})}    & \control{} &
        \end{quantikz}
    }%
    \caption{Extended circuit for BGK relaxation. The VQC outputs the \revtwo{normalised} equilibrium distribution \revtwo{$f_i^{eq}/\rho$}; a Fredkin gate, controlled by the ancilla qubit, mixes the pre-collision and equilibrium registers with weight $\tilde{\omega}$.}%
    \label{fig:relax_circuit}%
\end{figure}%
The encoding of $f_i$ is motivated by the Pauli-$Z$ measurement: applying $R_y(\theta)$ to $\ket{0}$ yields $\cos\!\left(\frac{\theta}{2}\right)\ket{0} + \sin\!\left(\frac{\theta}{2}\right)\ket{1}$, whose expectation satisfies $\langle Z \rangle = \cos(\theta)$. 
Setting \revtwo{$\theta_{f_i} = \arccos(f_i/\rho)$ maps the normalised population $f_i/\rho$ onto $\langle Z \rangle$, consistent with the normalised equilibrium learned by the VQC.}\par
The relaxation parameter \revthree{$\omega \in (0, 2)$} is normalised to \revthree{$\tilde{\omega} = \omega/2 \in (0, 1)$} so that it can serve as a \revthree{probability.} It is then encoded as $\theta_{\tilde{\omega}} = 2\arcsin\!\left(\sqrt{\tilde{\omega}}\right)$, preparing the ancilla state
\begin{equation}
    \ket{\psi_{\text{anc}}} = \sqrt{1-\tilde{\omega}}\,\ket{0} + \sqrt{\tilde{\omega}}\,\ket{1}.
\end{equation}
The Fredkin gate conditionally swaps the pre-collision and equilibrium registers controlled on this ancilla; tracing out the ancilla yields a reduced state whose expectation $\langle Z \rangle$ \revthree{equals the convex} mixture
\begin{equation}
    \revthree{\langle Z \rangle} = (1-\tilde{\omega})\,\revtwo{\frac{f_i}{\rho}} + \tilde{\omega}\,\revtwo{\frac{f_i^{\mathrm{eq}}}{\rho}}.
\end{equation}
The physical post-collision distribution is then recovered from the measurement outcome via the post-processing relation
\begin{equation}
    f_i^* = \revtwo{2\rho\,\langle Z \rangle - f_i}.
\end{equation}
\revthree{A convex mixture can only represent $\omega \le 1$, since over-relaxation requires a negative weight $1 - \omega$ of the pre-collision population.}
\revthree{Mixing with $\tilde{\omega} = \omega/2$ and applying the affine post-processing therefore recovers the full BGK update, $f_i^* = f_i + \omega\,(f_i^{\mathrm{eq}} - f_i)$, including $\omega > 1$.}
This approach contrasts with previous quantum LBM surrogates \cite{Itani.2025, Lacatus.2025}, which require retraining whenever the relaxation parameter changes; the present formulation \revthree{accepts arbitrary $\omega$ as a circuit input.}
\section{Numerical Results}
\label{chapter:Results}
The quantum surrogate replaces the collision operator in full LBM simulations, with conservation and symmetry corrections applied as post-processing at each time step.
\revthree{We assess its performance on the Taylor--Green vortex and the double shear layer benchmarks, using two- and three-dimensional surrogates.}
\revthree{The two-dimensional surrogate uses the PA architecture with SEL1 entangling layers (PA-SEL1-L6), while the three-dimensional one uses the SPA architecture with BEL3 entangling layers (SPA-BEL3-L6).}
\revthree{Both use six re-uploading layers.}
For training the \revthree{surrogates,} we retain the physical properties of the training dataset and the hyperparameters listed in \cref{tab:hyperparameters}.
\revthree{A complete performance analysis of the trained surrogate is provided in \cref{app:doc_symmetry,app:training_histories,app:surrogate_diagnostics}, including the relative error of each discrete velocity distribution, the symmetry error, and the effect of conservation and symmetry corrections.}
\subsection{\revthree{Taylor--Green Vortex}}
The \revthree{Taylor--Green vortex (TGV) at low Reynolds numbers} represents an unsteady, decaying vortex flow \revthree{governed by the incompressible Navier--Stokes equations.}
In three dimensions, the initial velocity field is defined as \cite{Brachet.1984}
\begin{equation} \label{eq:6.1}
    \begin{aligned}
    u_x(x,y,z) &= \revthree{u_0 \sin x \cos y \cos z} \text{,} \\
    u_y(x,y,z) &= \revthree{- u_0 \cos x \sin y \cos z} \text{,} \\
    \revone{u_z(x,y,z)} &= 0 \text{,}
    \end{aligned}
\end{equation} 
which satisfies the incompressibility condition $ \nabla \cdot \bm{u}= 0$ with $x,y,z \in [0, 2\pi]$.
For the TGV simulations, \revthree{this periodic domain is mapped onto a cube of edge length $1\,\text{m}$, discretised with} $20$ cells in each direction, \revthree{with} periodic boundary conditions.
\revthree{The initial velocity amplitude is set to} \revtwo{$u_0 = 0.1$} \revthree{in lattice units, and the} remaining physical parameters are chosen to achieve relaxation parameters \revthree{$\omega_1 = 0.7$} and \revthree{$\omega_2 = 1.3$.}\par
\begin{figure}[htb]%
    \centering%
    \resizebox{9cm}{!}{%
    \begin{tabular}{ c c }
        \includegraphics{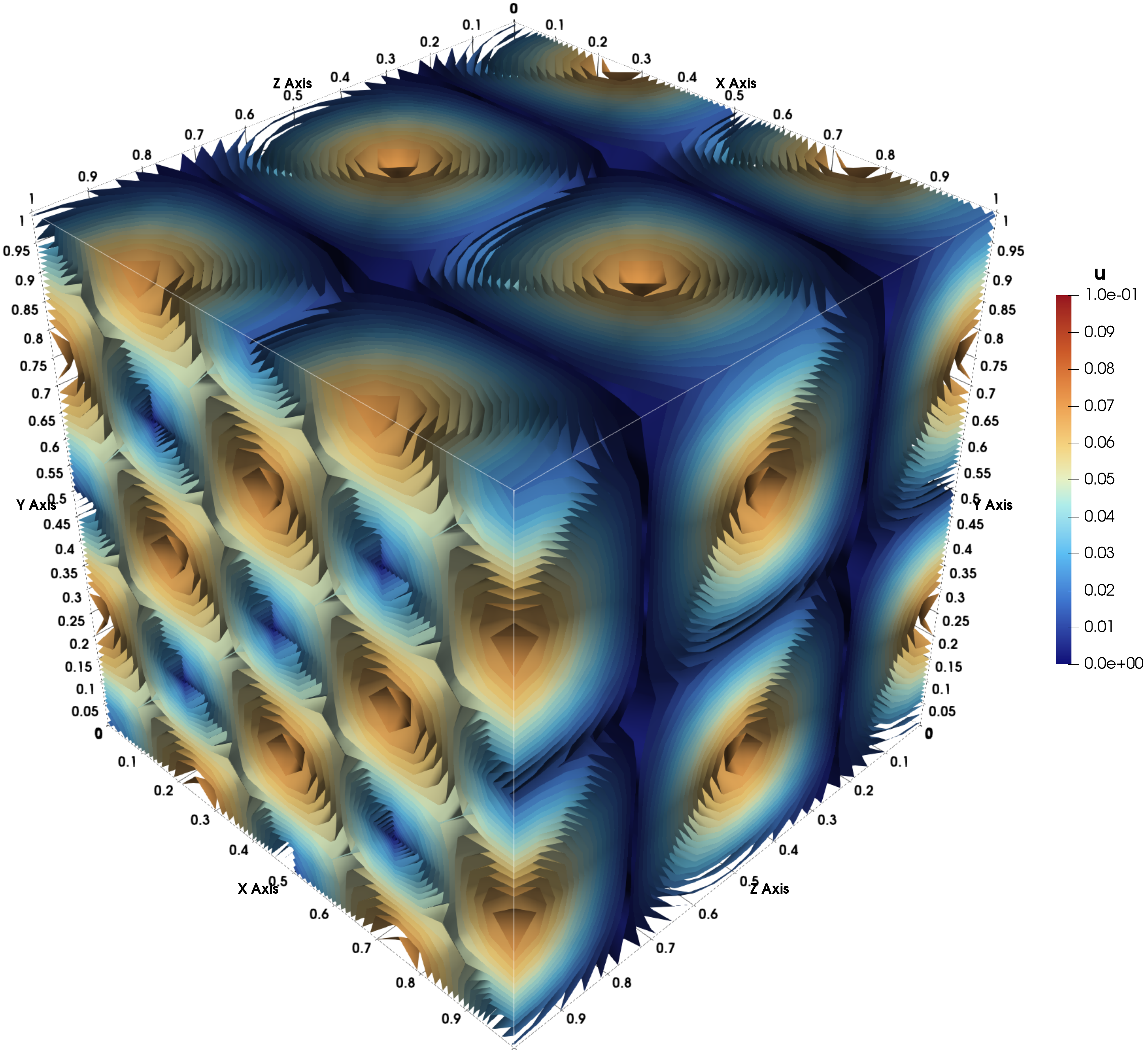} & \includegraphics{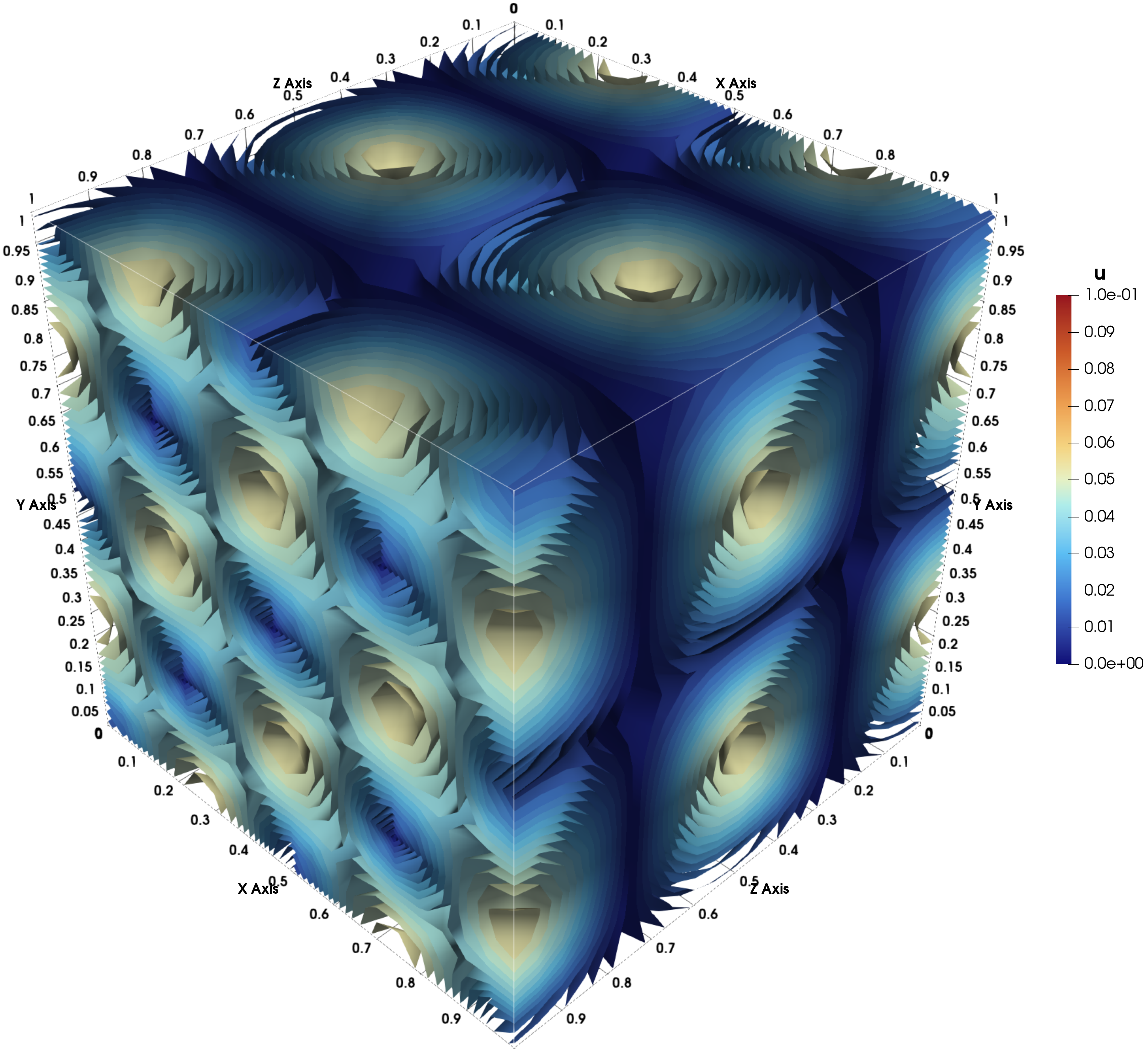} \\
        \includegraphics{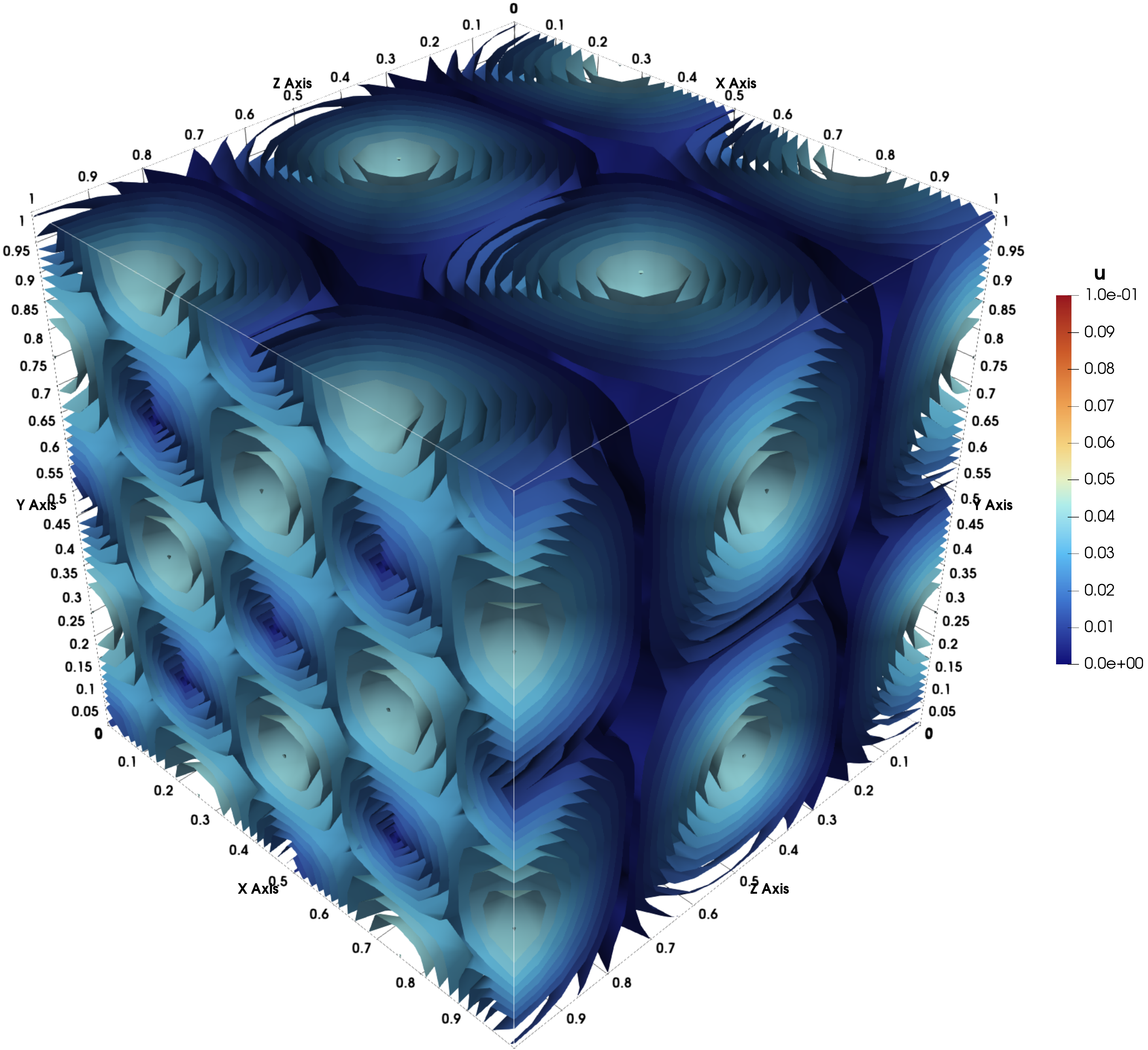} & \includegraphics{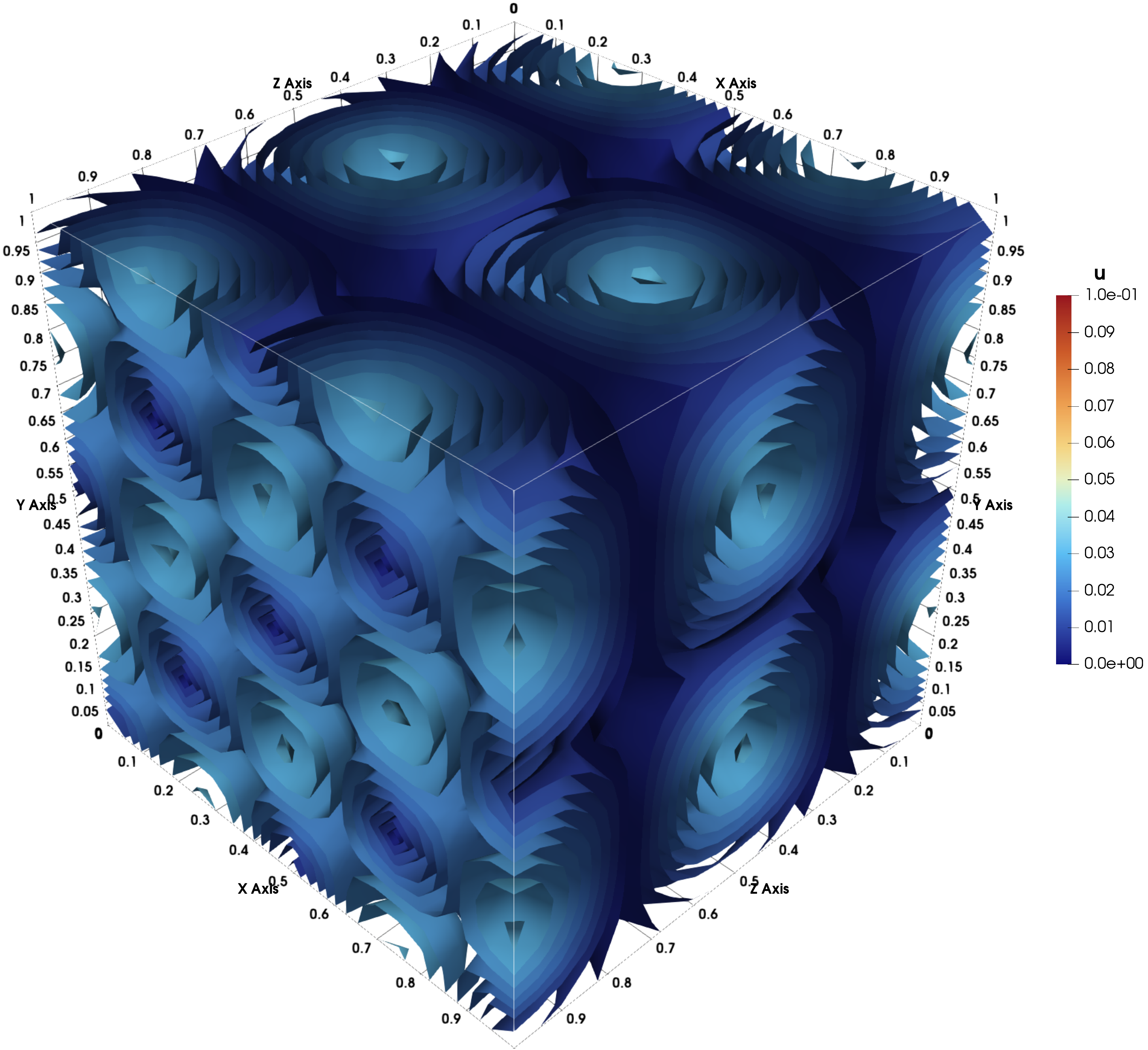}
    \end{tabular}
    }%
    \caption{Velocity field (\revthree{$\omega = 1.3$}) of a three-dimensional TGV at \revthree{$Re \approx 22$,} shown at time instances \revtwo{$t = 0.5\,\text{s}$, $t = 1.0\,\text{s}$, $t = 1.5\,\text{s}$, and $t = 2.0\,\text{s}$}.}%
    \label{fig:tgv_3d}%
\end{figure}%
As illustrated in \cref{fig:tgv_3d}, the surrogate collision model accurately captures the \revthree{symmetric} decay of the \revthree{periodic vortex array} over time. 
To quantitatively assess the surrogate's ability to reproduce vortex decay, we monitor the spatially averaged, non-dimensional kinetic energy, defined as
\begin{equation} \label{eq:6.2}
    \tilde{E}_k = \frac{1}{u_{0}^{2}} \left\langle \frac{1}{2}\,\rho(\bm{x}) \lvert \bm{u}(\bm{x}) \rvert^2 \right\rangle \text{,}
\end{equation}
where $u_0$ denotes the initial maximum velocity.\par
\revthree{\Cref{fig:tgv_kinetic_energy} compares $\tilde{E}_k$ of the surrogate without corrections (NAIVE), with the conservation correction (CONS), with symmetry averaging (SYM) and with both (CONS+SYM) with the BGK reference.}
\begin{figure}[htb]%
    \centering%
    \resizebox{0.9\textwidth}{!}{%
        \def\kepath{graphics/tgv_kinetic_energy/}%
        \input{graphics/tgv_kinetic_energy/kinetic_energy_tgv3d_log.tikz}%
    }%
    \caption{\revtwo{Non-dimensional kinetic energy $\tilde{E}_k$ (top) and its relative deviation from BGK in percent (bottom) over time for the three-dimensional Taylor--Green vortex at $\omega = 0.7$ ($Re = 6.5$) and $\omega = 1.3$ ($Re = 22.3$) for the NAIVE, CONS, SYM and CONS+SYM variants. Dashed lines denote one trained circuit per direction, dotted lines the DOC surrogate. The insets magnify the values at $t = 2\,\text{s}$.}}%
    \label{fig:tgv_kinetic_energy}%
\end{figure}%
\revtwo{
All variants reproduce the exponential decay of the kinetic energy, but their deviations from the BGK reference evolve differently.
The variants split into two groups, determined by whether mass and momentum are conserved.
Without the conservation correction (NAIVE and SYM), the deviation grows steadily over time, and the surrogate overestimates the kinetic energy.
Small violations of the collision invariants in every time step act as a spurious momentum source that counteracts viscous dissipation and accumulates.
With the conservation correction (CONS and CONS+SYM), this drift is suppressed, and the deviation remains orders of magnitude smaller throughout the simulation.
Symmetry averaging alone has a limited effect.
It reduces the deviation by up to half at $\omega = 0.7$, whereas SYM nearly coincides with NAIVE at $\omega = 1.3$.
Only at $\omega = 1.3$ does it further reduce the remaining deviation in combination with the conservation correction.
Conservation thus governs long-time stability, whereas symmetry averaging mainly improves the second-order moments, i.e. the momentum flux and the effective viscosity (\cref{app:surrogate_diagnostics}).
Both corrections are therefore retained for all subsequent simulations.\par
The dotted lines show the corresponding DOC surrogates.
DOC noticeably reduces the energy drift of the non-conservative variants, consistent with its lower symmetry error (\cref{fig:doc_symmetry}).
For the fully corrected CONS+SYM variant, the compressed and non-compressed surrogates perform equally well, so orbit compression reduces the number of trained circuits without loss of accuracy.\par
\begin{figure}[htbp]%
    \centering%
    \resizebox{0.7\textwidth}{!}{%
        \def\gridpath{graphics/GridRefinement/}%
        \input{graphics/GridRefinement/grid_refinement.tikz}%
    }%
    \caption{
        Grid refinement study for the two-dimensional TGV with the D2Q9 CONS+SYM surrogate. 
        The relative $L_2$ velocity error $\varepsilon_2$ separates the discretisation error of the BGK reference (squares), the surrogate's approximation error relative to BGK on the same grid (circles), and the surrogate's total error with respect to the analytical solution (triangles).
    }%
    \label{fig:grid_refinement}%
\end{figure}%
To separate the discretisation error from the surrogate error, we perform a grid refinement study for the two-dimensional TGV with the D2Q9 CONS+SYM surrogate under diffusive scaling (\cref{fig:grid_refinement}).
The BGK reference converges at second order towards the analytical solution, with $\varepsilon_2$ decreasing from $2\,\%$ at $N = 16$ to $0.03\,\%$ at $N = 128$.
The deviation between the surrogate and BGK remains approximately constant at ${\sim} \, 0.15 \, \%$, regardless of grid size.
Consequently, the surrogate follows the BGK operator's second-order convergence on coarse grids, while, on fine grids, its error saturates at the surrogate approximation level.
}
\subsection{\revthree{Double Shear Layer}}
\revthree{The double shear layer (DSL) captures the onset and growth of the Kelvin--Helmholtz instability in a periodic domain.}
The initial condition consists of two opposing hyperbolic-tangent shear layers centred at $y = 1/4$ and $y = 3/4$, seeded by a sinusoidal transverse perturbation
\begin{equation} \label{eq:6.3}
    \begin{aligned}
    u_x(x,y) &= u_0 \begin{cases} \tanh \, \bigl(80\,(y - \tfrac{1}{4})\bigr), & y < \tfrac{1}{2}, \\ \tanh \, \bigl(80\,(\tfrac{3}{4} - y)\bigr), & y \geq \tfrac{1}{2}, \end{cases} \\[4pt]
    u_y(x,y) &= \tfrac{1}{2}\, u_0 \sin \bigl(2\pi(x + \tfrac{1}{4})\bigr),
    \end{aligned}
\end{equation}
with $x \in [0,2]$, $y \in [0,1]$, $u_0 = 0.08\,\text{m/s}$, and uniform initial density $\rho = 1$.
\revthree{The domain of size $2\,\text{m} \times 1\,\text{m}$ is discretised with $200 \times 100$ grid cells.}
\revthree{We simulate Reynolds numbers $\text{Re} = 100$, $400$, $1000$ and $2000$, corresponding to relaxation parameters $\omega \approx 1.35$, $1.79$, $1.91$ and $1.95$.}
\par
\begin{figure}[!bt]%
    \centering%
    \resizebox{9cm}{!}{%
    \includegraphics[width=\textwidth]{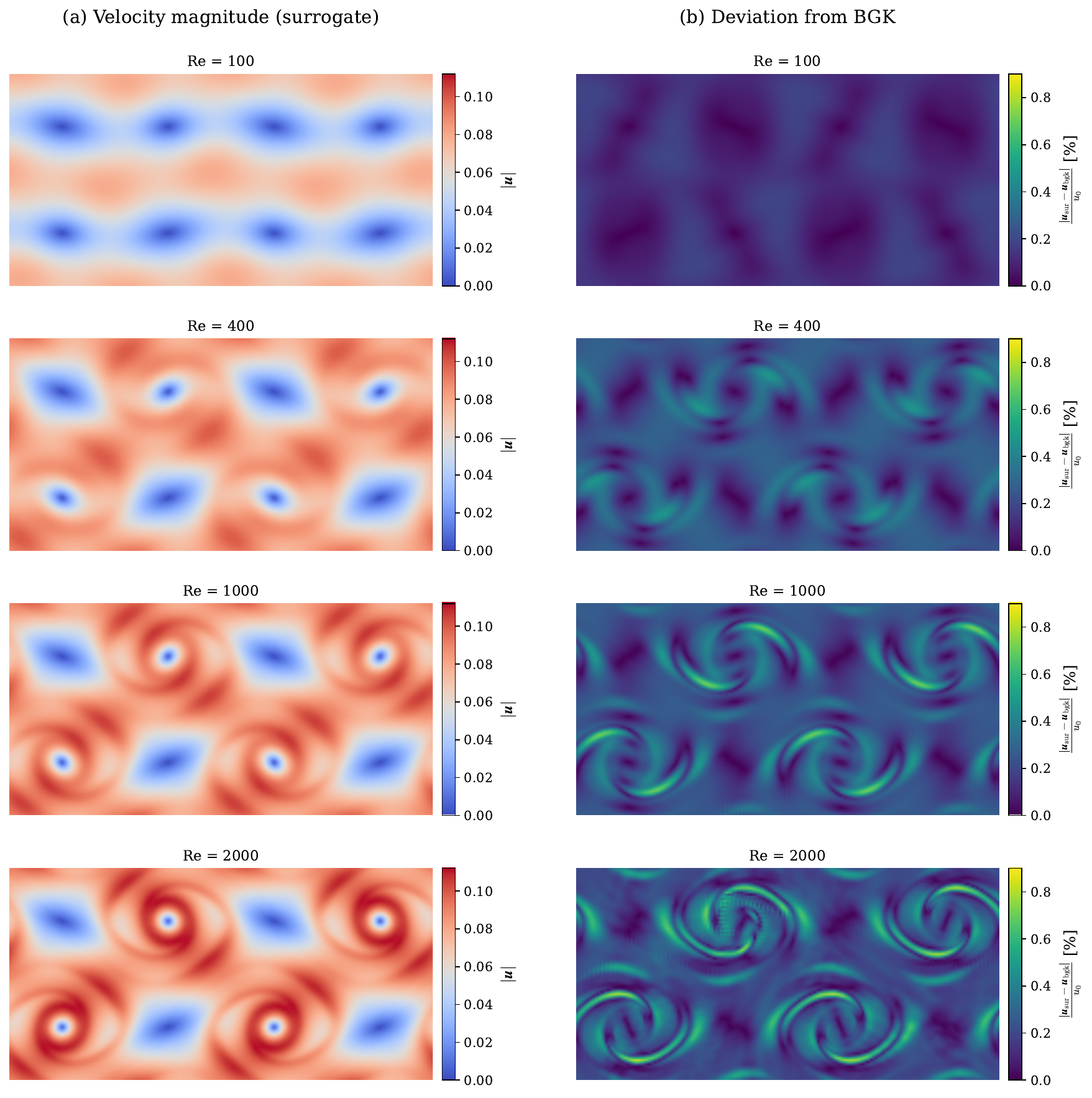}%
    }
    \caption{
        \revthree{DSL at $Re = 100$, $400$, $1000$ and $2000$, each shown at $t = 10\,\text{s}$.}
        \revthree{(a) Velocity magnitude $|\bm{u}|$ of the DOC surrogate.}
        \revthree{(b) Local velocity deviation from the BGK reference, $|\bm{u}_{\text{sur}} - \bm{u}_{\text{bgk}}|/u_0$, in percent.}
    }
    \label{fig:dsl_re}%
\end{figure}%
\revthree{\Cref{fig:dsl_re} shows that the surrogate reproduces the roll-up of the shear layers into Kelvin--Helmholtz vortices for all Reynolds numbers.}
\revthree{With increasing Re, the vortices wind up more tightly, and the shear layers between them thin into spiral filaments.}
\revthree{The deviation from BGK follows this structure.}
\revthree{At $\text{Re} = 100$, it is small and spread smoothly over the domain, whereas for $\text{Re} \geq 400$ it concentrates in the thin spiral filaments, where steep velocity gradients drive the populations furthest from equilibrium and it grows to $0.8\,\%$ of $u_0$ at $\text{Re} = 2000$.}
\revthree{At $\text{Re} = 2000$, weak grid-scale oscillations additionally appear inside the vortex cores, as local collision errors are barely damped at low viscosity, the same mechanism that limits the stability of BGK.\par}
\begin{figure}[htbp]%
    \centering%
    \resizebox{\textwidth}{!}{%
        \def\dslpath{graphics/dsl_ke_enstrophy/}%
        \def\shotpath{graphics/dsl_shot_noise/}%
        \input{graphics/dsl_ke_enstrophy/dsl_ke_enstrophy_shots.tikz}%
    }%
    \caption{
        \revthree{DSL simulation with the DOC surrogate (CONS+SYM).
        Relative deviation of (a) the non-dimensional kinetic energy $\tilde{E}_k$ and (b) the enstrophy $\tilde{\mathcal{E}}$ from the BGK reference over time; the dashed line marks zero deviation.}
        \revtwo{(c) Effect of shot noise: relative $L_2$ velocity error with respect to BGK at $t = 10\,\text{s}$ over the total number of shots per node and time step; the dotted lines with open markers show the noise-free surrogate.}
    }%
    \label{fig:dsl_ke_enstrophy}%
\end{figure}%
\revthree{\Cref{fig:dsl_ke_enstrophy} quantifies this agreement through the kinetic energy and the enstrophy.}
\revthree{For all Reynolds numbers, both deviations remain small and bounded over the entire simulation.}
\revthree{The kinetic energy is slightly underestimated, indicating a marginally increased dissipation of the surrogate.}
\revthree{The enstrophy deviates more strongly, as it depends on velocity gradients and is therefore most sensitive to the errors in the thin shear filaments.}
\revthree{Accordingly, its deviation is largest and positive at $Re = 2000$, where the small-scale oscillations in the vortex cores add spurious vorticity.\par}
\revtwo{
On quantum hardware, expectation values can only be estimated from a finite number of measurements.
To assess the effect of finite sampling, we compare the noise-free surrogate simulation, evaluated exactly from the state vector, with a noisy surrogate simulation in which every $\langle Z \rangle$ is replaced by the mean of $S$ single-shot outcomes,
\begin{equation}
    \langle Z \rangle_S = \frac{1}{S} \sum_{j=1}^{S} z_j \text{,} \qquad z_j \in \{-1, +1\} \text{,} \quad
    P(z_j = +1) = \frac{1 + \langle Z \rangle}{2} \text{,}
\end{equation}
which is unbiased with variance $(1 - \langle Z \rangle^2)/S$.
We sample each circuit evaluation independently at each node and time step, for a total of $72S$ shots per node and time step ($9$ populations $\times\, 8$ symmetry-transformed inputs).
\Cref{fig:dsl_ke_enstrophy}(c) compares the resulting velocity error with respect to BGK at $t = 10\,\text{s}$ of a noisy surrogate simulation as a function of the number of shots $S$.
The shot noise acts as a random perturbation of every collision and decays as $S^{-1/2}$, as expected for unbiased sampling.
It exceeds the approximation error of the surrogate by orders of magnitude for small $S$ and approaches it for $S \approx 10^{9}$, so that sampling rather than the learned collision operator limits the accuracy on hardware.
Although only the simulation at $Re = 2000$ diverges for the smallest shot number, numerical stability does not imply a physically meaningful solution.
When velocity errors rise above several percentage points, the Kelvin--Helmholtz dynamics are no longer reproduced.\par
Because this test assumes ideal sampling without gate errors, hardware noise would further increase the required number of shots by the overhead shown in \cref{fig:costs_train_meas}(b). 
Including gate errors, the PA-SEL1-L6 circuit with $n_{1\text{q}} = 92$ single- and $n_{2\text{q}} = 29$ two-qubit gates reaches a fidelity of $F \approx 0.90$ (\cref{eq:fidelity}) for the IBM Heron error rates used in \cref{fig:costs_train_meas}.
After rescaling the measured expectation values by $1/F$, \cref{eq:cost_shots} increases the required number of shots by a factor of $1/F^2 \approx 1.24$, shifting the curves in \cref{fig:dsl_ke_enstrophy}(c) upwards by $1/F \approx 11\,\%$ without changing their $S^{-1/2}$ decay.
}\par     
\section{Conclusion and Outlook}
\label{chapter:Conclusion}
We present a hybrid quantum-classical surrogate for the LBM collision operator that addresses the central obstacle for pure quantum solvers: the non-unitary, non-linear nature of the BGK collision step.
By decoupling equilibrium prediction, solved by a shallow angle-encoded variational quantum circuit, from BGK relaxation, which is embedded as a closed-form circuit extension, the surrogate encodes the relaxation parameter $\omega$ directly as a circuit input.
\revtwo{
This sets the approach apart from prior quantum LBM surrogates, which were demonstrated only for a fixed relaxation time ($\tau = 1$), and makes the learned equilibrium applicable across the range $\omega \in (0, 2)$ without additional training.\par
}
The circuit design was guided by a systematic analysis of expressibility, entanglement, and effective dimension, applied here for the first time to a fluid-dynamics VQC design problem.
Cross-validating these use-case-agnostic metrics against problem-specific accuracy showed that $L_{\text{SW}}$, the number of encoding repetitions, \revthree{which is the primary driver of expressibility, also governs the accuracy of the trained circuits.}
\revtwo{
Expressibility proved necessary but not sufficient for accuracy, whereas \revthree{the amount of entanglement did not determine it.}
The effective dimension proved most informative, as it reproduced the ranking of the architectures, but not the comparison between entangling layers.
\revthree{Beyond this ranking,} the architecture can be matched to the complexity of the target function.
For the simpler two-dimensional equilibrium, a lower-ranked architecture thus achieves errors comparable to those of the best-performing one in three dimensions.
The surrogate captures the full direction-dependent quadratic structure of the equilibrium, whereas the truncated Fourier series of the circuit does not extrapolate it reliably, so that the training range must cover the velocities of the simulation.
}
\revtwo{Orbit compression further reduces the number of independently trained circuits to one per symmetry orbit (three for both D2Q9 and D3Q19), lowering the training cost without loss of accuracy, while the number of circuit executions per node remains unchanged.}\par
Validation was performed on the three-dimensional Taylor--Green vortex and the two-dimensional double shear layer at Reynolds numbers up to \revthree{$Re = 2000$.}
Both benchmarks confirmed \revthree{low} surrogate errors, accurate reproduction of global integral quantities, and numerical stability over extended integration times.
\revtwo{A grid refinement study showed that the surrogate follows the second-order convergence of BGK until its error saturates at the approximation level of about $0.15\,\%$.}
\par
Several directions emerge from the present work and warrant further investigation.
First, the modular structure of the surrogate framework, in which the equilibrium prediction and the relaxation step are decoupled, naturally accommodates more complex collision models beyond the BGK model.
\revthree{Multi-relaxation-time (MRT) schemes, for instance, relax different moment orders with different rates, and entropic variants additionally impose an entropy condition on the relaxation step.}
\revthree{Since the relaxation is imposed analytically, such operators could reuse the learned equilibrium and would only require extending the relaxation step to moment-wise rates, without retraining.}
Second, incorporating additional physical constraints into the surrogate architecture itself, rather than as post-processing corrections, represents a promising avenue for a full-stack quantum LBM surrogate.
Equivariant circuit architectures that respect the lattice symmetry group by construction, for instance through symmetry-constrained parameterisations or group-theoretic ansätze, could eliminate the post-processing corrections entirely, yielding a physically consistent surrogate by design rather than by correction.
Third, as quantum hardware continues to mature, deploying the trained surrogates on physical quantum devices represents a crucial next step.
\revtwo{The shot-noise analysis shows that finite sampling, rather than the learned collision operator, then limits the accuracy.}
Noise mitigation strategies and hardware-specific circuit transpilation will be essential to preserve the accuracy demonstrated here in noiseless simulation.
\section*{CRediT authorship contribution statement}
\textbf{Lukas C. Birk}: Writing -- Original Draft, Visualisation, Software, Methodology, Formal analysis, Conceptualisation.
\textbf{David M. Wawrzyniak}: Writing -- Review \& Editing, Supervision, Software, Formal analysis, Conceptualisation. 
\textbf{Josef M. Winter}: Writing -- Review \& Editing, Supervision, Software, Formal analysis, Conceptualisation. 
\textbf{Steffen J. Schmidt}: Writing -- Review \& Editing, Supervision, Formal analysis, Conceptualisation. 
\textbf{Thomas Indinger}: Writing -- Review \& Editing, Supervision, Funding acquisition. 
\textbf{Christian F. Janßen}: Writing -- Review \& Editing, Supervision, Funding acquisition.
\textbf{Nikolaus A. Adams}: Writing -- Review \& Editing, Supervision, Project administration, Funding acquisition, Conceptualisation.

\section*{Declaration of competing interest}
The authors declare that they have no known competing financial
interests or personal relationships that could have appeared to influence
the work reported in this paper.

\section*{Data availability}
Data available to be shared. 
The algorithms required to reproduce the above findings are available to download from https://github.com/tumaer/qlbm.

\section*{Declaration of generative AI and AI-assisted technologies in the writing process}
During the preparation of this work, the authors used Grammarly and Claude to improve readability. 
After using these tools, the authors reviewed and edited the content as needed and took full responsibility for the content of the publication.

\section*{Acknowledgements}
The authors gratefully acknowledge funding for this research from Siemens Digital Industries Software.

\appendix
\setcounter{figure}{0}
\section{\revthree{Symmetry Error of Directional Orbit Compression}}
\label{app:doc_symmetry}
\revtwo{
Directional orbit compression (DOC) shares one circuit among all directions of an orbit, but this alone does not make the learned model equivariant.
In D2Q9, for example, the circuit shared by the positive $x$ and $y$ directions may still weight $u_x$ and $u_y$ differently and thereby violate the symmetry relating these two directions.
To assess how strongly such violations occur in practice, we compare each learned population with its symmetry-transformed counterpart.
An exactly equivariant model satisfies $f_{g(i)}(g\bm{u}) = f_i(\bm{u})$ for every lattice symmetry $g$, and we quantify deviations by the relative symmetry error $|f_{g(i)}(g\bm{u}) - f_i(\bm{u})|/f_i(\bm{u})$.\par
\begin{figure}[htb]%
    \centering%
    \resizebox{13cm}{!}{%
        \def\docpath{graphics/SymmetryErrorDOC/data/}%
        \input{graphics/SymmetryErrorDOC/doc_vs_19.tikz}%
    }%
    \caption{
        Cumulative distribution of the symmetry error of the learned D3Q19 equilibrium (SPA-BEL3-L6, trained on $|\bm{u}| \le 0.1$, without symmetry averaging and conservation projection) for DOC (dashed) and 19 independently trained circuits (dotted), shown for (a) axis and (b) diagonal directions.
        Errors are evaluated for all 47 non-trivial lattice symmetries $g$ and 500 random velocities with $|\bm{u}| \le 0.1$ ($\rho = 1$).
    }%
    \label{fig:doc_symmetry}%
\end{figure}%
\Cref{fig:doc_symmetry} shows its cumulative distribution for the D3Q19 model without symmetry averaging.
For both axis and diagonal directions, DOC yields lower errors than 19 separately trained circuits.
With independent circuits, each population is fitted separately, so their approximation errors are uncorrelated and no symmetry relation holds exactly.
With DOC, all directions of an orbit evaluate the same circuit on transformed inputs, so every symmetry $g$ that maps one direction to another by this input transformation is satisfied exactly, independent of training.
The remaining symmetries, in particular reflections, are not imposed by parameter sharing and hold only approximately.
Orbit compression must therefore be combined with the symmetry averaging of \cref{eq:5.9} to enforce the full lattice symmetry exactly.
}   
\setcounter{figure}{0}
\section{\revthree{Training Histories}}
\label{app:training_histories}
\revtwo{\Cref{fig:training_histories} shows the validation histories of the final surrogate models used in \cref{chapter:Results}.
\begin{figure}[htbp]%
    \centering%
    \resizebox{0.8\textwidth}{!}{%
        \def\histpath{graphics/TrainingHistories/}%
        \input{graphics/TrainingHistories/training_histories.tikz}%
    }%
    \caption{
        Validation MSE of the circuit output per epoch for the final surrogate models: (a) D2Q9 DOC (PA-SEL1-L6), (b) D3Q19 DOC (SPA-BEL3-L6) and (c) D3Q19 with one circuit per direction (SPA-BEL3-L6). 
        Colours denote the orbit of the direction (rest, axis, diagonal). 
        The learning rate is reduced by a factor of 10 every three epochs (shaded phases).
    }%
    \label{fig:training_histories}%
\end{figure}%
In all cases, the validation MSE decreases by two to three orders of magnitude within nine epochs, with sharp drops after each learning-rate reduction at epochs $4$ and $7$.
Isolated spikes appear in the per-direction histories only at the high learning rate and do not affect the final models, since we keep the weights of the epoch with the lowest validation loss.
With orbit compression, the circuits reach the same validation accuracy as the 19 individually trained circuits.
On the test set, the final models achieve mean relative errors between $0.003\,\%$ and $0.06\,\%$.
}
\section{\revthree{Surrogate Accuracy and Flow Diagnostics}}
\label{app:surrogate_diagnostics}
\setcounter{figure}{0}
\revthree{\Cref{fig:loss_d3q19,fig:loss_d2q9} report the per-direction accuracy of the D3Q19 and D2Q9 surrogates.}
\begin{figure}[htbp]%
    \centering%
    \resizebox{0.8\textwidth}{!}{%
        \def\stressdatapath{graphics/D3Q19_analysis/data/}%
        \input{graphics/D3Q19_analysis/stress_viscosity_tgv_doc.tikz}%
    }%
    \caption{\revthree{Accuracy of the D3Q19 surrogate (SPA-BEL3-L6, DOC, trained on $8 \times 10^4$ samples) for the variants}
    \revthree{NAIVE, CONS, SYM and CONS+SYM, evaluated on $10^4$ unseen synthetic samples (open boxes) and at all $8 \times 10^3$ nodes}
    \revthree{of the TGV ($20^3$ cells, $\text{Re} = 22.3$, $\omega = 1.3$, $t = 0.5\,\text{s}$; hatched boxes).}}
    \label{fig:loss_d3q19}%
\end{figure}%
\begin{figure}[htbp]%
    \centering%
    \resizebox{0.8\textwidth}{!}{%
        \def\stressdatapath{graphics/D2Q9_analysis/data/}%
        \input{graphics/D2Q9_analysis/stress_viscosity_dsl1000_doc.tikz}%
    }%
    \caption{\revthree{Accuracy of the D2Q9 surrogate (PA-SEL1-L6, DOC, trained on $8 \times 10^4$ samples) for the variants}
    \revthree{NAIVE, CONS, SYM and CONS+SYM, evaluated on $10^4$ unseen synthetic samples (open boxes) and at all $2 \times 10^4$ nodes}
    \revthree{of the DSL ($200 \times 100$ cells, $\text{Re} = 1000$, $\omega = 1.91$, $t = 10\,\text{s}$; hatched boxes).}}
    \label{fig:loss_d2q9}%
\end{figure}%
\revtwo{
They further assess whether the learned collision preserves the hydrodynamically relevant moments and the effects of the post-collision corrections.\par
Panel (a) shows that both surrogates reproduce the BGK post-collision populations.
For both models, the medians lie between $0.001\%$ and $0.08\%$, with the smallest errors for the rest population and the largest for the diagonal directions, whose small populations ($w_i = 1/36$) amplify the relative error.\par
Panel (b) evaluates the collision increment $\Delta f_i = f_i^* - f_i$, i.e. the change of the populations during collision, at every node of the flow, normalised by the increment at the same node.
This quantity directly tests whether the collision has been learned.
For D3Q19 in the TGV, the median errors lie between $0.04\%$ and $0.43\%$, whereas for D2Q9 in the DSL at $\text{Re} = 1000$ they reach $1$--$6\,\%$ across all variants and $1$--$3\,\%$ with both corrections.
As the equilibrium momentum flux in panel (c) is reproduced with at least comparable accuracy in both cases, this difference stems from the flow rather than the surrogate.
In the DSL, weak velocity gradients between the thin shear layers yield small collision increments, amplifying small absolute errors in the relative measure, whereas the TGV exhibits velocity gradients throughout the domain.\par
Panel (c) evaluates the error in the deviatoric equilibrium momentum flux $\mathrm{dev}(\rho\bm{u}\bm{u})$, which removes the dominant isotropic pressure contribution $\rho c_s^2 \, \mathbf{I}$ and retains the anisotropic part of the advective momentum flux.
Since the normalisation $|\operatorname{dev}(\rho\bm{u}\bm{u})| \propto |\bm{u}|^2$ vanishes at stagnation points, nodes with $|\bm{u}| < 0.1 \, |\bm{u}|_{\max}$ are excluded.
The conservation correction leaves this error unchanged, since it only adjusts the populations along the mass and momentum vectors, which carry no deviatoric second moment.
Symmetry averaging, in contrast, removes error components that violate the lattice symmetry, such as unequal treatment of the velocity components, which manifest directly as anisotropic errors in the momentum flux.
With symmetry averaging, it drops to $0.25\%$ (D3Q19) and $0.3\%$ (D2Q9), becomes independent of the data set, and is confined to a narrow band.\par
Panel (d) evaluates the relative error in the post-collision non-equilibrium stress $\bm{\Pi}^{*,\text{neq}} = \sum_i \bm{c}_i \bm{c}_i \bigl(f_i^* - f_i^{\text{eq}}\bigr)$.
Since the surrogate modifies only the equilibrium populations, the difference between the post-collision non-equilibrium stresses of the surrogate and BGK equals $\omega$ times the equilibrium-stress error in (c).
Normalising by the collision-induced change of BGK, $\omega\,|\operatorname{dev} \bm{\Pi}^{\text{neq}}|_F$, cancels the factor $\omega$, so the error in (d) equals the error in (c) rescaled by $|\operatorname{dev}(\rho \bm{u}\bm{u})|_F / |\operatorname{dev} \bm{\Pi}^{\text{neq}}|_F$.
This ratio is larger in the DSL, where nearly uniform regions between the shear layers combine high velocities with weak velocity gradients, yielding small non-equilibrium stresses.\par
Panel (e) evaluates the effective viscosity of the surrogate collision.
The effective relaxation rate follows from projecting the post-collision onto the pre-collision non-equilibrium stress and defines the effective viscosity $\nu_{\text{eff}} = c_s^2 \bigl(1/\omega_{\text{eff}} - 1/2\bigr)$, with $\omega_{\text{eff}} = \omega$ and $\nu_{\text{eff}} = \nu$ for exact BGK.
The plotted quantity is the ratio of effective to prescribed relaxation time, $\tau_{\text{eff}}/\tau = (1 - \omega/2)\,\nu_{\text{eff}}/\nu + \omega/2$.
It equals one when the surrogate reproduces the prescribed viscosity.
In both cases, the medians coincide with one, indicating no systematic viscosity bias, while the remaining node-wise scatter corresponds to the component of the error in (d) parallel to $\bm{\Pi}^{\text{neq}}$, which changes the effective relaxation and hence the viscosity.
Symmetry averaging narrows this scatter for D3Q19, while for D2Q9 it remains at about $\pm 0.3\%$, consistent with its small effect in (c) and (d).\par
}  
\par



\clearpage 
\bibliographystyle{elsarticle-num}
\bibliography{literature.bib}
 
\end{document}

%% file: graphics/InterExtraPolation/data/inter_extra_polation.tikz
\pgfplotsset{compat=1.18}%
\providecommand{\combinedpath}{data/}%
\begin{tikzpicture}
\begin{groupplot}[group style={group size=3 by 1, horizontal sep=0.8cm},
    scale only axis, width=4.4cm, height=3.6cm, xmin=-0.2, xmax=0.2, xtick={-0.2,-0.1,0,0.1,0.2},
    ymin=-0.1, ymax=0.18, ytick={-0.1,-0.05,0,0.05,0.1,0.15}, set layers,
    scaled ticks=false, xticklabel style={/pgf/number format/fixed, /pgf/number format/precision=2},
    yticklabel style={/pgf/number format/fixed, /pgf/number format/precision=2},
    title style={font=\footnotesize}, label style={font=\small}, tick label style={font=\footnotesize},
    legend style={at={(0.5,0.97)}, anchor=north, draw=none, font=\footnotesize, fill=none},
    legend cell align=left]
\nextgroupplot[title={(b) along $\bm{c}_1$}, xlabel={$u_x$},
    ylabel={$\dfrac{f_1}{\rho w_1} - 1 - 3\,\bm{c}_1\!\cdot\!\bm{u}$}]
    \addplot[fill=gray!12, draw=none, forget plot, on layer=axis background] coordinates {(-0.1,-0.1) (0.1,-0.1) (0.1,0.18) (-0.1,0.18)} \closedcycle;
    \addplot[black, dashed, thick] table[x=s, y expr=18*\thisrow{par_exact} - 1 - 3*\thisrow{s}] {\combinedpath lines_LionDog.dat};
    \addlegendentry{exact}
    \addplot[colour1, very thick] table[x=s, y expr=18*\thisrow{par_pred} - 1 - 3*\thisrow{s}] {\combinedpath lines_LionDog.dat};
    \addlegendentry{learned}
\nextgroupplot[title={(c) perpendicular to $\bm{c}_1$}, xlabel={$u_y$}, yticklabels={}]
    \addplot[fill=gray!12, draw=none, forget plot, on layer=axis background] coordinates {(-0.1,-0.1) (0.1,-0.1) (0.1,0.18) (-0.1,0.18)} \closedcycle;
    \addplot[black, dashed, thick] table[x=s, y expr=18*\thisrow{perp_exact} - 1] {\combinedpath lines_LionDog.dat};
    \addplot[colour1, very thick] table[x=s, y expr=18*\thisrow{perp_pred} - 1] {\combinedpath lines_LionDog.dat};
\nextgroupplot[title={(d) along the space diagonal}, xlabel={$\pm|\bm{u}|$}, yticklabels={}]
    \addplot[fill=gray!12, draw=none, forget plot, on layer=axis background] coordinates {(-0.1,-0.1) (0.1,-0.1) (0.1,0.18) (-0.1,0.18)} \closedcycle;
    \addplot[black, dashed, thick] table[x=s, y expr=18*\thisrow{diag_exact} - 1 - 1.7320508*\thisrow{s}] {\combinedpath lines_LionDog.dat};
    \addplot[colour1, very thick] table[x=s, y expr=18*\thisrow{diag_pred} - 1 - 1.7320508*\thisrow{s}] {\combinedpath lines_LionDog.dat};
\end{groupplot}
\begin{axis}[at={($(group c1r1.north west)+(0,2.6cm)$)}, anchor=south west, scale only axis,
    width=14.8cm, height=4.2cm, xmin=0, xmax=0.2, ymode=log, ymin=1e-6, ymax=1e-1, set layers,
    scaled x ticks=false, xtick={0,0.02,...,0.2},
    xticklabel style={/pgf/number format/fixed, /pgf/number format/precision=2},
    title={(a) $f_1$, SPA-BEL3-L6, trained range $|\bm{u}| \le 0.1$}, title style={font=\small},
    label style={font=\small}, tick label style={font=\footnotesize},
    xlabel={$|\bm{u}|$}, ylabel={$\frac{\mid f_1^{\text{pred}} - f_1^{\text{true}} \mid}{f_1^{\text{true}}}$},
    legend style={at={(0.98,0.2)}, anchor=south east, draw=none, font=\footnotesize, fill=none}, legend columns=3,
    legend cell align=left]
    \addplot[fill=gray!12, draw=none, forget plot, on layer=axis background] coordinates {(0,1e-6) (0.1,1e-6) (0.1,1e-1) (0,1e-1)} \closedcycle;
    \addplot[name path=p95, draw=none, forget plot] table[x=s, y=p95] {\combinedpath bands_LionDog.dat};
    \addplot[name path=p05, draw=none, forget plot] table[x=s, y=p05] {\combinedpath bands_LionDog.dat};
    \addplot[colour1!25] fill between[of=p95 and p05];
    \addlegendentry{5--95\,\%}
    \addplot[name path=p75, draw=none, forget plot] table[x=s, y=p75] {\combinedpath bands_LionDog.dat};
    \addplot[name path=p25, draw=none, forget plot] table[x=s, y=p25] {\combinedpath bands_LionDog.dat};
    \addplot[colour1!55] fill between[of=p75 and p25];
    \addlegendentry{25--75\,\%}
    \addplot[colour1, very thick] table[x=s, y=median] {\combinedpath bands_LionDog.dat};
    \addlegendentry{median}
    \draw[black, dashed, thick] (axis cs:0.1,1e-6) -- (axis cs:0.1,1e-1);
    \node[anchor=south, font=\small] at (axis cs:0.05,2e-6) {interpolation};
    \node[anchor=south, font=\small] at (axis cs:0.15,2e-6) {extrapolation};
\end{axis}
\end{tikzpicture}%

%% file: graphics/tgv_kinetic_energy/kinetic_energy_tgv3d_log.tikz
\small%
\pgfplotsset{compat=1.18}
\providecommand{\kepath}{}%

\begin{tikzpicture}[scale=1]%
    \begin{axis}[width = 6cm, height = 6cm, scale only axis=true
                 ,title={$\omega_1 = 0.7$}
                 ,title style={font=\small}
                 ,xmin = 0, xmax = 2
                 ,ymin = 1e-07
                 ,ymax = 0.125
                 ,ymode=log
                 ,ytickten={-7,-6,-5,-4,-3,-2,-1,0}
                 ,legend style={at={(1.05,0.1333)}, anchor=west, draw=none, font=\normalsize}
                 ,legend cell align={left}
                 ,tick label style={font=\small}
                 ,label style={font=\small}
                 ,xtick={0, 0.25, 0.5, 0.75, 1.0, 1.25, 1.5, 1.75, 2.0}
                 ,xticklabels={}
                 ,ylabel = {$\tilde{E}_k (t)$}
                ]%
        \addplot [black, ultra thick] table [x=time, y=bgk] {\kepath kinetic_energy_tgv3d_w0.7.dat};
        \addplot [colour1, very thick, dashed] table [x=time, y=naive] {\kepath kinetic_energy_tgv3d_w0.7.dat};
        \addplot [colour1, very thick, dotted] table [x=time, y=naive_doc] {\kepath kinetic_energy_tgv3d_w0.7.dat};
        \addplot [colour3, very thick, dashed] table [x=time, y=cons] {\kepath kinetic_energy_tgv3d_w0.7.dat};
        \addplot [colour3, very thick, dotted] table [x=time, y=cons_doc] {\kepath kinetic_energy_tgv3d_w0.7.dat};
        \addplot [colour4, very thick, dashed] table [x=time, y=sym] {\kepath kinetic_energy_tgv3d_w0.7.dat};
        \addplot [colour4, very thick, dotted] table [x=time, y=sym_doc] {\kepath kinetic_energy_tgv3d_w0.7.dat};
        \addplot [colour5, very thick, dashed] table [x=time, y=symcons] {\kepath kinetic_energy_tgv3d_w0.7.dat};
        \addplot [colour5, very thick, dotted] table [x=time, y=symcons_doc] {\kepath kinetic_energy_tgv3d_w0.7.dat};
        \draw[black, thick] (axis cs:1.995,1.3946e-04) rectangle (axis cs:2,1.4310e-04);
        \coordinate (zoomtop0) at (axis cs:1.995,1.4310e-04);
        \coordinate (zoombot0) at (axis cs:1.995,1.3946e-04);
    \end{axis}%

    \begin{axis}[width=2.8cm, height=2.8cm, scale only axis=true, ticks=none, name=inset0,
        xmin=1.995, xmax=2, ymin=1.3946e-04, ymax=1.4310e-04, xshift=0.5cm, yshift=0.5cm,
        restrict x to domain=1.949:2.001]  
        \addplot [black, semithick] table [x=time, y=bgk] {\kepath kinetic_energy_tgv3d_w0.7.dat};
        \addplot [colour1, semithick, dashed] table [x=time, y=naive] {\kepath kinetic_energy_tgv3d_w0.7.dat};
        \addplot [colour1, semithick, dotted] table [x=time, y=naive_doc] {\kepath kinetic_energy_tgv3d_w0.7.dat};
        \addplot [colour3, semithick, dashed] table [x=time, y=cons] {\kepath kinetic_energy_tgv3d_w0.7.dat};
        \addplot [colour3, semithick, dotted] table [x=time, y=cons_doc] {\kepath kinetic_energy_tgv3d_w0.7.dat};
        \addplot [colour4, semithick, dashed] table [x=time, y=sym] {\kepath kinetic_energy_tgv3d_w0.7.dat};
        \addplot [colour4, semithick, dotted] table [x=time, y=sym_doc] {\kepath kinetic_energy_tgv3d_w0.7.dat};
        \addplot [colour5, semithick, dashed] table [x=time, y=symcons] {\kepath kinetic_energy_tgv3d_w0.7.dat};
        \addplot [colour5, semithick, dotted] table [x=time, y=symcons_doc] {\kepath kinetic_energy_tgv3d_w0.7.dat};
    \end{axis}
    \draw[black] (zoomtop0) -- (inset0.north east);
    \draw[black] (zoombot0) -- (inset0.south east);

    \begin{axis}[width = 6cm, height = 4cm, scale only axis=true
                 ,xlabel = {time in $s$}
                 ,xmin = 0, xmax = 2
                 ,tick label style={font=\small}
                 ,label style={font=\small}
                 ,xtick={0, 0.25, 0.5, 0.75, 1.0, 1.25, 1.5, 1.75, 2.0}
                 ,xticklabels={0, , 0.5, , 1.0, , 1.5, , 2.0}
                 ,xshift = 0cm, yshift = -4.4cm
                 ,ymode=log, ymin = 0.0001, ymax = 100
                 ,ytickten={-4,-3,-2,-1,0,1,2}
                 ,unbounded coords=discard, filter discard warning=false
                 ,ylabel = {$\dfrac{|\tilde{E}_k^{\mathrm{sur}} - \tilde{E}_k^{\mathrm{bgk}}|}{\tilde{E}_k^{\mathrm{bgk}}}$ [\%]}
                ]%
        \addplot [colour1, very thick, dashed] table [x=time, y=abspct_naive] {\kepath kinetic_energy_tgv3d_w0.7.dat};
        \addplot [colour1, very thick, dotted] table [x=time, y=abspct_naive_doc] {\kepath kinetic_energy_tgv3d_w0.7.dat};
        \addplot [colour3, very thick, dashed] table [x=time, y=abspct_cons] {\kepath kinetic_energy_tgv3d_w0.7.dat};
        \addplot [colour3, very thick, dotted] table [x=time, y=abspct_cons_doc] {\kepath kinetic_energy_tgv3d_w0.7.dat};
        \addplot [colour4, very thick, dashed] table [x=time, y=abspct_sym] {\kepath kinetic_energy_tgv3d_w0.7.dat};
        \addplot [colour4, very thick, dotted] table [x=time, y=abspct_sym_doc] {\kepath kinetic_energy_tgv3d_w0.7.dat};
        \addplot [colour5, very thick, dashed] table [x=time, y=abspct_symcons] {\kepath kinetic_energy_tgv3d_w0.7.dat};
        \addplot [colour5, very thick, dotted] table [x=time, y=abspct_symcons_doc] {\kepath kinetic_energy_tgv3d_w0.7.dat};
    \end{axis}%

    \begin{axis}[width = 6cm, height = 6cm, scale only axis=true
                 ,title={$\omega_2 = 1.3$}
                 ,title style={font=\small}
                 ,xmin = 0, xmax = 2
                 ,ymin = 0.001
                 ,ymax = 0.125
                 ,ymode=log
                 ,ytickten={-3,-2,-1,0}
                 ,legend style={at={(1.05,0.1333)}, anchor=west, draw=none, font=\normalsize}
                 ,legend cell align={left}
                 ,tick label style={font=\small}
                 ,label style={font=\small}
                 ,xtick={0, 0.25, 0.5, 0.75, 1.0, 1.25, 1.5, 1.75, 2.0}
                 ,xticklabels={}
                 ,xshift = 7.5cm
                ]%
        \addplot [black, ultra thick] table [x=time, y=bgk] {\kepath kinetic_energy_tgv3d_w1.3.dat};
        \addlegendentry{BGK}%
        \addplot [colour1, very thick, dashed] table [x=time, y=naive] {\kepath kinetic_energy_tgv3d_w1.3.dat};
        \addlegendentry{Sur. NAIVE}%
        \addplot [colour1, very thick, dotted] table [x=time, y=naive_doc] {\kepath kinetic_energy_tgv3d_w1.3.dat};
        \addlegendentry{Sur. NAIVE DOC}%
        \addplot [colour3, very thick, dashed] table [x=time, y=cons] {\kepath kinetic_energy_tgv3d_w1.3.dat};
        \addlegendentry{Sur. CONS}%
        \addplot [colour3, very thick, dotted] table [x=time, y=cons_doc] {\kepath kinetic_energy_tgv3d_w1.3.dat};
        \addlegendentry{Sur. CONS DOC}%
        \addplot [colour4, very thick, dashed] table [x=time, y=sym] {\kepath kinetic_energy_tgv3d_w1.3.dat};
        \addlegendentry{Sur. SYM}%
        \addplot [colour4, very thick, dotted] table [x=time, y=sym_doc] {\kepath kinetic_energy_tgv3d_w1.3.dat};
        \addlegendentry{Sur. SYM DOC}%
        \addplot [colour5, very thick, dashed] table [x=time, y=symcons] {\kepath kinetic_energy_tgv3d_w1.3.dat};
        \addlegendentry{Sur. CONS+SYM}%
        \addplot [colour5, very thick, dotted] table [x=time, y=symcons_doc] {\kepath kinetic_energy_tgv3d_w1.3.dat};
        \addlegendentry{Sur. CONS+SYM DOC}%
        \draw[black, thick] (axis cs:1.995,1.3624e-02) rectangle (axis cs:2,1.3740e-02);
        \coordinate (zoomtop1) at (axis cs:1.995,1.3740e-02);
        \coordinate (zoombot1) at (axis cs:1.995,1.3624e-02);
    \end{axis}%

    \begin{axis}[width=2.8cm, height=2.8cm, scale only axis=true, ticks=none, name=inset1,
        xmin=1.995, xmax=2, ymin=1.3624e-02, ymax=1.3740e-02, xshift=8cm, yshift=0.5cm,
        restrict x to domain=1.949:2.001]  
        \addplot [black, semithick] table [x=time, y=bgk] {\kepath kinetic_energy_tgv3d_w1.3.dat};
        \addplot [colour1, semithick, dashed] table [x=time, y=naive] {\kepath kinetic_energy_tgv3d_w1.3.dat};
        \addplot [colour1, semithick, dotted] table [x=time, y=naive_doc] {\kepath kinetic_energy_tgv3d_w1.3.dat};
        \addplot [colour3, semithick, dashed] table [x=time, y=cons] {\kepath kinetic_energy_tgv3d_w1.3.dat};
        \addplot [colour3, semithick, dotted] table [x=time, y=cons_doc] {\kepath kinetic_energy_tgv3d_w1.3.dat};
        \addplot [colour4, semithick, dashed] table [x=time, y=sym] {\kepath kinetic_energy_tgv3d_w1.3.dat};
        \addplot [colour4, semithick, dotted] table [x=time, y=sym_doc] {\kepath kinetic_energy_tgv3d_w1.3.dat};
        \addplot [colour5, semithick, dashed] table [x=time, y=symcons] {\kepath kinetic_energy_tgv3d_w1.3.dat};
        \addplot [colour5, semithick, dotted] table [x=time, y=symcons_doc] {\kepath kinetic_energy_tgv3d_w1.3.dat};
    \end{axis}
    \draw[black] (zoomtop1) -- (inset1.north east);
    \draw[black] (zoombot1) -- (inset1.south east);

    \begin{axis}[width = 6cm, height = 4cm, scale only axis=true
                 ,xlabel = {time in $s$}
                 ,xmin = 0, xmax = 2
                 ,tick label style={font=\small}
                 ,label style={font=\small}
                 ,xtick={0, 0.25, 0.5, 0.75, 1.0, 1.25, 1.5, 1.75, 2.0}
                 ,xticklabels={0, , 0.5, , 1.0, , 1.5, , 2.0}
                 ,xshift = 7.5cm, yshift = -4.4cm
                 ,ymode=log, ymin = 0.0001, ymax = 100
                 ,ytickten={-4,-3,-2,-1,0,1,2}
                 ,unbounded coords=discard, filter discard warning=false
                ]%
        \addplot [colour1, very thick, dashed] table [x=time, y=abspct_naive] {\kepath kinetic_energy_tgv3d_w1.3.dat};
        \addplot [colour1, very thick, dotted] table [x=time, y=abspct_naive_doc] {\kepath kinetic_energy_tgv3d_w1.3.dat};
        \addplot [colour3, very thick, dashed] table [x=time, y=abspct_cons] {\kepath kinetic_energy_tgv3d_w1.3.dat};
        \addplot [colour3, very thick, dotted] table [x=time, y=abspct_cons_doc] {\kepath kinetic_energy_tgv3d_w1.3.dat};
        \addplot [colour4, very thick, dashed] table [x=time, y=abspct_sym] {\kepath kinetic_energy_tgv3d_w1.3.dat};
        \addplot [colour4, very thick, dotted] table [x=time, y=abspct_sym_doc] {\kepath kinetic_energy_tgv3d_w1.3.dat};
        \addplot [colour5, very thick, dashed] table [x=time, y=abspct_symcons] {\kepath kinetic_energy_tgv3d_w1.3.dat};
        \addplot [colour5, very thick, dotted] table [x=time, y=abspct_symcons_doc] {\kepath kinetic_energy_tgv3d_w1.3.dat};
    \end{axis}%

\end{tikzpicture}%

%% file: graphics/GridRefinement/grid_refinement.tikz
\pgfplotsset{compat=1.18}%
\providecommand{\gridpath}{}%
\begin{tikzpicture}
\begin{groupplot}[group style={group size=1 by 1, horizontal sep=0.5cm},
    width=6.5cm, height=6cm, xmode=log, ymode=log, log basis x=2, ymin=1e-2, ymax=5,
    xlabel={grid size $N$}, title style={font=\small}, label style={font=\small},
    tick label style={font=\footnotesize},
    legend style={at={(1.04,1)}, anchor=north west, draw=none, font=\small}, legend cell align=left]
\nextgroupplot[title={Re = 20, $\omega = 1.445$}, ylabel={$\varepsilon_2(\bm{a}, \bm{b})$ [\%]}, xtick={16,32,64,128}, xticklabels={16,32,64,128}]
    \addplot[colour1, very thick, mark=square*, mark size=2.2pt, mark options={solid, draw=colour1, fill=colour1}] table[x=N, y=BGK] {\gridpath grid_refinement_re20.dat};
    \addlegendentry{$\varepsilon_2(\bm{u}_{\mathrm{BGK}}, \bm{u}_{\mathrm{exact}})$}
    \addplot[colour2, very thick, mark=triangle*, mark size=2.8pt, mark options={solid, draw=colour2, fill=colour2}] table[x=N, y=CONSSYM] {\gridpath grid_refinement_re20.dat};
    \addlegendentry{$\varepsilon_2(\bm{u}_{\mathrm{CONS+SYM}}, \bm{u}_{\mathrm{exact}})$}
    \addplot[colour3, very thick, mark=*, mark size=2.2pt, mark options={solid, draw=colour3, fill=colour3}] table[x=N, y=CONSSYM_vs_BGK] {\gridpath grid_refinement_re20.dat};
    \addlegendentry{$\varepsilon_2(\bm{u}_{\mathrm{CONS+SYM}}, \bm{u}_{\mathrm{BGK}})$}
\end{groupplot}
\node[anchor=south west, font=\small] at ($(group c1r1.south east)+(0.25cm,0.2cm)$)
    {$\varepsilon_2(\bm{a}, \bm{b}) = \dfrac{\sqrt{\sum_{n=1}^{N^2} \mid \bm{a}_n - \bm{b}_n \mid^2}}{\sqrt{\sum_{n=1}^{N^2} \mid \bm{b}_n \mid^2}}$};
\end{tikzpicture}%

%% file: graphics/dsl_ke_enstrophy/dsl_ke_enstrophy_shots.tikz
\pgfplotsset{compat=1.18}%
\providecommand{\dslpath}{}%
\providecommand{\shotpath}{}%
\begin{tikzpicture}
\begin{groupplot}[group style={group size=3 by 1, horizontal sep=2.1cm},
    scale only axis, width=4.6cm, height=4.2cm, ylabel near ticks,
    title style={font=\small, yshift=1.1cm}, label style={font=\small}, ylabel style={font=\scriptsize}, tick label style={font=\footnotesize}]
\nextgroupplot[title={(a) Kinetic energy}, xmin=0, xmax=10, xlabel={$t$ [s]},
    scaled y ticks=false, yticklabel style={/pgf/number format/fixed, /pgf/number format/precision=3}, ytick distance=0.005,
    ylabel={$\dfrac{\tilde{E}_k^{\mathrm{sur}} - \tilde{E}_k^{\mathrm{bgk}}}{\tilde{E}_k^{\mathrm{bgk}}}$ [\%]}]
    \addplot[black, dashed, forget plot] coordinates {(0,0) (10,0)};
    \addplot[colour1, thick, solid, mark=*, mark size=1.8pt, mark options={solid, draw=colour1, fill=colour1}] table[x=time, y=ke_re100] {\dslpath dsl_corgi_rel_deviation.dat};
    \addplot[colour2, thick, solid, mark=square*, mark size=1.8pt, mark options={solid, draw=colour2, fill=colour2}] table[x=time, y=ke_re400] {\dslpath dsl_corgi_rel_deviation.dat};
    \addplot[colour3, thick, solid, mark=triangle*, mark size=1.8pt, mark options={solid, draw=colour3, fill=colour3}] table[x=time, y=ke_re1000] {\dslpath dsl_corgi_rel_deviation.dat};
    \addplot[colour4, thick, solid, mark=diamond*, mark size=1.8pt, mark options={solid, draw=colour4, fill=colour4}] table[x=time, y=ke_re2000] {\dslpath dsl_corgi_rel_deviation.dat};
\nextgroupplot[title={(b) Enstrophy}, xmin=0, xmax=10, xlabel={$t$ [s]},
    scaled y ticks=false, yticklabel style={/pgf/number format/fixed, /pgf/number format/precision=2}, ytick distance=0.05,
    ylabel={$\dfrac{\tilde{\mathcal{E}}^{\mathrm{sur}} - \tilde{\mathcal{E}}^{\mathrm{bgk}}}{\tilde{\mathcal{E}}^{\mathrm{bgk}}}$ [\%]}]
    \addplot[black, dashed, forget plot] coordinates {(0,0) (10,0)};
    \addplot[colour1, thick, solid, mark=*, mark size=1.8pt, mark options={solid, draw=colour1, fill=colour1}] table[x=time, y=enstrophy_re100] {\dslpath dsl_corgi_rel_deviation.dat};
    \addplot[colour2, thick, solid, mark=square*, mark size=1.8pt, mark options={solid, draw=colour2, fill=colour2}] table[x=time, y=enstrophy_re400] {\dslpath dsl_corgi_rel_deviation.dat};
    \addplot[colour3, thick, solid, mark=triangle*, mark size=1.8pt, mark options={solid, draw=colour3, fill=colour3}] table[x=time, y=enstrophy_re1000] {\dslpath dsl_corgi_rel_deviation.dat};
    \addplot[colour4, thick, solid, mark=diamond*, mark size=1.8pt, mark options={solid, draw=colour4, fill=colour4}] table[x=time, y=enstrophy_re2000] {\dslpath dsl_corgi_rel_deviation.dat};
\nextgroupplot[title={(c) Shot noise},
    xmode=log, ymode=log, xmin=240000, xmax=2.16e+11, clip=false,
    xlabel={Total shots per node and time step $72S$},
    ylabel={$\dfrac{\|\bm{u}_{\mathrm{sur}} - \bm{u}_{\mathrm{bgk}}\|_2}{\|\bm{u}_{\mathrm{bgk}}\|_2}$ [\%]},
    legend style={at={(-0.957,-0.33)}, anchor=north, draw=none, font=\small, legend columns=6, /tikz/every even column/.append style={column sep=0.3cm}}, legend cell align=left]
    \addplot[colour1, very thick, mark=*, mark size=2.2pt, mark options={solid, draw=colour1, fill=colour1}, unbounded coords=jump, error bars/.cd, y dir=both, y explicit, error bar style={colour1, thick}] table[x=total, y=l2_re100, y error minus=l2_lo_re100, y error plus=l2_hi_re100] {\shotpath shots_dsl.dat};
    \addlegendentry{Re $= 100$}
    \addplot[colour1, thick, densely dotted, forget plot] coordinates {(240000,1.729904e-01) (2.16e+11,1.729904e-01)};
    \addplot[only marks, mark=o, mark size=2.2pt, mark options={solid, draw=colour1, fill=white, thick}, forget plot] coordinates {(720000,1.729904e-01) (7.2e+06,1.729904e-01) (7.2e+07,1.729904e-01) (7.2e+08,1.729904e-01) (7.2e+09,1.729904e-01) (7.2e+10,1.729904e-01)};
    \addplot[colour2, very thick, mark=square*, mark size=2.2pt, mark options={solid, draw=colour2, fill=colour2}, unbounded coords=jump, error bars/.cd, y dir=both, y explicit, error bar style={colour2, thick}] table[x=total, y=l2_re400, y error minus=l2_lo_re400, y error plus=l2_hi_re400] {\shotpath shots_dsl.dat};
    \addlegendentry{Re $= 400$}
    \addplot[colour2, thick, densely dotted, forget plot] coordinates {(240000,2.635623e-01) (2.16e+11,2.635623e-01)};
    \addplot[only marks, mark=square, mark size=2.2pt, mark options={solid, draw=colour2, fill=white, thick}, forget plot] coordinates {(720000,2.635623e-01) (7.2e+06,2.635623e-01) (7.2e+07,2.635623e-01) (7.2e+08,2.635623e-01) (7.2e+09,2.635623e-01) (7.2e+10,2.635623e-01)};
    \addplot[colour3, very thick, mark=triangle*, mark size=2.2pt, mark options={solid, draw=colour3, fill=colour3}, unbounded coords=jump, error bars/.cd, y dir=both, y explicit, error bar style={colour3, thick}] table[x=total, y=l2_re1000, y error minus=l2_lo_re1000, y error plus=l2_hi_re1000] {\shotpath shots_dsl.dat};
    \addlegendentry{Re $= 1000$}
    \addplot[colour3, thick, densely dotted, forget plot] coordinates {(240000,2.675144e-01) (2.16e+11,2.675144e-01)};
    \addplot[only marks, mark=triangle, mark size=2.2pt, mark options={solid, draw=colour3, fill=white, thick}, forget plot] coordinates {(720000,2.675144e-01) (7.2e+06,2.675144e-01) (7.2e+07,2.675144e-01) (7.2e+08,2.675144e-01) (7.2e+09,2.675144e-01) (7.2e+10,2.675144e-01)};
    \addplot[colour4, very thick, mark=diamond*, mark size=2.2pt, mark options={solid, draw=colour4, fill=colour4}, unbounded coords=jump, error bars/.cd, y dir=both, y explicit, error bar style={colour4, thick}] table[x=total, y=l2_re2000, y error minus=l2_lo_re2000, y error plus=l2_hi_re2000] {\shotpath shots_dsl.dat};
    \addlegendentry{Re $= 2000$}
    \addplot[colour4, thick, densely dotted, forget plot] coordinates {(240000,2.756385e-01) (2.16e+11,2.756385e-01)};
    \addplot[only marks, mark=diamond, mark size=2.2pt, mark options={solid, draw=colour4, fill=white, thick}, forget plot] coordinates {(720000,2.756385e-01) (7.2e+06,2.756385e-01) (7.2e+07,2.756385e-01) (7.2e+08,2.756385e-01) (7.2e+09,2.756385e-01) (7.2e+10,2.756385e-01)};
    \addplot[only marks, mark=x, mark size=4pt, very thick, colour4, forget plot] coordinates {(720000,300)};
    \node[anchor=west, font=\scriptsize, text=colour4!80!black] at (axis cs:936000,300) {diverged};
    \addlegendimage{black, dashed}\addlegendentry{BGK (reference), (a) and (b)}
    \addlegendimage{black, thick, densely dotted, mark=o, mark options={solid, fill=white}}\addlegendentry{noise-free surrogate, (c)}
\end{groupplot}
\begin{axis}[at={(group c3r1.south west)}, anchor=south west, scale only axis, width=4.6cm, height=4.2cm, xmode=log, xmin=3333.33, xmax=3e+09, axis x line*=top, axis y line=none, ymin=0, ymax=1, tick label style={font=\footnotesize}, label style={font=\footnotesize}, xlabel={Shots per circuit execution $S$}]
\end{axis}
\end{tikzpicture}%

%% file: graphics/SymmetryErrorDOC/doc_vs_19.tikz
\pgfplotsset{compat=1.18}%
\providecommand{\docpath}{data/}%
\begin{tikzpicture}
\begin{axis}[name=pa, scale only axis, width=6.6cm, height=4.6cm, xmode=log, xmin=1e-10, xmax=0.1,
    ymin=0, ymax=102, ytick={0,20,40,60,80,100}, xtick={1e-10,1e-8,1e-6,1e-4,1e-2}, ylabel={cumulative share of tests [\%]}, 
    xlabel={symmetry error $|f_{g(i)}(g\bm{u}) - f_i(\bm{u})| / f_i(\bm{u})$},
    title={(a) axis directions $\bm{c}_1, \dots, \bm{c}_6$}, title style={font=\small}, label style={font=\small}, tick label style={font=\footnotesize},
    legend style={at={(0.03,0.97)}, anchor=north west, draw=none, fill=none, font=\footnotesize}, legend cell align=left]
    \addplot[colour1, very thick, densely dotted] table[x=x, y expr=100*\thisrow{19_axis}] {\docpath symmetry_cdf.dat};
    \addlegendentry{trained per direction}
    \addplot[colour1, very thick, dashed] table[x=x, y expr=100*\thisrow{DOC_axis}] {\docpath symmetry_cdf.dat};
    \addlegendentry{DOC}
\end{axis}
\begin{axis}[at={($(pa.south east)+(1.0cm,0)$)}, anchor=south west, scale only axis, width=6.6cm, height=4.6cm, xmode=log, xmin=1e-10, xmax=0.1,
    ymin=0, ymax=102, ytick={0,20,40,60,80,100}, xtick={1e-10,1e-8,1e-6,1e-4,1e-2}, yticklabels={}, 
    xlabel={symmetry error $|f_{g(i)}(g\bm{u}) - f_i(\bm{u})| / f_i(\bm{u})$},
    title={(b) diagonal directions $\bm{c}_7, \dots, \bm{c}_{18}$}, title style={font=\small}, label style={font=\small}, tick label style={font=\footnotesize},
    legend style={at={(0.03,0.97)}, anchor=north west, draw=none, fill=none, font=\footnotesize}, legend cell align=left]
    \addplot[colour1, very thick, densely dotted] table[x=x, y expr=100*\thisrow{19_diag}] {\docpath symmetry_cdf.dat};
    
    \addplot[colour1, very thick, dashed] table[x=x, y expr=100*\thisrow{DOC_diag}] {\docpath symmetry_cdf.dat};
    
\end{axis}
\end{tikzpicture}%

%% file: graphics/TrainingHistories/training_histories.tikz
\pgfplotsset{compat=1.18}%
\providecommand{\histpath}{}%
\begin{tikzpicture}
\begin{axis}[scale only axis, ymode=log, xmin=0.5, xmax=9.5, xtick={1,...,9}, xlabel={epoch}, set layers, title style={font=\small}, label style={font=\small}, tick label style={font=\footnotesize}, legend cell align=left, width=6.0cm, height=3.6cm, ymin=1e-11, ymax=0.0001, title={(a) D2Q9 PA SEL1 L6 DOC}, name=pa, ylabel={MSE}]
    \addplot[fill=gray!10, draw=none, forget plot, on layer=axis background] coordinates {(0.5,1e-11) (3.5,1e-11) (3.5,0.0001) (0.5,0.0001)} \closedcycle;
    \addplot[fill=gray!10, draw=none, forget plot, on layer=axis background] coordinates {(6.5,1e-11) (9.5,1e-11) (9.5,0.0001) (6.5,0.0001)} \closedcycle;
    \node[anchor=north, font=\scriptsize, text=gray] at (axis cs:2,0.0001) {lr $10^{-3}$};
    \node[anchor=north, font=\scriptsize, text=gray] at (axis cs:5,0.0001) {lr $10^{-4}$};
    \node[anchor=north, font=\scriptsize, text=gray] at (axis cs:8,0.0001) {lr $10^{-5}$};
    \addplot[colour1, very thick, solid, mark=*, mark size=2.2pt, mark options={solid, draw=colour1, fill=colour1}, forget plot] table[x=epoch, y=val_loss] {\histpath Corgi_DOC/Corgi_BGK_Parallel_rest_SEL1_L6.dat};
    \addplot[colour3, very thick, solid, mark=square*, mark size=2.2pt, mark options={solid, draw=colour3, fill=colour3}, forget plot] table[x=epoch, y=val_loss] {\histpath Corgi_DOC/Corgi_BGK_Parallel_axis_SEL1_L6.dat};
    \addplot[colour5, very thick, solid, mark=triangle*, mark size=2.2pt, mark options={solid, draw=colour5, fill=colour5}, forget plot] table[x=epoch, y=val_loss] {\histpath Corgi_DOC/Corgi_BGK_Parallel_diag_SEL1_L6.dat};
\end{axis}
\begin{axis}[scale only axis, ymode=log, xmin=0.5, xmax=9.5, xtick={1,...,9}, xlabel={epoch}, set layers, title style={font=\small}, label style={font=\small}, tick label style={font=\footnotesize}, legend cell align=left, width=6.0cm, height=3.6cm, ymin=1e-11, ymax=0.0001, title={(b) D3Q19 SPA BEL3 L6 DOC}, name=pb, at={($(pa.north east)+(0.6cm,0)$)}, anchor=north west, yticklabels={}, legend style={at={(1.04,1)}, anchor=north west, draw=none, font=\footnotesize}]
    \addplot[fill=gray!10, draw=none, forget plot, on layer=axis background] coordinates {(0.5,1e-11) (3.5,1e-11) (3.5,0.0001) (0.5,0.0001)} \closedcycle;
    \addplot[fill=gray!10, draw=none, forget plot, on layer=axis background] coordinates {(6.5,1e-11) (9.5,1e-11) (9.5,0.0001) (6.5,0.0001)} \closedcycle;
    \node[anchor=north, font=\scriptsize, text=gray] at (axis cs:2,0.0001) {lr $10^{-3}$};
    \node[anchor=north, font=\scriptsize, text=gray] at (axis cs:5,0.0001) {lr $10^{-4}$};
    \node[anchor=north, font=\scriptsize, text=gray] at (axis cs:8,0.0001) {lr $10^{-5}$};
    \addplot[colour1, very thick, solid, mark=*, mark size=2.2pt, mark options={solid, draw=colour1, fill=colour1}] table[x=epoch, y=val_loss] {\histpath LionDog_DOC/LionDog_BGK_SuperParallel_rest_BEL3_L6.dat};
    \addlegendentry{rest}
    \addplot[colour3, very thick, solid, mark=square*, mark size=2.2pt, mark options={solid, draw=colour3, fill=colour3}] table[x=epoch, y=val_loss] {\histpath LionDog_DOC/LionDog_BGK_SuperParallel_axis_BEL3_L6.dat};
    \addlegendentry{axis}
    \addplot[colour5, very thick, solid, mark=triangle*, mark size=2.2pt, mark options={solid, draw=colour5, fill=colour5}] table[x=epoch, y=val_loss] {\histpath LionDog_DOC/LionDog_BGK_SuperParallel_diag_BEL3_L6.dat};
    \addlegendentry{diagonal}
\end{axis}
\begin{axis}[scale only axis, ymode=log, xmin=0.5, xmax=9.5, xtick={1,...,9}, xlabel={epoch}, set layers, title style={font=\small}, label style={font=\small}, tick label style={font=\footnotesize}, legend cell align=left, width=12.6cm, height=7.0cm, ymin=3e-11, ymax=1e-05, name=pc, at={($(pa.south west)+(0,-2.2cm)$)}, anchor=north west, ylabel={MSE}, title={(c) D3Q19 SPA BEL3 L6 (19 circuits)}, legend style={at={(1.04,1)}, anchor=north west, draw=none, font=\footnotesize}, legend columns=1, legend style={row sep=-1.5pt}]
    \addplot[fill=gray!10, draw=none, forget plot, on layer=axis background] coordinates {(0.5,3e-11) (3.5,3e-11) (3.5,1e-05) (0.5,1e-05)} \closedcycle;
    \addplot[fill=gray!10, draw=none, forget plot, on layer=axis background] coordinates {(6.5,3e-11) (9.5,3e-11) (9.5,1e-05) (6.5,1e-05)} \closedcycle;
    \node[anchor=north, font=\scriptsize, text=gray] at (axis cs:2,1e-05) {lr $10^{-3}$};
    \node[anchor=north, font=\scriptsize, text=gray] at (axis cs:5,1e-05) {lr $10^{-4}$};
    \node[anchor=north, font=\scriptsize, text=gray] at (axis cs:8,1e-05) {lr $10^{-5}$};
    \addplot[colour1, semithick, solid, mark=*, mark size=1.8pt, mark options={solid, draw=colour1, fill=colour1}] table[x expr=\thisrow{epoch}-0.2000, y=val_loss] {\histpath LionDog_19dir/LionDog_BGK_SuperParallel_d3q0_BEL3_L6.dat};
    \addlegendentry{$f_{0}$}
    \addplot[colour3, semithick, solid, mark=square*, mark size=1.8pt, mark options={solid, draw=colour3, fill=colour3}] table[x expr=\thisrow{epoch}-0.1778, y=val_loss] {\histpath LionDog_19dir/LionDog_BGK_SuperParallel_d3q1_BEL3_L6.dat};
    \addlegendentry{$f_{1}$}
    \addplot[colour3, semithick, solid, mark=triangle*, mark size=1.8pt, mark options={solid, draw=colour3, fill=colour3}] table[x expr=\thisrow{epoch}-0.1556, y=val_loss] {\histpath LionDog_19dir/LionDog_BGK_SuperParallel_d3q2_BEL3_L6.dat};
    \addlegendentry{$f_{2}$}
    \addplot[colour3, semithick, solid, mark=diamond*, mark size=1.8pt, mark options={solid, draw=colour3, fill=colour3}] table[x expr=\thisrow{epoch}-0.1333, y=val_loss] {\histpath LionDog_19dir/LionDog_BGK_SuperParallel_d3q3_BEL3_L6.dat};
    \addlegendentry{$f_{3}$}
    \addplot[colour3, semithick, solid, mark=pentagon*, mark size=1.8pt, mark options={solid, draw=colour3, fill=colour3}] table[x expr=\thisrow{epoch}-0.1111, y=val_loss] {\histpath LionDog_19dir/LionDog_BGK_SuperParallel_d3q4_BEL3_L6.dat};
    \addlegendentry{$f_{4}$}
    \addplot[colour3, semithick, solid, mark=triangle*, mark size=1.8pt, mark options={solid, draw=colour3, fill=colour3, rotate=180}] table[x expr=\thisrow{epoch}-0.0889, y=val_loss] {\histpath LionDog_19dir/LionDog_BGK_SuperParallel_d3q5_BEL3_L6.dat};
    \addlegendentry{$f_{5}$}
    \addplot[colour3, semithick, solid, mark=asterisk, mark size=1.8pt, mark options={solid, draw=colour3, fill=white}] table[x expr=\thisrow{epoch}-0.0667, y=val_loss] {\histpath LionDog_19dir/LionDog_BGK_SuperParallel_d3q6_BEL3_L6.dat};
    \addlegendentry{$f_{6}$}
    \addplot[colour5, semithick, dashed, mark=o, mark size=1.8pt, mark options={solid, draw=colour5, fill=white}] table[x expr=\thisrow{epoch}-0.0444, y=val_loss] {\histpath LionDog_19dir/LionDog_BGK_SuperParallel_d3q7_BEL3_L6.dat};
    \addlegendentry{$f_{7}$}
    \addplot[colour5, semithick, dashed, mark=square, mark size=1.8pt, mark options={solid, draw=colour5, fill=white}] table[x expr=\thisrow{epoch}-0.0222, y=val_loss] {\histpath LionDog_19dir/LionDog_BGK_SuperParallel_d3q8_BEL3_L6.dat};
    \addlegendentry{$f_{8}$}
    \addplot[colour5, semithick, dashed, mark=triangle, mark size=1.8pt, mark options={solid, draw=colour5, fill=white}] table[x=epoch, y=val_loss] {\histpath LionDog_19dir/LionDog_BGK_SuperParallel_d3q9_BEL3_L6.dat};
    \addlegendentry{$f_{9}$}
    \addplot[colour5, semithick, dashed, mark=diamond, mark size=1.8pt, mark options={solid, draw=colour5, fill=white}] table[x expr=\thisrow{epoch}+0.0222, y=val_loss] {\histpath LionDog_19dir/LionDog_BGK_SuperParallel_d3q10_BEL3_L6.dat};
    \addlegendentry{$f_{10}$}
    \addplot[colour5, semithick, dashed, mark=pentagon, mark size=1.8pt, mark options={solid, draw=colour5, fill=white}] table[x expr=\thisrow{epoch}+0.0444, y=val_loss] {\histpath LionDog_19dir/LionDog_BGK_SuperParallel_d3q11_BEL3_L6.dat};
    \addlegendentry{$f_{11}$}
    \addplot[colour5, semithick, dashed, mark=triangle, mark size=1.8pt, mark options={solid, draw=colour5, fill=white, rotate=180}] table[x expr=\thisrow{epoch}+0.0667, y=val_loss] {\histpath LionDog_19dir/LionDog_BGK_SuperParallel_d3q12_BEL3_L6.dat};
    \addlegendentry{$f_{12}$}
    \addplot[colour5, semithick, densely dotted, mark=oplus, mark size=1.8pt, mark options={solid, draw=colour5, fill=white}] table[x expr=\thisrow{epoch}+0.0889, y=val_loss] {\histpath LionDog_19dir/LionDog_BGK_SuperParallel_d3q13_BEL3_L6.dat};
    \addlegendentry{$f_{13}$}
    \addplot[colour5, semithick, densely dotted, mark=otimes, mark size=1.8pt, mark options={solid, draw=colour5, fill=white}] table[x expr=\thisrow{epoch}+0.1111, y=val_loss] {\histpath LionDog_19dir/LionDog_BGK_SuperParallel_d3q14_BEL3_L6.dat};
    \addlegendentry{$f_{14}$}
    \addplot[colour5, semithick, densely dotted, mark=star, mark size=1.8pt, mark options={solid, draw=colour5, fill=white}] table[x expr=\thisrow{epoch}+0.1333, y=val_loss] {\histpath LionDog_19dir/LionDog_BGK_SuperParallel_d3q15_BEL3_L6.dat};
    \addlegendentry{$f_{15}$}
    \addplot[colour5, semithick, densely dotted, mark=Mercedes star, mark size=1.8pt, mark options={solid, draw=colour5, fill=white}] table[x expr=\thisrow{epoch}+0.1556, y=val_loss] {\histpath LionDog_19dir/LionDog_BGK_SuperParallel_d3q16_BEL3_L6.dat};
    \addlegendentry{$f_{16}$}
    \addplot[colour5, semithick, densely dotted, mark=x, mark size=1.8pt, mark options={solid, draw=colour5, fill=white}] table[x expr=\thisrow{epoch}+0.1778, y=val_loss] {\histpath LionDog_19dir/LionDog_BGK_SuperParallel_d3q17_BEL3_L6.dat};
    \addlegendentry{$f_{17}$}
    \addplot[colour5, semithick, densely dotted, mark=+, mark size=1.8pt, mark options={solid, draw=colour5, fill=white}] table[x expr=\thisrow{epoch}+0.2000, y=val_loss] {\histpath LionDog_19dir/LionDog_BGK_SuperParallel_d3q18_BEL3_L6.dat};
    \addlegendentry{$f_{18}$}
\end{axis}
\end{tikzpicture}%

%% file: graphics/D3Q19_analysis/stress_viscosity_tgv_doc.tikz
\small%
\pgfplotsset{
    boxbase/.style={
        boxplot/every box/.style={draw=black, solid},
        boxplot/every whisker/.style={black, solid},
        boxplot/every median/.style={black, solid, thick},
        no markers, forget plot,
    },
    vNAIVE/.style={boxbase, fill=colour1},
    vCONS/.style={boxbase, fill=colour3},
    vSYM/.style={boxbase, fill=colour4},
    vCONSSYM/.style={boxbase, fill=colour5},
    srcSYN/.style={},
    srcTGV/.style={boxplot/every box/.append style={postaction={pattern=north east lines, pattern color=black}}},
    panel/.style={
        scale only axis=true, title style={font=\footnotesize},
        every axis label/.append style={font=\scriptsize},
        every tick label/.append style={font=\scriptsize},
        tick align=inside, boxplot/draw direction=y,
    },
    variantaxis/.style={
        panel, width=2.93cm, height=3.6cm, xmin=0.4, xmax=4.6,
        xtick={1,2,3,4}, xticklabels={NAIVE, CONS, SYM, CONS+SYM},
        x tick label style={rotate=40, anchor=north east, font=\scriptsize},
    },
    directionaxis/.style={
        panel, width=13.2cm, height=3.4cm, xmin=0.4, xmax=19.6,
        xtick={1,...,19}, xticklabels={0,1,2,3,4,5,6,7,8,9,10,11,12,13,14,15,16,17,18}, xlabel={$c_i$},
    },
}%
\providecommand{\databox}[5]{%
    \pgfplotstablegetelem{#2}{lw}\of#1\let\dblw\pgfplotsretval
    \pgfplotstablegetelem{#2}{q1}\of#1\let\dbqa\pgfplotsretval
    \pgfplotstablegetelem{#2}{med}\of#1\let\dbmed\pgfplotsretval
    \pgfplotstablegetelem{#2}{q3}\of#1\let\dbqc\pgfplotsretval
    \pgfplotstablegetelem{#2}{uw}\of#1\let\dbuw\pgfplotsretval
    \edef\dbplot{\noexpand\addplot[#3, boxplot prepared={draw position=#4, box extend=#5,
        lower whisker=\dblw, lower quartile=\dbqa, median=\dbmed, upper quartile=\dbqc,
        upper whisker=\dbuw}] coordinates {};}%
    \dbplot
}%
\providecommand{\stressdatapath}{data/}%
\pgfplotstableread{\stressdatapath a_populations.dat}\tabapopulations
\pgfplotstableread{\stressdatapath b_increment.dat}\tabbincrement
\pgfplotstableread{\stressdatapath c_eq_stress.dat}\tabceqstress
\pgfplotstableread{\stressdatapath d_neq_stress.dat}\tabdneqstress
\pgfplotstableread{\stressdatapath e_viscosity.dat}\tabeviscosity
\pgfplotstableread{\stressdatapath e_inset.dat}\tabeinset
\begin{tikzpicture}
\begin{axis}[directionaxis, ymode=log, log basis y={10}, name=a,
    title={(a) post-collision populations $f_i^*$},
    ymin=1e-10, ymax=0.01, ytick={1e-10,1e-8,1e-6,1e-4,1e-2},
    ylabel={$\frac{|f_i^{*,\mathrm{sur}} - f_i^{*,\mathrm{bgk}}|}{|f_i^{*,\mathrm{bgk}}|}$},
]
    \databox{\tabapopulations}{0}{vNAIVE, srcSYN}{0.7000}{0.170}
    \databox{\tabapopulations}{1}{vNAIVE, srcSYN}{1.7000}{0.170}
    \databox{\tabapopulations}{2}{vNAIVE, srcSYN}{2.7000}{0.170}
    \databox{\tabapopulations}{3}{vNAIVE, srcSYN}{3.7000}{0.170}
    \databox{\tabapopulations}{4}{vNAIVE, srcSYN}{4.7000}{0.170}
    \databox{\tabapopulations}{5}{vNAIVE, srcSYN}{5.7000}{0.170}
    \databox{\tabapopulations}{6}{vNAIVE, srcSYN}{6.7000}{0.170}
    \databox{\tabapopulations}{7}{vNAIVE, srcSYN}{7.7000}{0.170}
    \databox{\tabapopulations}{8}{vNAIVE, srcSYN}{8.7000}{0.170}
    \databox{\tabapopulations}{9}{vNAIVE, srcSYN}{9.7000}{0.170}
    \databox{\tabapopulations}{10}{vNAIVE, srcSYN}{10.7000}{0.170}
    \databox{\tabapopulations}{11}{vNAIVE, srcSYN}{11.7000}{0.170}
    \databox{\tabapopulations}{12}{vNAIVE, srcSYN}{12.7000}{0.170}
    \databox{\tabapopulations}{13}{vNAIVE, srcSYN}{13.7000}{0.170}
    \databox{\tabapopulations}{14}{vNAIVE, srcSYN}{14.7000}{0.170}
    \databox{\tabapopulations}{15}{vNAIVE, srcSYN}{15.7000}{0.170}
    \databox{\tabapopulations}{16}{vNAIVE, srcSYN}{16.7000}{0.170}
    \databox{\tabapopulations}{17}{vNAIVE, srcSYN}{17.7000}{0.170}
    \databox{\tabapopulations}{18}{vNAIVE, srcSYN}{18.7000}{0.170}
    \databox{\tabapopulations}{19}{vCONS, srcSYN}{0.9000}{0.170}
    \databox{\tabapopulations}{20}{vCONS, srcSYN}{1.9000}{0.170}
    \databox{\tabapopulations}{21}{vCONS, srcSYN}{2.9000}{0.170}
    \databox{\tabapopulations}{22}{vCONS, srcSYN}{3.9000}{0.170}
    \databox{\tabapopulations}{23}{vCONS, srcSYN}{4.9000}{0.170}
    \databox{\tabapopulations}{24}{vCONS, srcSYN}{5.9000}{0.170}
    \databox{\tabapopulations}{25}{vCONS, srcSYN}{6.9000}{0.170}
    \databox{\tabapopulations}{26}{vCONS, srcSYN}{7.9000}{0.170}
    \databox{\tabapopulations}{27}{vCONS, srcSYN}{8.9000}{0.170}
    \databox{\tabapopulations}{28}{vCONS, srcSYN}{9.9000}{0.170}
    \databox{\tabapopulations}{29}{vCONS, srcSYN}{10.9000}{0.170}
    \databox{\tabapopulations}{30}{vCONS, srcSYN}{11.9000}{0.170}
    \databox{\tabapopulations}{31}{vCONS, srcSYN}{12.9000}{0.170}
    \databox{\tabapopulations}{32}{vCONS, srcSYN}{13.9000}{0.170}
    \databox{\tabapopulations}{33}{vCONS, srcSYN}{14.9000}{0.170}
    \databox{\tabapopulations}{34}{vCONS, srcSYN}{15.9000}{0.170}
    \databox{\tabapopulations}{35}{vCONS, srcSYN}{16.9000}{0.170}
    \databox{\tabapopulations}{36}{vCONS, srcSYN}{17.9000}{0.170}
    \databox{\tabapopulations}{37}{vCONS, srcSYN}{18.9000}{0.170}
    \databox{\tabapopulations}{38}{vSYM, srcSYN}{1.1000}{0.170}
    \databox{\tabapopulations}{39}{vSYM, srcSYN}{2.1000}{0.170}
    \databox{\tabapopulations}{40}{vSYM, srcSYN}{3.1000}{0.170}
    \databox{\tabapopulations}{41}{vSYM, srcSYN}{4.1000}{0.170}
    \databox{\tabapopulations}{42}{vSYM, srcSYN}{5.1000}{0.170}
    \databox{\tabapopulations}{43}{vSYM, srcSYN}{6.1000}{0.170}
    \databox{\tabapopulations}{44}{vSYM, srcSYN}{7.1000}{0.170}
    \databox{\tabapopulations}{45}{vSYM, srcSYN}{8.1000}{0.170}
    \databox{\tabapopulations}{46}{vSYM, srcSYN}{9.1000}{0.170}
    \databox{\tabapopulations}{47}{vSYM, srcSYN}{10.1000}{0.170}
    \databox{\tabapopulations}{48}{vSYM, srcSYN}{11.1000}{0.170}
    \databox{\tabapopulations}{49}{vSYM, srcSYN}{12.1000}{0.170}
    \databox{\tabapopulations}{50}{vSYM, srcSYN}{13.1000}{0.170}
    \databox{\tabapopulations}{51}{vSYM, srcSYN}{14.1000}{0.170}
    \databox{\tabapopulations}{52}{vSYM, srcSYN}{15.1000}{0.170}
    \databox{\tabapopulations}{53}{vSYM, srcSYN}{16.1000}{0.170}
    \databox{\tabapopulations}{54}{vSYM, srcSYN}{17.1000}{0.170}
    \databox{\tabapopulations}{55}{vSYM, srcSYN}{18.1000}{0.170}
    \databox{\tabapopulations}{56}{vSYM, srcSYN}{19.1000}{0.170}
    \databox{\tabapopulations}{57}{vCONSSYM, srcSYN}{1.3000}{0.170}
    \databox{\tabapopulations}{58}{vCONSSYM, srcSYN}{2.3000}{0.170}
    \databox{\tabapopulations}{59}{vCONSSYM, srcSYN}{3.3000}{0.170}
    \databox{\tabapopulations}{60}{vCONSSYM, srcSYN}{4.3000}{0.170}
    \databox{\tabapopulations}{61}{vCONSSYM, srcSYN}{5.3000}{0.170}
    \databox{\tabapopulations}{62}{vCONSSYM, srcSYN}{6.3000}{0.170}
    \databox{\tabapopulations}{63}{vCONSSYM, srcSYN}{7.3000}{0.170}
    \databox{\tabapopulations}{64}{vCONSSYM, srcSYN}{8.3000}{0.170}
    \databox{\tabapopulations}{65}{vCONSSYM, srcSYN}{9.3000}{0.170}
    \databox{\tabapopulations}{66}{vCONSSYM, srcSYN}{10.3000}{0.170}
    \databox{\tabapopulations}{67}{vCONSSYM, srcSYN}{11.3000}{0.170}
    \databox{\tabapopulations}{68}{vCONSSYM, srcSYN}{12.3000}{0.170}
    \databox{\tabapopulations}{69}{vCONSSYM, srcSYN}{13.3000}{0.170}
    \databox{\tabapopulations}{70}{vCONSSYM, srcSYN}{14.3000}{0.170}
    \databox{\tabapopulations}{71}{vCONSSYM, srcSYN}{15.3000}{0.170}
    \databox{\tabapopulations}{72}{vCONSSYM, srcSYN}{16.3000}{0.170}
    \databox{\tabapopulations}{73}{vCONSSYM, srcSYN}{17.3000}{0.170}
    \databox{\tabapopulations}{74}{vCONSSYM, srcSYN}{18.3000}{0.170}
    \databox{\tabapopulations}{75}{vCONSSYM, srcSYN}{19.3000}{0.170}
    \draw[gray, dotted] (axis cs:1.5,\pgfkeysvalueof{/pgfplots/ymin}) -- (axis cs:1.5,\pgfkeysvalueof{/pgfplots/ymax});
    \draw[gray, dotted] (axis cs:7.5,\pgfkeysvalueof{/pgfplots/ymin}) -- (axis cs:7.5,\pgfkeysvalueof{/pgfplots/ymax});
\end{axis}

\begin{axis}[directionaxis, ymode=log, log basis y={10}, name=b, at={($(a.south west)-(0,1.55cm)$)}, anchor=north west,
    title={(b) collision increment $\Delta f_i = f_i^* - f_i$},
    ymin=1e-08, ymax=0.1, ytick={1e-8,1e-6,1e-4,1e-2},
    ylabel={$\frac{|\Delta f_i^{\mathrm{sur}}(\bm{x}) - \Delta f_i^{\mathrm{bgk}}(\bm{x})|}{\|\Delta \bm{f}^{\mathrm{bgk}}(\bm{x})\|_2}$},
]
    \databox{\tabbincrement}{0}{vNAIVE, srcTGV}{0.7000}{0.170}
    \databox{\tabbincrement}{1}{vNAIVE, srcTGV}{1.7000}{0.170}
    \databox{\tabbincrement}{2}{vNAIVE, srcTGV}{2.7000}{0.170}
    \databox{\tabbincrement}{3}{vNAIVE, srcTGV}{3.7000}{0.170}
    \databox{\tabbincrement}{4}{vNAIVE, srcTGV}{4.7000}{0.170}
    \databox{\tabbincrement}{5}{vNAIVE, srcTGV}{5.7000}{0.170}
    \databox{\tabbincrement}{6}{vNAIVE, srcTGV}{6.7000}{0.170}
    \databox{\tabbincrement}{7}{vNAIVE, srcTGV}{7.7000}{0.170}
    \databox{\tabbincrement}{8}{vNAIVE, srcTGV}{8.7000}{0.170}
    \databox{\tabbincrement}{9}{vNAIVE, srcTGV}{9.7000}{0.170}
    \databox{\tabbincrement}{10}{vNAIVE, srcTGV}{10.7000}{0.170}
    \databox{\tabbincrement}{11}{vNAIVE, srcTGV}{11.7000}{0.170}
    \databox{\tabbincrement}{12}{vNAIVE, srcTGV}{12.7000}{0.170}
    \databox{\tabbincrement}{13}{vNAIVE, srcTGV}{13.7000}{0.170}
    \databox{\tabbincrement}{14}{vNAIVE, srcTGV}{14.7000}{0.170}
    \databox{\tabbincrement}{15}{vNAIVE, srcTGV}{15.7000}{0.170}
    \databox{\tabbincrement}{16}{vNAIVE, srcTGV}{16.7000}{0.170}
    \databox{\tabbincrement}{17}{vNAIVE, srcTGV}{17.7000}{0.170}
    \databox{\tabbincrement}{18}{vNAIVE, srcTGV}{18.7000}{0.170}
    \databox{\tabbincrement}{19}{vCONS, srcTGV}{0.9000}{0.170}
    \databox{\tabbincrement}{20}{vCONS, srcTGV}{1.9000}{0.170}
    \databox{\tabbincrement}{21}{vCONS, srcTGV}{2.9000}{0.170}
    \databox{\tabbincrement}{22}{vCONS, srcTGV}{3.9000}{0.170}
    \databox{\tabbincrement}{23}{vCONS, srcTGV}{4.9000}{0.170}
    \databox{\tabbincrement}{24}{vCONS, srcTGV}{5.9000}{0.170}
    \databox{\tabbincrement}{25}{vCONS, srcTGV}{6.9000}{0.170}
    \databox{\tabbincrement}{26}{vCONS, srcTGV}{7.9000}{0.170}
    \databox{\tabbincrement}{27}{vCONS, srcTGV}{8.9000}{0.170}
    \databox{\tabbincrement}{28}{vCONS, srcTGV}{9.9000}{0.170}
    \databox{\tabbincrement}{29}{vCONS, srcTGV}{10.9000}{0.170}
    \databox{\tabbincrement}{30}{vCONS, srcTGV}{11.9000}{0.170}
    \databox{\tabbincrement}{31}{vCONS, srcTGV}{12.9000}{0.170}
    \databox{\tabbincrement}{32}{vCONS, srcTGV}{13.9000}{0.170}
    \databox{\tabbincrement}{33}{vCONS, srcTGV}{14.9000}{0.170}
    \databox{\tabbincrement}{34}{vCONS, srcTGV}{15.9000}{0.170}
    \databox{\tabbincrement}{35}{vCONS, srcTGV}{16.9000}{0.170}
    \databox{\tabbincrement}{36}{vCONS, srcTGV}{17.9000}{0.170}
    \databox{\tabbincrement}{37}{vCONS, srcTGV}{18.9000}{0.170}
    \databox{\tabbincrement}{38}{vSYM, srcTGV}{1.1000}{0.170}
    \databox{\tabbincrement}{39}{vSYM, srcTGV}{2.1000}{0.170}
    \databox{\tabbincrement}{40}{vSYM, srcTGV}{3.1000}{0.170}
    \databox{\tabbincrement}{41}{vSYM, srcTGV}{4.1000}{0.170}
    \databox{\tabbincrement}{42}{vSYM, srcTGV}{5.1000}{0.170}
    \databox{\tabbincrement}{43}{vSYM, srcTGV}{6.1000}{0.170}
    \databox{\tabbincrement}{44}{vSYM, srcTGV}{7.1000}{0.170}
    \databox{\tabbincrement}{45}{vSYM, srcTGV}{8.1000}{0.170}
    \databox{\tabbincrement}{46}{vSYM, srcTGV}{9.1000}{0.170}
    \databox{\tabbincrement}{47}{vSYM, srcTGV}{10.1000}{0.170}
    \databox{\tabbincrement}{48}{vSYM, srcTGV}{11.1000}{0.170}
    \databox{\tabbincrement}{49}{vSYM, srcTGV}{12.1000}{0.170}
    \databox{\tabbincrement}{50}{vSYM, srcTGV}{13.1000}{0.170}
    \databox{\tabbincrement}{51}{vSYM, srcTGV}{14.1000}{0.170}
    \databox{\tabbincrement}{52}{vSYM, srcTGV}{15.1000}{0.170}
    \databox{\tabbincrement}{53}{vSYM, srcTGV}{16.1000}{0.170}
    \databox{\tabbincrement}{54}{vSYM, srcTGV}{17.1000}{0.170}
    \databox{\tabbincrement}{55}{vSYM, srcTGV}{18.1000}{0.170}
    \databox{\tabbincrement}{56}{vSYM, srcTGV}{19.1000}{0.170}
    \databox{\tabbincrement}{57}{vCONSSYM, srcTGV}{1.3000}{0.170}
    \databox{\tabbincrement}{58}{vCONSSYM, srcTGV}{2.3000}{0.170}
    \databox{\tabbincrement}{59}{vCONSSYM, srcTGV}{3.3000}{0.170}
    \databox{\tabbincrement}{60}{vCONSSYM, srcTGV}{4.3000}{0.170}
    \databox{\tabbincrement}{61}{vCONSSYM, srcTGV}{5.3000}{0.170}
    \databox{\tabbincrement}{62}{vCONSSYM, srcTGV}{6.3000}{0.170}
    \databox{\tabbincrement}{63}{vCONSSYM, srcTGV}{7.3000}{0.170}
    \databox{\tabbincrement}{64}{vCONSSYM, srcTGV}{8.3000}{0.170}
    \databox{\tabbincrement}{65}{vCONSSYM, srcTGV}{9.3000}{0.170}
    \databox{\tabbincrement}{66}{vCONSSYM, srcTGV}{10.3000}{0.170}
    \databox{\tabbincrement}{67}{vCONSSYM, srcTGV}{11.3000}{0.170}
    \databox{\tabbincrement}{68}{vCONSSYM, srcTGV}{12.3000}{0.170}
    \databox{\tabbincrement}{69}{vCONSSYM, srcTGV}{13.3000}{0.170}
    \databox{\tabbincrement}{70}{vCONSSYM, srcTGV}{14.3000}{0.170}
    \databox{\tabbincrement}{71}{vCONSSYM, srcTGV}{15.3000}{0.170}
    \databox{\tabbincrement}{72}{vCONSSYM, srcTGV}{16.3000}{0.170}
    \databox{\tabbincrement}{73}{vCONSSYM, srcTGV}{17.3000}{0.170}
    \databox{\tabbincrement}{74}{vCONSSYM, srcTGV}{18.3000}{0.170}
    \databox{\tabbincrement}{75}{vCONSSYM, srcTGV}{19.3000}{0.170}
    \draw[gray, dotted] (axis cs:1.5,\pgfkeysvalueof{/pgfplots/ymin}) -- (axis cs:1.5,\pgfkeysvalueof{/pgfplots/ymax});
    \draw[gray, dotted] (axis cs:7.5,\pgfkeysvalueof{/pgfplots/ymin}) -- (axis cs:7.5,\pgfkeysvalueof{/pgfplots/ymax});
\end{axis}

\begin{axis}[variantaxis, name=c, at={($(b.south west)-(0,1.55cm)$)}, anchor=north west,
    title={(c) equilibrium momentum flux}, ymode=log, log basis y={10}, ymin=0.0001, ymax=1, ytick={1e-4,1e-3,1e-2,1e-1,1},
    ylabel={$\frac{\left\|\mathrm{dev}\!(\rho\bm{u}\bm{u})^{\mathrm{sur}}-(\rho\bm{u}\bm{u})^{\mathrm{bgk}}\right\|_F}{\left\|\mathrm{dev}(\rho\bm{u}\bm{u})^{\mathrm{bgk}}\right\|_F}$},
]
    \databox{\tabceqstress}{0}{vNAIVE, srcSYN}{0.8000}{0.34}
    \databox{\tabceqstress}{1}{vNAIVE, srcTGV}{1.2000}{0.34}
    \databox{\tabceqstress}{2}{vCONS, srcSYN}{1.8000}{0.34}
    \databox{\tabceqstress}{3}{vCONS, srcTGV}{2.2000}{0.34}
    \databox{\tabceqstress}{4}{vSYM, srcSYN}{2.8000}{0.34}
    \databox{\tabceqstress}{5}{vSYM, srcTGV}{3.2000}{0.34}
    \databox{\tabceqstress}{6}{vCONSSYM, srcSYN}{3.8000}{0.34}
    \databox{\tabceqstress}{7}{vCONSSYM, srcTGV}{4.2000}{0.34}
    \addplot[black, dashed, forget plot] coordinates {(0.4,1) (4.6,1)};
\end{axis}

\begin{axis}[variantaxis, name=d, at={($(c.north east)+(2.3cm,0)$)}, anchor=north west,
    title={(d) non-eq. momentum flux}, ymode=log, log basis y={10}, ymin=1e-06, ymax=0.1, ytick={1e-6,1e-5,1e-4,1e-3,1e-2,1e-1},
    ylabel={$\frac{\|\mathrm{dev}(\bm{\Pi}^{\mathrm{D},*,\mathrm{neq,sur}}-\bm{\Pi}^{\mathrm{D},*,\mathrm{neq,bgk}})\|_F}{\omega\,\|\mathrm{dev}\bm{\Pi}^{\mathrm{D,neq}}\|_F}$},
]
    \databox{\tabdneqstress}{0}{vNAIVE, srcTGV}{1.0000}{0.6}
    \databox{\tabdneqstress}{1}{vCONS, srcTGV}{2.0000}{0.6}
    \databox{\tabdneqstress}{2}{vSYM, srcTGV}{3.0000}{0.6}
    \databox{\tabdneqstress}{3}{vCONSSYM, srcTGV}{4.0000}{0.6}
\end{axis}

\begin{axis}[variantaxis, name=e, at={($(d.north east)+(2.1cm,0)$)}, anchor=north west,
    title={(e) effective viscosity}, ymin=0.99, ymax=1.01, ytick={0.99,0.995,1,1.005,1.01},
    scaled y ticks=false, yticklabel style={/pgf/number format/fixed, /pgf/number format/precision=3},
    ylabel={$\left(1 - \frac{\omega}{2}\right)\frac{\nu_{\mathrm{eff}}}{\nu} + \frac{\omega}{2}$},
]
    \databox{\tabeviscosity}{0}{vNAIVE, srcTGV}{1.0000}{0.6}
    \databox{\tabeviscosity}{1}{vCONS, srcTGV}{2.0000}{0.6}
    \databox{\tabeviscosity}{2}{vSYM, srcTGV}{3.0000}{0.6}
    \databox{\tabeviscosity}{3}{vCONSSYM, srcTGV}{4.0000}{0.6}
    \addplot[black, dashed, forget plot] coordinates {(0.4,1) (4.6,1)};
\end{axis}
\begin{axis}[panel, at={($(e.south east)+(-0.08cm,0.08cm)$)}, anchor=south east, width=1.3cm, height=1.25cm,
    xmin=0.4, xmax=2.6, xtick=\empty, ymin=0.9996, ymax=1.0004, ytick={0.9997,1,1.0003},
    yticklabel pos=right, yticklabel style={font=\tiny, /pgf/number format/fixed, /pgf/number format/precision=4},
    scaled y ticks=false,
    axis background/.style={fill=white},
]
    \databox{\tabeinset}{0}{vSYM, srcTGV}{1.0000}{0.5}
    \databox{\tabeinset}{1}{vCONSSYM, srcTGV}{2.0000}{0.5}
    \addplot[black, dashed, forget plot] coordinates {(0.4,1) (2.6,1)};
\end{axis}
\coordinate (legendpos) at ({$(a.south west)!0.5!(a.south east)$} |- c.south);
\begin{axis}[hide axis, scale only axis, width=1pt, height=1pt, xmin=0, xmax=1, ymin=0, ymax=1,
    at={(legendpos)}, anchor=north, yshift=-1.9cm,
    legend columns=6, legend cell align=left,
    legend style={at={(0.5,1)}, anchor=north, draw=none, font=\scriptsize,
                  /tikz/every even column/.append style={column sep=0.3cm}},
]
    \addlegendimage{area legend, draw=black, fill=colour1}\addlegendentry{NAIVE}
    \addlegendimage{area legend, draw=black, fill=colour3}\addlegendentry{CONS}
    \addlegendimage{area legend, draw=black, fill=colour4}\addlegendentry{SYM}
    \addlegendimage{area legend, draw=black, fill=colour5}\addlegendentry{CONS+SYM}
    \addlegendimage{area legend, draw=black, fill=white}\addlegendentry{syn.\ data}
    \addlegendimage{area legend, draw=black, fill=white, postaction={pattern=north east lines}}\addlegendentry{TGV ($t = 0.5\,$s, $Re \approx 22$)}
\end{axis}
\end{tikzpicture}%

%% file: graphics/D2Q9_analysis/stress_viscosity_dsl1000_doc.tikz
\small%
\pgfplotsset{
    boxbase/.style={
        boxplot/every box/.style={draw=black, solid},
        boxplot/every whisker/.style={black, solid},
        boxplot/every median/.style={black, solid, thick},
        no markers, forget plot,
    },
    vNAIVE/.style={boxbase, fill=colour1},
    vCONS/.style={boxbase, fill=colour3},
    vSYM/.style={boxbase, fill=colour4},
    vCONSSYM/.style={boxbase, fill=colour5},
    srcSYN/.style={},
    srcFLOW/.style={boxplot/every box/.append style={postaction={pattern=north east lines, pattern color=black}}},
    panel/.style={
        scale only axis=true, title style={font=\footnotesize},
        every axis label/.append style={font=\scriptsize},
        every tick label/.append style={font=\scriptsize},
        tick align=inside, boxplot/draw direction=y,
    },
    variantaxis/.style={
        panel, width=2.93cm, height=3.6cm, xmin=0.4, xmax=4.6,
        xtick={1,2,3,4}, xticklabels={NAIVE, CONS, SYM, CONS+SYM},
        x tick label style={rotate=40, anchor=north east, font=\scriptsize},
    },
    directionaxis/.style={
        panel, width=13.2cm, height=3.4cm, xmin=0.4, xmax=9.6,
        xtick={1,...,9}, xticklabels={0,1,2,3,4,5,6,7,8}, xlabel={$c_i$},
    },
}%
\providecommand{\databox}[5]{%
    \pgfplotstablegetelem{#2}{lw}\of#1\let\dblw\pgfplotsretval
    \pgfplotstablegetelem{#2}{q1}\of#1\let\dbqa\pgfplotsretval
    \pgfplotstablegetelem{#2}{med}\of#1\let\dbmed\pgfplotsretval
    \pgfplotstablegetelem{#2}{q3}\of#1\let\dbqc\pgfplotsretval
    \pgfplotstablegetelem{#2}{uw}\of#1\let\dbuw\pgfplotsretval
    \edef\dbplot{\noexpand\addplot[#3, boxplot prepared={draw position=#4, box extend=#5,
        lower whisker=\dblw, lower quartile=\dbqa, median=\dbmed, upper quartile=\dbqc,
        upper whisker=\dbuw}] coordinates {};}%
    \dbplot
}%
\providecommand{\stressdatapath}{data/}%
\pgfplotstableread{\stressdatapath a_populations.dat}\tabapopulations
\pgfplotstableread{\stressdatapath b_increment.dat}\tabbincrement
\pgfplotstableread{\stressdatapath c_eq_stress.dat}\tabceqstress
\pgfplotstableread{\stressdatapath d_neq_stress.dat}\tabdneqstress
\pgfplotstableread{\stressdatapath e_viscosity.dat}\tabeviscosity
\pgfplotstableread{\stressdatapath e_inset.dat}\tabeinset
\begin{tikzpicture}
\begin{axis}[directionaxis, ymode=log, log basis y={10}, name=a,
    title={(a) post-collision populations $f_i^*$},
    ymin=1e-12, ymax=0.01, ytick={1e-12,1e-10,1e-8,1e-6,1e-4,1e-2},
    ylabel={$\frac{|f_i^{*,\mathrm{sur}} - f_i^{*,\mathrm{bgk}}|}{|f_i^{*,\mathrm{bgk}}|}$},
]
    \databox{\tabapopulations}{0}{vNAIVE, srcSYN}{0.7000}{0.170}
    \databox{\tabapopulations}{1}{vNAIVE, srcSYN}{1.7000}{0.170}
    \databox{\tabapopulations}{2}{vNAIVE, srcSYN}{2.7000}{0.170}
    \databox{\tabapopulations}{3}{vNAIVE, srcSYN}{3.7000}{0.170}
    \databox{\tabapopulations}{4}{vNAIVE, srcSYN}{4.7000}{0.170}
    \databox{\tabapopulations}{5}{vNAIVE, srcSYN}{5.7000}{0.170}
    \databox{\tabapopulations}{6}{vNAIVE, srcSYN}{6.7000}{0.170}
    \databox{\tabapopulations}{7}{vNAIVE, srcSYN}{7.7000}{0.170}
    \databox{\tabapopulations}{8}{vNAIVE, srcSYN}{8.7000}{0.170}
    \databox{\tabapopulations}{9}{vCONS, srcSYN}{0.9000}{0.170}
    \databox{\tabapopulations}{10}{vCONS, srcSYN}{1.9000}{0.170}
    \databox{\tabapopulations}{11}{vCONS, srcSYN}{2.9000}{0.170}
    \databox{\tabapopulations}{12}{vCONS, srcSYN}{3.9000}{0.170}
    \databox{\tabapopulations}{13}{vCONS, srcSYN}{4.9000}{0.170}
    \databox{\tabapopulations}{14}{vCONS, srcSYN}{5.9000}{0.170}
    \databox{\tabapopulations}{15}{vCONS, srcSYN}{6.9000}{0.170}
    \databox{\tabapopulations}{16}{vCONS, srcSYN}{7.9000}{0.170}
    \databox{\tabapopulations}{17}{vCONS, srcSYN}{8.9000}{0.170}
    \databox{\tabapopulations}{18}{vSYM, srcSYN}{1.1000}{0.170}
    \databox{\tabapopulations}{19}{vSYM, srcSYN}{2.1000}{0.170}
    \databox{\tabapopulations}{20}{vSYM, srcSYN}{3.1000}{0.170}
    \databox{\tabapopulations}{21}{vSYM, srcSYN}{4.1000}{0.170}
    \databox{\tabapopulations}{22}{vSYM, srcSYN}{5.1000}{0.170}
    \databox{\tabapopulations}{23}{vSYM, srcSYN}{6.1000}{0.170}
    \databox{\tabapopulations}{24}{vSYM, srcSYN}{7.1000}{0.170}
    \databox{\tabapopulations}{25}{vSYM, srcSYN}{8.1000}{0.170}
    \databox{\tabapopulations}{26}{vSYM, srcSYN}{9.1000}{0.170}
    \databox{\tabapopulations}{27}{vCONSSYM, srcSYN}{1.3000}{0.170}
    \databox{\tabapopulations}{28}{vCONSSYM, srcSYN}{2.3000}{0.170}
    \databox{\tabapopulations}{29}{vCONSSYM, srcSYN}{3.3000}{0.170}
    \databox{\tabapopulations}{30}{vCONSSYM, srcSYN}{4.3000}{0.170}
    \databox{\tabapopulations}{31}{vCONSSYM, srcSYN}{5.3000}{0.170}
    \databox{\tabapopulations}{32}{vCONSSYM, srcSYN}{6.3000}{0.170}
    \databox{\tabapopulations}{33}{vCONSSYM, srcSYN}{7.3000}{0.170}
    \databox{\tabapopulations}{34}{vCONSSYM, srcSYN}{8.3000}{0.170}
    \databox{\tabapopulations}{35}{vCONSSYM, srcSYN}{9.3000}{0.170}
    \draw[gray, dotted] (axis cs:1.5,\pgfkeysvalueof{/pgfplots/ymin}) -- (axis cs:1.5,\pgfkeysvalueof{/pgfplots/ymax});
    \draw[gray, dotted] (axis cs:5.5,\pgfkeysvalueof{/pgfplots/ymin}) -- (axis cs:5.5,\pgfkeysvalueof{/pgfplots/ymax});
\end{axis}

\begin{axis}[directionaxis, ymode=log, log basis y={10}, name=b, at={($(a.south west)-(0,1.55cm)$)}, anchor=north west,
    title={(b) collision increment $\Delta f_i = f_i^* - f_i$},
    ymin=1e-08, ymax=1, ytick={1e-8,1e-6,1e-4,1e-2,1},
    ylabel={$\frac{|\Delta f_i^{\mathrm{sur}}(\bm{x}) - \Delta f_i^{\mathrm{bgk}}(\bm{x})|}{\|\Delta \bm{f}^{\mathrm{bgk}}(\bm{x})\|_2}$},
]
    \databox{\tabbincrement}{0}{vNAIVE, srcFLOW}{0.7000}{0.170}
    \databox{\tabbincrement}{1}{vNAIVE, srcFLOW}{1.7000}{0.170}
    \databox{\tabbincrement}{2}{vNAIVE, srcFLOW}{2.7000}{0.170}
    \databox{\tabbincrement}{3}{vNAIVE, srcFLOW}{3.7000}{0.170}
    \databox{\tabbincrement}{4}{vNAIVE, srcFLOW}{4.7000}{0.170}
    \databox{\tabbincrement}{5}{vNAIVE, srcFLOW}{5.7000}{0.170}
    \databox{\tabbincrement}{6}{vNAIVE, srcFLOW}{6.7000}{0.170}
    \databox{\tabbincrement}{7}{vNAIVE, srcFLOW}{7.7000}{0.170}
    \databox{\tabbincrement}{8}{vNAIVE, srcFLOW}{8.7000}{0.170}
    \databox{\tabbincrement}{9}{vCONS, srcFLOW}{0.9000}{0.170}
    \databox{\tabbincrement}{10}{vCONS, srcFLOW}{1.9000}{0.170}
    \databox{\tabbincrement}{11}{vCONS, srcFLOW}{2.9000}{0.170}
    \databox{\tabbincrement}{12}{vCONS, srcFLOW}{3.9000}{0.170}
    \databox{\tabbincrement}{13}{vCONS, srcFLOW}{4.9000}{0.170}
    \databox{\tabbincrement}{14}{vCONS, srcFLOW}{5.9000}{0.170}
    \databox{\tabbincrement}{15}{vCONS, srcFLOW}{6.9000}{0.170}
    \databox{\tabbincrement}{16}{vCONS, srcFLOW}{7.9000}{0.170}
    \databox{\tabbincrement}{17}{vCONS, srcFLOW}{8.9000}{0.170}
    \databox{\tabbincrement}{18}{vSYM, srcFLOW}{1.1000}{0.170}
    \databox{\tabbincrement}{19}{vSYM, srcFLOW}{2.1000}{0.170}
    \databox{\tabbincrement}{20}{vSYM, srcFLOW}{3.1000}{0.170}
    \databox{\tabbincrement}{21}{vSYM, srcFLOW}{4.1000}{0.170}
    \databox{\tabbincrement}{22}{vSYM, srcFLOW}{5.1000}{0.170}
    \databox{\tabbincrement}{23}{vSYM, srcFLOW}{6.1000}{0.170}
    \databox{\tabbincrement}{24}{vSYM, srcFLOW}{7.1000}{0.170}
    \databox{\tabbincrement}{25}{vSYM, srcFLOW}{8.1000}{0.170}
    \databox{\tabbincrement}{26}{vSYM, srcFLOW}{9.1000}{0.170}
    \databox{\tabbincrement}{27}{vCONSSYM, srcFLOW}{1.3000}{0.170}
    \databox{\tabbincrement}{28}{vCONSSYM, srcFLOW}{2.3000}{0.170}
    \databox{\tabbincrement}{29}{vCONSSYM, srcFLOW}{3.3000}{0.170}
    \databox{\tabbincrement}{30}{vCONSSYM, srcFLOW}{4.3000}{0.170}
    \databox{\tabbincrement}{31}{vCONSSYM, srcFLOW}{5.3000}{0.170}
    \databox{\tabbincrement}{32}{vCONSSYM, srcFLOW}{6.3000}{0.170}
    \databox{\tabbincrement}{33}{vCONSSYM, srcFLOW}{7.3000}{0.170}
    \databox{\tabbincrement}{34}{vCONSSYM, srcFLOW}{8.3000}{0.170}
    \databox{\tabbincrement}{35}{vCONSSYM, srcFLOW}{9.3000}{0.170}
    \draw[gray, dotted] (axis cs:1.5,\pgfkeysvalueof{/pgfplots/ymin}) -- (axis cs:1.5,\pgfkeysvalueof{/pgfplots/ymax});
    \draw[gray, dotted] (axis cs:5.5,\pgfkeysvalueof{/pgfplots/ymin}) -- (axis cs:5.5,\pgfkeysvalueof{/pgfplots/ymax});
\end{axis}

\begin{axis}[variantaxis, name=c, at={($(b.south west)-(0,1.55cm)$)}, anchor=north west,
    title={(c) equilibrium momentum flux}, ymode=log, log basis y={10}, ymin=1e-05, ymax=0.1, ytick={1e-5,1e-4,1e-3,1e-2,1e-1},
    ylabel={$\frac{\left\|\mathrm{dev}\!(\rho\bm{u}\bm{u})^{\mathrm{sur}}-(\rho\bm{u}\bm{u})^{\mathrm{bgk}}\right\|_F}{\left\|\mathrm{dev}(\rho\bm{u}\bm{u})^{\mathrm{bgk}}\right\|_F}$},
]
    \databox{\tabceqstress}{0}{vNAIVE, srcSYN}{0.8000}{0.34}
    \databox{\tabceqstress}{1}{vNAIVE, srcFLOW}{1.2000}{0.34}
    \databox{\tabceqstress}{2}{vCONS, srcSYN}{1.8000}{0.34}
    \databox{\tabceqstress}{3}{vCONS, srcFLOW}{2.2000}{0.34}
    \databox{\tabceqstress}{4}{vSYM, srcSYN}{2.8000}{0.34}
    \databox{\tabceqstress}{5}{vSYM, srcFLOW}{3.2000}{0.34}
    \databox{\tabceqstress}{6}{vCONSSYM, srcSYN}{3.8000}{0.34}
    \databox{\tabceqstress}{7}{vCONSSYM, srcFLOW}{4.2000}{0.34}
    \addplot[black, dashed, forget plot] coordinates {(0.4,1) (4.6,1)};
\end{axis}

\begin{axis}[variantaxis, name=d, at={($(c.north east)+(2.3cm,0)$)}, anchor=north west,
    title={(d) non-eq.\ momentum flux}, ymode=log, log basis y={10}, ymin=1e-06, ymax=0.1, ytick={1e-6,1e-5,1e-4,1e-3,1e-2,1e-1},
    ylabel={$\frac{\|\mathrm{dev}(\bm{\Pi}^{\mathrm{D},*,\mathrm{neq,sur}}-\bm{\Pi}^{\mathrm{D},*,\mathrm{neq,bgk}})\|_F}{\omega\,\|\mathrm{dev}\bm{\Pi}^{\mathrm{D,neq}}\|_F}$}
]
    \databox{\tabdneqstress}{0}{vNAIVE, srcFLOW}{1.0000}{0.6}
    \databox{\tabdneqstress}{1}{vCONS, srcFLOW}{2.0000}{0.6}
    \databox{\tabdneqstress}{2}{vSYM, srcFLOW}{3.0000}{0.6}
    \databox{\tabdneqstress}{3}{vCONSSYM, srcFLOW}{4.0000}{0.6}
\end{axis}

\begin{axis}[variantaxis, name=e, at={($(d.north east)+(2.1cm,0)$)}, anchor=north west,
    title={(e) effective viscosity}, ymin=0.98, ymax=1.02, ytick={0.98,0.99,1,1.01,1.02},
    scaled y ticks=false, yticklabel style={/pgf/number format/fixed, /pgf/number format/precision=2},
    ylabel={$\left(1 - \frac{\omega}{2}\right)\frac{\nu_{\mathrm{eff}}}{\nu} + \frac{\omega}{2}$},
]
    \databox{\tabeviscosity}{0}{vNAIVE, srcFLOW}{1.0000}{0.6}
    \databox{\tabeviscosity}{1}{vCONS, srcFLOW}{2.0000}{0.6}
    \databox{\tabeviscosity}{2}{vSYM, srcFLOW}{3.0000}{0.6}
    \databox{\tabeviscosity}{3}{vCONSSYM, srcFLOW}{4.0000}{0.6}
    \addplot[black, dashed, forget plot] coordinates {(0.4,1) (4.6,1)};
\end{axis}
\coordinate (legendpos) at ({$(a.south west)!0.5!(a.south east)$} |- c.south);
\begin{axis}[hide axis, scale only axis, width=1pt, height=1pt, xmin=0, xmax=1, ymin=0, ymax=1,
    at={(legendpos)}, anchor=north, yshift=-1.9cm,
    legend columns=6, legend cell align=left,
    legend style={at={(0.5,1)}, anchor=north, draw=none, font=\scriptsize,
                  /tikz/every even column/.append style={column sep=0.3cm}},
]
    \addlegendimage{area legend, draw=black, fill=colour1}\addlegendentry{NAIVE}
    \addlegendimage{area legend, draw=black, fill=colour3}\addlegendentry{CONS}
    \addlegendimage{area legend, draw=black, fill=colour4}\addlegendentry{SYM}
    \addlegendimage{area legend, draw=black, fill=colour5}\addlegendentry{CONS+SYM}
    \addlegendimage{area legend, draw=black, fill=white}\addlegendentry{syn.\ data}
    \addlegendimage{area legend, draw=black, fill=white, postaction={pattern=north east lines}}\addlegendentry{DSL ($t = 10\,$s, $Re=1000$)}
\end{axis}
\end{tikzpicture}%